\documentclass{jfm}

\usepackage{graphicx}
\usepackage{natbib}
\usepackage{epstopdf, epsfig}
\usepackage{amsmath}
\usepackage{amssymb}
\usepackage{latexsym}
\usepackage{wasysym}
\usepackage{multirow}
\usepackage{color}
\usepackage{subfigure}
\usepackage{pdflscape}
\usepackage{multirow}
\usepackage{dashrule}
\usepackage{tikz}
\usepackage{cancel}
\usetikzlibrary{arrows}

\def\includefigs{\let\ifincfigs=\iftrue}
\def\noincludefigs{\let\ifincfigs=\iffalse}
\includefigs

\newbox\epsfvertlab
\newbox\epsfhorlab
\newbox\epsffiglab

\newdimen\epsfvlabsize
\newdimen\scott

\def\setvlabel#1{\setbox\epsfvertlab=\vbox{\hbox{#1}}}%
\def\sethlabel#1{\setbox\epsfhorlab=\vbox{\hbox{#1}}}%
\def\figlab#1 #2 #3{\setbox\epsffiglab=\vbox to 0pt{%
\ifvoid\epsffiglab\else\box\epsffiglab\fi\vss\hbox to 0pt{\raise #2 \hbox{\hskip #1 #3}\hss}}}

\newdimen\fighor
\newdimen\figver
\newbox\rotbox
\long\def\lrlap#1{\hbox to 0pt{#1\hss}}
\long\def\verttex#1#2#3{{\fighor = #1\figver = #2\vbox to \figver{\vss%
\hbox to \fighor{\hfill\hsize=\fighor%
\lrlap{\rotstart{-90 rotate}\vbox to \fighor{#3\vfil}\rotfinish}}}}}

\def\dvipsvspec#1{\special{ps:#1}}%  passes #1 verbatim to the output
\def\dvipsrotstart#1{\dvipsvspec{gsave currentpoint currentpoint translate
   #1 neg exch neg exch translate}}% #1 can be any origin-fixing transformation
\def\dvipsrotfinish{\dvipsvspec{currentpoint grestore moveto}}% gets back in synch

\def\rotstart#1{\dvipsrotstart{#1}}
\def\rotfinish{\dvipsrotfinish}

\def\epsfsetlab{%
\ifvoid\epsfvertlab%
\else%
\verttex{\epsfvlabsize}{\epsfysize}%
{\hbox to \epsfysize{\hss\box\epsfvertlab\hss}}%
\fi%
\ifvoid\epsfhorlab%
\else%
\scott=\epsfxsize%
\advance\scott by \epsfvlabsize%
\rlap{\vtop{\hrule height0pt\hbox to \scott{\hss\box\epsfhorlab\hss}}}%
\fi%
}

\def\epsfsetover{\ifvoid\epsffiglab\else\box\epsffiglab\fi}

\newread\epsffilein    % file to \read
\newif\ifepsffileok    % continue looking for the bounding box?
\newif\ifepsfbbfound   % success?
\newif\ifepsfverbose   % report what you're making?
\newdimen\epsfxsize    % horizontal size after scaling
\newdimen\epsfysize    % vertical size after scaling
\newdimen\epsftsize    % horizontal size before scaling
\newdimen\epsfrsize    % vertical size before scaling
\newdimen\epsftmp      % register for arithmetic manipulation
\newdimen\pspoints     % conversion factor
\def\epsfbox#1{
   \ifvoid\epsfvertlab%
   \else\epsfvlabsize=\ht\epsfvertlab \advance\epsfvlabsize by \dp\epsfvertlab\fi%
   \leavevmode\global\def\epsfllx{72}\global\def\epsflly{72}%
   \global\def\epsfurx{540}\global\def\epsfury{720}%
   \def\lbracket{[}\def\testit{#1}\ifx\testit\lbracket
   \let\next=\epsfgetlitbb\else\let\next=\epsfnormal\fi\next{#1}}%
\def\epsfgetlitbb#1#2 #3 #4 #5]#6{\epsfgrab #2 #3 #4 #5 .\\%
   \epsfsetgraph{#6}}%
\def\epsfnormal#1{\epsfgetbb{#1}\epsfsetgraph{#1}}%
\def\epsfgetbb#1{%
\openin\epsffilein=#1
\ifeof\epsffilein\errmessage{I couldn't open #1, will ignore it}\else
   {\epsffileoktrue \chardef\other=12
    \def\do##1{\catcode`##1=\other}\dospecials \catcode`\ =10
    \loop
       \read\epsffilein to \epsffileline
       \ifeof\epsffilein\epsffileokfalse\else
          \expandafter\epsfaux\epsffileline:. \\%
       \fi
   \ifepsffileok\repeat
   \ifepsfbbfound\else
    \ifepsfverbose\message{No bounding box comment in #1; using defaults}\fi\fi
   }\closein\epsffilein\fi}%
\def\epsfsetgraph#1{%
   \epsfrsize=\epsfury\pspoints
   \advance\epsfrsize by-\epsflly\pspoints
   \epsftsize=\epsfurx\pspoints
   \advance\epsftsize by-\epsfllx\pspoints
   \epsfxsize\epsfsize\epsftsize\epsfrsize
   \ifnum\epsfxsize=0 \ifnum\epsfysize=0
      \epsfxsize=\epsftsize \epsfysize=\epsfrsize
     \else\epsftmp=\epsftsize \divide\epsftmp\epsfrsize
       \epsfxsize=\epsfysize \multiply\epsfxsize\epsftmp
       \multiply\epsftmp\epsfrsize \advance\epsftsize-\epsftmp
       \epsftmp=\epsfysize
       \loop \advance\epsftsize\epsftsize \divide\epsftmp 2
       \ifnum\epsftmp>0
          \ifnum\epsftsize<\epsfrsize\else
             \advance\epsftsize-\epsfrsize \advance\epsfxsize\epsftmp \fi
       \repeat
     \fi
   \else\epsftmp=\epsfrsize \divide\epsftmp\epsftsize
     \epsfysize=\epsfxsize \multiply\epsfysize\epsftmp   
     \multiply\epsftmp\epsftsize \advance\epsfrsize-\epsftmp
     \epsftmp=\epsfxsize
     \loop \advance\epsfrsize\epsfrsize \divide\epsftmp 2
     \ifnum\epsftmp>0
        \ifnum\epsfrsize<\epsftsize\else
           \advance\epsfrsize-\epsftsize \advance\epsfysize\epsftmp \fi
     \repeat     
   \fi
   \ifepsfverbose\message{#1: width=\the\epsfxsize, height=\the\epsfysize}\fi
   \epsftmp=10\epsfxsize \divide\epsftmp\pspoints
   \epsfsetlab%
   \ifincfigs%
     \vbox to\epsfysize{\vfil\hbox to\epsfxsize{%
        \includegraphics{#1}%
        \epsfsetover\hfil}}%
   \else%
     \epsfsetover%
     \vbox to\epsfysize{\hrule\vss\hbox to\epsfxsize{\vrule height
                        \epsfysize\hfil\vrule}\vss\hrule}%
   \fi%
\epsfxsize=0pt\epsfysize=0pt}%

{\catcode`\%=12 \global\let\epsfpercent=%\global\def\epsfbblit{%BoundingBox}}%
\long\def\epsfaux#1#2:#3\\{\ifx#1\epsfpercent
   \def\testit{#2}\ifx\testit\epsfbblit
      \epsfgrab #3 . . . \\%
      \epsffileokfalse
      \global\epsfbbfoundtrue
   \fi\else\ifx#1\par\else\epsffileokfalse\fi\fi}%
\def\epsfgrab #1 #2 #3 #4 #5\\{%
   \global\def\epsfllx{#1}\ifx\epsfllx\empty
      \epsfgrab #2 #3 #4 #5 .\\\else
   \global\def\epsflly{#2}%
   \global\def\epsfurx{#3}\global\def\epsfury{#4}\fi}%
\def\epsfsize#1#2{\epsfxsize}
\def\ifspace{\ifcat\issp.\else~\fi}
\def\tspace{\futurelet\issp\ifspace}
\def\a{({\it a\kern 1pt})\tspace}
\def\b{({\it b\kern 1pt})\tspace}
\def\c{({\it c\kern 1pt})\tspace}
\def\d{({\it d\kern 1pt})\tspace}
\def\e{({\it e\kern 1pt})\tspace}
\def\f{({\it f\kern 1pt})\tspace}
\def\g{({\it g\kern 1pt})\tspace}
\def\h{({\it h\kern 1pt})\tspace}
\def\i{({\it i\kern 1pt})\tspace}
\def\j{({\it j\kern 1pt})\tspace}
\def\abc#1{({\it #1\kern 1pt})\tspace}

\newcount\ndots
\def\drawline#1#2{\raise 2.5pt\vbox{\hrule width #1pt height #2pt}}

\def\trian{\raise 1.25pt\hbox{$\scriptscriptstyle\triangle$}\nobreak\ }
\def\solidtrian{\raise 1.25pt
\hbox to 3bp{
\hfill}\nobreak\ }
\def\dsolidtrian{\raise 1.25pt
\hbox to 3bp{
\hfill}\nobreak\ }
\def\soliddiamond{\raise 1.25pt
\hbox to 4bp{
\hfill}\nobreak\ }

\def\square{${\vcenter{\hrule height .4pt 
              \hbox{\vrule width .4pt height 3pt \kern 3pt \vrule width .4pt}
          \hrule height .4pt}}$\nobreak\ }

\def\plus{\raise 1.25pt \hbox{$\scriptscriptstyle +$}\nobreak\ }
\def\x{\raise 1.25pt \hbox{$\scriptscriptstyle \times$}\nobreak\ }
\def\legendtable#1{\vbox{\baselineskip=10pt\tabskip=0pt\let\\=\cr\halign{\hfil##\hskip 3pt&##\hfil\cr#1\crcr}}}
\def\lllegend#1 #2 #3{\figlab {#1} {#2} {\legendtable{#3}}}
\def\lrlegend#1 #2 #3{\figlab {#1} {#2} {\llap{\legendtable{#3}}}}
\def\ullegend#1 #2 #3{\figlab {#1} {#2} {\vtop{\hrule height 0pt\legendtable{#3}}}}
\def\urlegend#1 #2 #3{\figlab {#1} {#2} {\llap{\vtop{\hrule height 0pt\legendtable{#3}}}}}

\newdimen\xorigon
\newdimen\yorigon
\newdimen\scaleval
\newdimen\scaleorigon

\def\setxscale#1 #2 #3 #4 #5 {%
    \xorigon=#1\yorigon=#3%
    \scaleval=#2\advance\scaleval by -\xorigon%
    \tempdimen=#5 pt\advance\tempdimen by -#4pt%
    \divide\tempdimen by 1000%
    \divide\scaleval by \tempdimen%
    \scaleorigon=-#4pt\divide\scaleorigon by 1000%
    \multiply\scaleorigon by \scaleval}
\def\xtickup#1 #2{\tempdimen=#1pt\divide\tempdimen by 1000%
    \multiply\tempdimen by \scaleval\advance\tempdimen by \scaleorigon%
    \advance\tempdimen by \xorigon%
    \figlab {\tempdimen} {\yorigon} {\vbox {\hbox to 0pt{\hss #2\hss}%
        \baselineskip=8pt\lineskiplimit=-5pt%
        \hbox to 0pt{\hss \vrule height 3pt\hss}}}}
\def\xtickdown#1 #2{\tempdimen=#1pt\divide\tempdimen by 1000%
    \multiply\tempdimen by \scaleval\advance\tempdimen by \scaleorigon%
    \advance\tempdimen by \xorigon%
    \figlab {\tempdimen} {\yorigon} {\vbox to 0pt {\hbox to 0pt{\hss \vrule height 3pt\hss}%
        \nointerlineskip\vskip 3pt%
        \hbox to 0pt{\hss #2\hss}\vss}}}
\def\nofig#1#2{\leavevmode{\vbox {\hrule \hbox to #1{\vrule height #2 \hfill \vrule} \hrule}} }

\DeclareGraphicsExtensions{.pdf}

\newcommand{\bea}{\begin{equation}\begin{aligned}}
\newcommand{\eea}{\end{aligned}\end{equation}}

\shorttitle{Wall-bounded flow over a realistically rough SHS}
\shortauthor{K. Alam{\'e} and K. Mahesh}

\title{Wall-bounded flow over a realistically rough superhydrophobic surface} 

\author{Karim Alam{\'e}\aff{1}
\and Krishnan Mahesh\aff{1}
\corresp{\email{kmahesh@umn.edu}}}

\affiliation{\aff{1}Department of Aerospace Engineering and Mechanics, University of Minnesota, Minneapolis, MN 55455, USA}

\begin{document}

\maketitle

\begin{abstract}
Direct numerical simulation (DNS) is performed for two wall-bounded flow configurations: laminar Couette flow at $Re=740$ and turbulent channel flow at $Re_{\tau}=180$, where $\tau$ is the shear stress at the wall. The top wall is smooth and the bottom wall is a realistically rough superhydrophobic surface (SHS), generated from a three-dimensional surface profile measurement. The air--water interface, which is assumed to be flat, is simulated using the volume-of-fluid (VOF) approach. The two flow cases are studied with varying interface heights $h$ to understand its effect on slip and drag reduction ($DR$). For the laminar Couette flow case, the presence of the surface roughness is felt up to $40\%$ of the channel height in the wall-normal direction. Nonlinear dependence of $DR$ on $h$ is observed with three distinct regions. A nonlinear curve fit is obtained for gas fraction $\phi_g$ as a function of $h$, where $\phi_g$ determines the amount of slip area exposed to the flow. A power law fit is obtained from the data for the effective slip length as a function of $\phi_g$ and is compared to those derived for structured geometry. For the turbulent channel flow, statistics of the flow field are compared to that of a smooth wall to understand the effects of roughness and $h$. Four cases are simulated ranging from fully wetted to fully covered and two intermediate regions in between. Scaling laws for slip length, slip velocity, roughness function and $DR$ are obtained for different penetration depths and are compared to past work for structured geometry. $DR$ is shown to depend on a competing effect between slip velocity and turbulent losses due to the Reynolds shear stress contribution. Presence of trapped air in the cavities significantly alters near-wall flow physics where we examine near-wall structures and propose a physical mechanism for their behaviour. The fully wetted roughness increases the peak value of turbulent intensities, whereas the presence of the interface suppresses them. The pressure fluctuations have competing contributions between turbulent pressure fluctuations and stagnation due to asperities, the near-wall structure is altered and breaks down with increasing slip. Overall, there exists a competing effect between the interface and the asperities, the interface suppresses turbulence whereas the asperities enhance them. The present work demonstrates DNS over a realistic SHS for the first time, to the best of our knowledge.  

\end{abstract}

\section{Introduction}
\label{sec:intro}
Superhydrophobicity is a property attributed to surface roughness (ridges, grooves, posts or random textures) and surface chemistry which maintains large contact angles for sessile drops, thus producing low wettability, known as the Cassie--Baxter state \citep{casbax}. The interface meniscus creates an air mattress that acts like a lubricant for the outer flow \citep{rothstein2010}. When the interface fails, the liquid fills the surface cavities and the superhydrophobic effect is lost. This is referred to as the Wenzel state \citep{wenzel1936}. 

Nature provides numerous examples of superhydrophobic surfaces (SHS), which can be exploited for practical applications. For example, the lotus leaf is believed to take advantage of superhydrophobicity for a self-cleaning mechanism \citep{barthnein_1997}. Frictional drag reduction is central to the performance of marine vessels, and anti-biofouling, anti-icing and microfluidic devices \citep{furstetal_2005,genzerefimenko_2006,fangetal_2008,jungetal_2011}. Any impact on skin friction drag reduction substantially improves the overall performance and yields savings in fuel cost \citep{choikim_2006}. In the present work we focus on drag reduction using SHS in two canonical flow configurations: laminar Couette flow and turbulent channel flow.
 
With recent developments in three-dimensional printing and microfabrication processes, it is possible to create surfaces exhibiting superhydrophobic characteristics when coupled with chemical treatments. Laminar flows over SHS have been studied both numerically and experimentally. SHS have been shown to achieve drag reduction \citep{ouetal_2004,ourothstein_2005,josephetal_2006,choikim_2006,maynesetal_2007,Woolford2009,emami2011}. Analytical models relate the slip lengths to various surface parameters such as groove width, pitch and height \citep{lauga2003,ybertetal_2007} or the slip velocities to geometry \citep{Seo2016}. In general the SHS are considered to be simple grooved geometries, and numerically the interface is typically assumed to be flat and represented using zero-shear boundary conditions. Others have included the effect of viscosity on the interface \citep{vinogradova_1995,belyaevetal_2010,nizkaya_2014}. Several authors have investigated the effect of the curvature due to the meniscus and modified the analytical solutions to take curvature into account \citep{cottin2003,sbragaglia2007,teokhoo_2014,lietal_2017}.

Turbulent flows over textured surfaces have been studied extensively in the past. Experimentally, it becomes difficult to conduct measurements near the wall and to maintain a stable interface, but drag reduction and slip lengths have been investigated. Some past work reported that SHS had no effect on turbulent statistics \citep{zhao_2007,peguero_2009}, while others reported otherwise \citep{gogteetal_2005,henoch2006,Daniello2009,jung_2010,aljallis_2013,bidkar2014,Park2014,Srinivasan2015}. Investigation of interface stability was studied using post-processed pressure fluctuations \citep{Seo2015} . \cite{Rosenberg2016} showed that the turbulent skin friction is reduced over air- and liquid-impregnated surfaces (SLIPS) for Taylor-Couette flows. Numerically, the interface is assumed to be flat and modelled using zero-shear boundary conditions \citep{Martell2009,Martell2010,frohnapfel_2010,parketal_2013,Jelly2014,Turk2014} and homogenised slip length models instead of zero-shear boundary conditions \citep{minkim_2004,Fukagata2006,Busse2012}. The profiles of mean velocity, turbulence intensities and Reynolds shear stresses were characterised for the inner part of turbulent boundary layers over several SHS with varying textures and a range of $Re_{\tau}$ \citep{ling_katz_2016}. \cite{Jung2016} studied the effect of anisotropy in the slip-length models for different interface heights in idealised SHS. Recently, \cite{arenas2017} used the level-set method to study SHS with varying viscosity ratios over spanwise and streamwise grooves. \cite{rastegari2018} applied the Boltzmann method in their DNS to study the effect of the liquid--gas interface in longitudinal grooves by modelling it as a stationary, curved and free shear boundary; the meniscus shape was determined using the Young-Laplace equation. The sustainable pressure bounds of SHS were further investigated by \cite{rastegari2019}. \cite{fairhall2019} showed that drag reduction is proportional to the difference between the virtual origin of the mean flow and the virtual origin of the overlying turbulence.

%All past work on flow over SHS have considered simple geometries such as longitudinal and transverse grooves or posts. Realistically rough surfaces are devoid of any such regular pattern. \cite{prf_seo2018} studied the effect of texture randomization in turbulent flows over SHS using posts. 
Most past work on flow over SHS has considered idealised geometries such as grooves or posts. \cite{prf_seo2018} recently studied turbulent flow over SHS idealised as random slip/no-slip patches. To the best of our knowledge, none of the past numerical work has simulated  a multiphase flow over realistically rough surfaces as presented in this paper. The main goal of the present work is to perform DNS of (i) laminar Couette flow and (ii) turbulent channel flow, where the bottom wall is a realistically rough surface. We aim to explore the effect of interface height on slip, drag reduction, near-wall flow field and turbulence statistics. The rest of the paper is organised as follows: \S \ref{sec:simdetail} describes simulation details including the numerical method, parameters and problem formulation. Results are described in \S \ref{sec:res} which include flow visualisations, steady and mean flow field properties and drag reduction in laminar Couette flow. The mean flow statistics, scaling laws and flow structure are also presented for turbulent flow. Finally, the work is summarised in \S \ref{sec:summ}.     

\section{Simulation details}
\label{sec:simdetail}

\subsection{Numerical method}

Direct numerical simulation (DNS) is performed using a mass-conserving volume-of-fluid (VOF) methodology on structured grids to study the effect of an air--water interface over a realistically rough surface. The governing equations are solved using the finite-volume algorithm developed by \citet{mahesh2004} for the incompressible Navier-Stokes equations. The governing equations for the momentum and continuity are given by the Navier-Stokes equations:
\begin{equation}
\label{eq:momentum}
\frac{\partial u_i}{\partial t} + \frac{\partial}{\partial x_j}(u_iu_j)=-\frac{1}{\rho}\frac{\partial p}{\partial x_i} + \frac{1}{\rho}\frac{\partial}{\partial x_j} \left[\mu\left(\frac{\partial u_i}{\partial x_j}+\frac{\partial u_j}{\partial x_i}\right)\right]+F_{st,i}+\delta_{i1}K_i,
\end{equation}
\begin{equation}
\label{eq:continuity}
 \frac{\partial u_i}{\partial x_i} = 0,
\end{equation}
where $u_i$ and $x_i$ are the $i$\textit{th} component of the velocity and position vectors respectively, $p$ denotes pressure, $\rho$ is density and $\mu$ is the viscosity of the fluid. The fluids are assumed to be immiscible. Additionally in eq.~\ref{eq:momentum}, $\delta_{i1}$ is the Kroenecker delta, $K_i$ is the body force which is only active in the liquid phase and $F_{st,i}$ the surface tension force. 
%The surface tension force is modeled as a continuum surface force as proposed by \cite{brackbill1992}:
%\begin{equation}
%F_{st,i} = \sigma\kappa\frac{\partial c}{\partial x_i},
%\end{equation}
%where $\sigma$ is the surface tension constant, and $\kappa$ is the curvature calculated using the height function which has been shown to significantly reduce numerical errors that are associated with surface tension.
%$F_{st,i}$ disappears from eq. \ref{eq:momentum} since we assume infinite surface tension such that the interface is flat leading to zero curvature. 
The algorithm is robust and emphasises discrete kinetic energy conservation in the inviscid limit which enables it to simulate high-Reynolds-number flows without adding numerical dissipation. The solution is advanced in time by an implicit scheme using successive over-relaxation (SOR). A predictor--corrector methodology is used where the velocities are first predicted using the momentum equation and then corrected using the pressure gradient obtained from the Poisson equation yielded by the continuity equation. The Poisson equation is solved using a multigrid pre-conditioned conjugate gradient method (CGM) using the Trilinos libraries (Sandia National Labs). The multigrid pre-conditioner uses a Chebyshev smoother with a third-order polynomial and a maximum number of levels set to $4$. The implicit time advancement uses the Crank-Nicholson discretisation with a linearisation of the convection terms.

The volume fraction is represented by a colour function $c$ to keep track of two different fluids. The colour function $c$ varies between the constant value of one in a filled cell to zero in an empty cell, with an intermediate value between zero and one to define an interface cell where $0 \leq c \leq 1$ . The reconstruction and advection steps are based on a set of analytic relations proposed by \cite{scardovelli2000}. The governing equations for the colour function material derivative are given by
\begin{equation}
\label{eq:coloradvect}
\frac{\partial c}{\partial t} + \cancelto{0}{u_j\frac{\partial c}{\partial x_j}} = 0,
\end{equation}
where the advection term is neglected in the following simulations since we assume that the interface is stationary due to an infinite surface tension. 
The density and viscosity are evaluated as
\begin{equation}
\label{eq:density}
 {\rho = \rho_g + (\rho_l - \rho_g)c},
\end{equation} 
\begin{equation}
\label{eq:viscosity}
 {\mu = \mu_g + (\mu_l - \mu_g)c},
\end{equation}
where the subscript `$l$' denotes the liquid phase and `$g$' the gas phase. 
The surface is represented by obstacle cells which are masked out. At the beginning of a simulation run, the fluid and obstacle cells are flagged accordingly:
\begin{equation}
mask = \begin{cases}
             1, &\text{ if fluid cell }\\
             0, &\text{ if obstacle cell} 
            \end{cases}
            ;
\end{equation}
this step is performed once.
The wetted masked cells (cells that share a face between a fluid and obstacle cell) enforce a zero face-normal velocity ${v}_N|_{mask}=0$. The cell-centred velocities satisfy a no-slip boundary condition, with the exception of corner cells that take a weighted average of the neighbouring cell-centred values.  
The algorithm has been validated with experimental results for a variety of flows involving superhydrophobicity \citep{lietal_2016,lietal_2017} and fully wetted roughness \citep{ma_aiaa2019}. In this study, we enforce a zero face-normal velocity at the interface ${v}_N|_{interface}=0$. The condition models a high-surface-tension regime with a stable flat interface. This was done to focus on the effect of varying interface heights $h$ for a finite-viscosity lubricant. The assumption made is valid for flow regimes where the interfacial surface tension dominates the interface dynamics. Further discussion of the validity of our assumptions is presented in appendix \ref{appA}. The statistics of the turbulent channel flow were averaged over a period of 300 flow-through times after the discharge had reached a steady-state value. 

\subsection{Surface generation}
\label{sec:surfgen}

\begin{figure}
\centering{
\includegraphics[scale=0.5, trim={2cm 2cm 2cm 2cm}, clip]{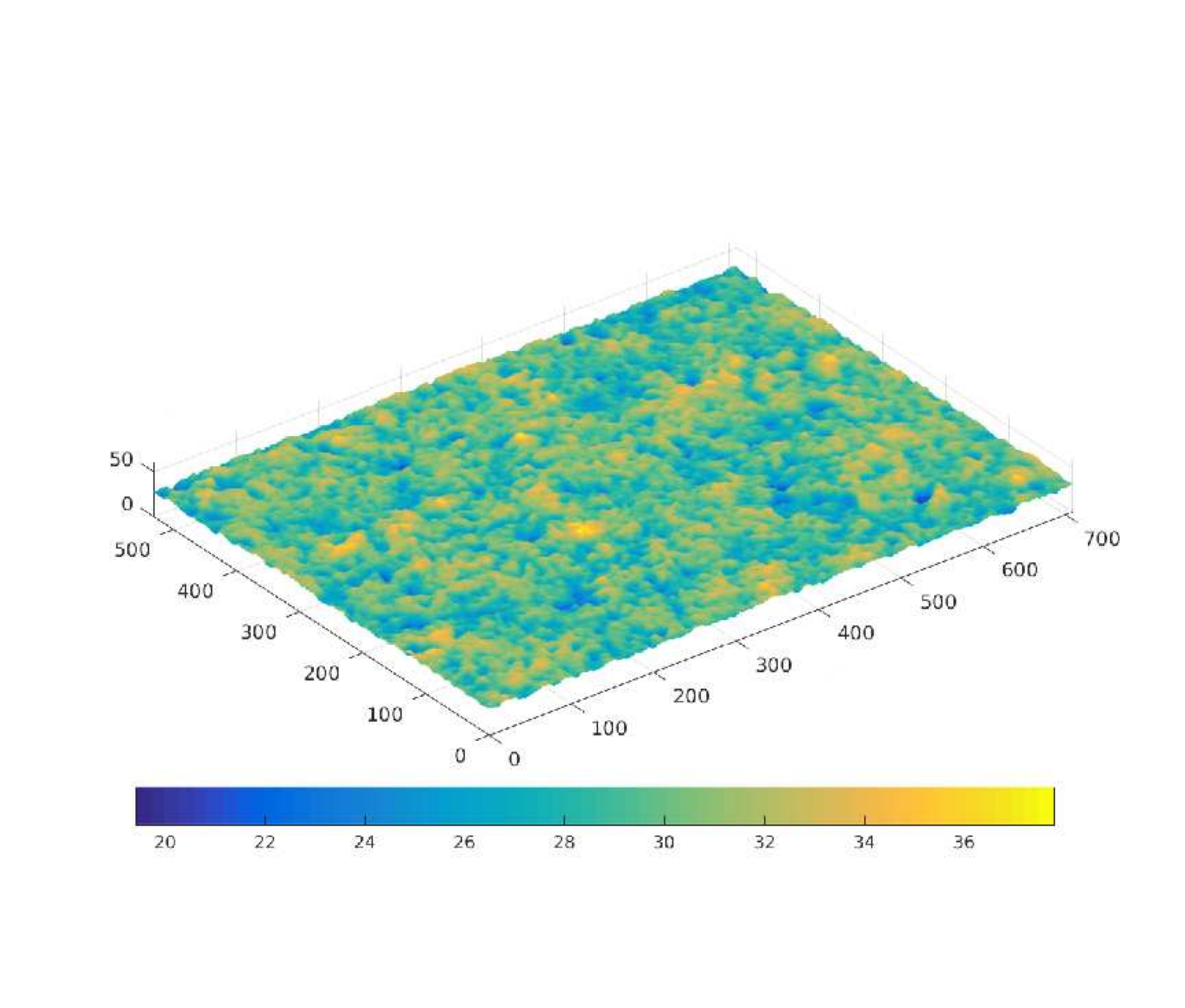}
\put(-75,60){$x(\mu m)$}
\put(-250,65){$z(\mu m)$}
\put(-280,135){$h_s(\mu m)$}
}
\caption{Illustration of the real rough surface. The contour legend describes the height of the surface profile.}
\label{fig:surface_plot}
\end{figure}

The roughness used in the present work is obtained from a real surface manufactured at UT Dallas (courtesy Professor Wonjae Choi), with a three-dimensional (3-D) surface profile measurement using a 20X objective lens obtained from MIT (courtesy Professor Gareth McKinley). The sample is aluminium 6061 sandblasted using 150 grit, etched for 25 s, boehmetized for 30 min and hydrophobised using Ultra Ever Dry top coat in isopropanol. Figure \ref{fig:surface_plot} provides an illustration of the scanned surface data coloured with height. The surface statistics and power spectral density (PSD) of the surface height are provided in appendix \ref{appB}. 

We begin with a pre-processing step by reading the scanned surface data. The number of pixels in the scan width and height are stored as the number of nodes in the streamwise and spanwise directions respectively. The values of the roughness height and spatial location are then interpolated to cell centres given our domain of choice. The cell centre values are then written to a new file with a structured data format. Any obstacle cell which shares an edge with a fluid cell is tagged as a boundary cell. Boundary cells can either be an edge cell (if the boundary cell borders exactly one fluid cell) or a corner cell (if the boundary cell shares a corner with two or more fluid cells). The discretised surface is checked with the original data and the errors in the surface statistics are presented in table \ref{tab:scaled_masked_stats}, appendix \ref{appB}. The momentum equations are solved inside the fluid domain while the pressure is solved everywhere. The weighted average applied at the corner cells does not affect the pressure equation since we use collocated grids where the face-normal velocities are set to zero at the boundaries independent of the cell centre value. This ensures a proper pressure jump recovery at the obstacle walls where the values inside the obstacle domain do not affect the pressure values in the fluid domain. 

\subsection{Problem description}

Simulations are performed for two canonical problems: (i) laminar Couette flow and (ii) turbulent channel flow, where the surface described in \S \ref{sec:surfgen} is used as the bottom wall. In the experiments performed by \cite{ling_katz_2016}, the tunnel pressure (which controls the interface location) is increased which compresses the air layer into the SHS and in turn expose more asperities, thereby reducing the extent of drag reduction. The aim of this paper is to model this effect over an idealised flat interface numerically by progressively increasing the height $h$ and measuring flow properties for each interface location in different flow regimes. The maximum interface height is non-physical in a realistic scenario, but it serves the purpose of providing the largest amount of slip that is theoretically achievable. It also helps describe the trend between limiting cases. The problem description is given in the following sections. 

\subsubsection{Laminar Couette}

\begin{figure}
\centering{
\includegraphics[width=90mm]{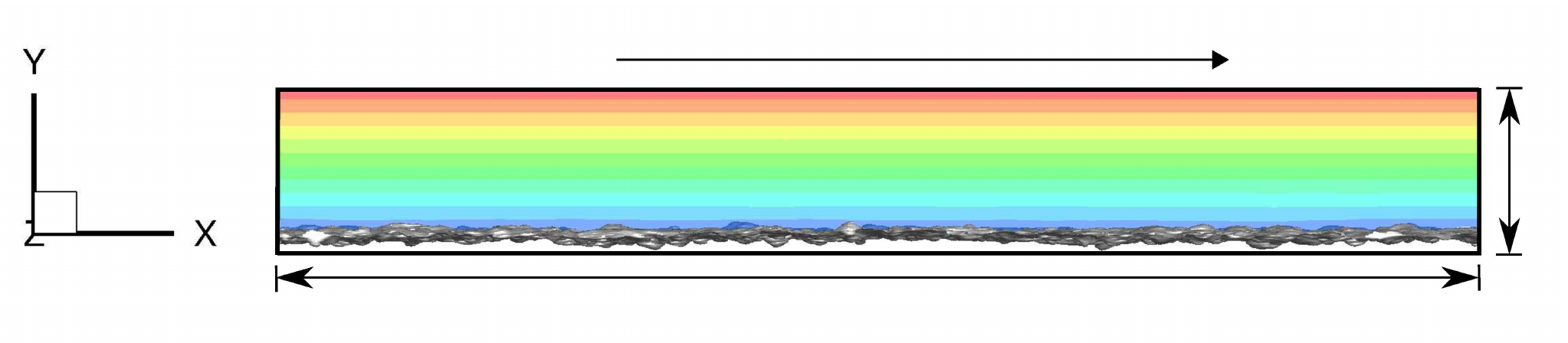}
\put(-115,0){$L_x$}
\put(0,25){$L_y$}
\put(-115,50){$U_{\infty}$}
}
\caption{Illustration of the computational domain for the laminar Couette simulation, roughness and the instantaneous velocity field }
\label{fig:lamcouette_domain}
\end{figure}
\begin{table}
 \begin{center}
%\begin{tabular}{l|{c}|{c}|{c}|{c}|{c}|{c}|{c}|{c}|{c}|{r}}
\begin{tabular}{l|{c}|{c}{c}{c}{c}{c}{c}{r}}
& Case & h & $L_{x}$ & $L_{y}$ & $L_{z}$ & $N_{x}\times N_{y}\times N_{z}$ & $\Delta y_{min}$ & $\Delta y_{max}$ \\
\hline
Laminar Couette Flow    & 1-18  & $S_v$ - $S_p$ & $8H$ & $H$ & $6H$ & $341 \times 128 \times 256$ & $0.006$ & $0.05$\\
\hline
			& L-S   & - & $8H$ & $H$ & $6H$ & $341 \times 128 \times 256$ & $0.006$ & $0.05$\\
			& L-RFW & $S_v$ & $8H$ & $H$ & $6H$ & $341 \times 128 \times 256$ & $0.006$ & $0.05$\\
			& L-RI1 & $0$ & $8H$ & $H$ & $6H$ & $341 \times 128 \times 256$ & $0.006$ & $0.05$\\
			& L-RI2 & $S_q$ & $8H$ & $H$ & $6H$ & $341 \times 128 \times 256$ & $0.006$ & $0.05$\\
			& L-RI3 & $S_p$ & $8H$ & $H$ & $6H$ & $341 \times 128 \times 256$ & $0.006$ & $0.05$\\
\hline
\end{tabular}
\caption{Case names, interface location, domain extents and grid resolution for the laminar Couette flow problem. L denotes the laminar cases. S and R denote a smooth and rough wall respectively. Fully wetted roughness is denoted by FW. I1, I2 and I3 represent the interface height at three locations: $0$, $S_q$ and $S_p$ respectively.}
\label{tab:gridres_lamcouette}
\end{center}
\end{table}

The height $H$ of the top wall was chosen such that the root-mean-square (RMS) roughness height $S_q$ is around 2\% of $H$. The original surface is scaled to achieve the roughness height ratios described above. The reference system is chosen such that the origin coincides with the arithmetic mean elevation of the roughness. The schematic diagram shown in figure \ref{fig:lamcouette_domain} illustrates the flow domain. No-slip boundary conditions are prescribed on the bottom surface and a constant velocity $U_{\infty}$ in the streamwise $x$-direction is prescribed at the top wall. The streamwise ($x$) and spanwise ($z$) directions are periodic; a non-uniform grid is used in the wall-normal ($y$) direction with clustering in the rough wall region. The interface location was varied from the maximum valley depth $S_v$ all the way up to the maximum peak height of the roughness $S_p$ over $18$ increments. Table \ref{tab:gridres_lamcouette} gives the grid details. The Reynolds number $Re=U_{\infty}H/\nu=740$, where $U_{\infty}$ and $H$ are taken to be unity and the liquid phase being the reference material property. A smooth planar Couette flow (Case  L-S) is used as a baseline such that the reference shear stress  $\tau_{o}=\mu U_{\infty}/H = \mu_{w}$, where $\mu_{w}$ is the reference viscosity in the water phase. First, a fully wetted case (L-RFW) is simulated to baseline the effect of roughness on drag when compared to the smooth wall. Case L-RI1 denotes $h=0$, Case L-RI2 denotes $h=Sq$ and Case L-RI3 denotes $h=S_p$. The viscosity ratio $\mu_{r}=\mu_{a}/\mu_{w}=1/50$ is used to represent an air--water interface. The change in shear stress due to the roughness and $h$ is used to compute the drag reduction defined using the following relation:
\begin{equation}
 DR(\%) = \frac{(\tau_o-\tau)}{\tau_o} \times 100.
 \label{eq:dr}
\end{equation}

\subsubsection{Turbulent channel}
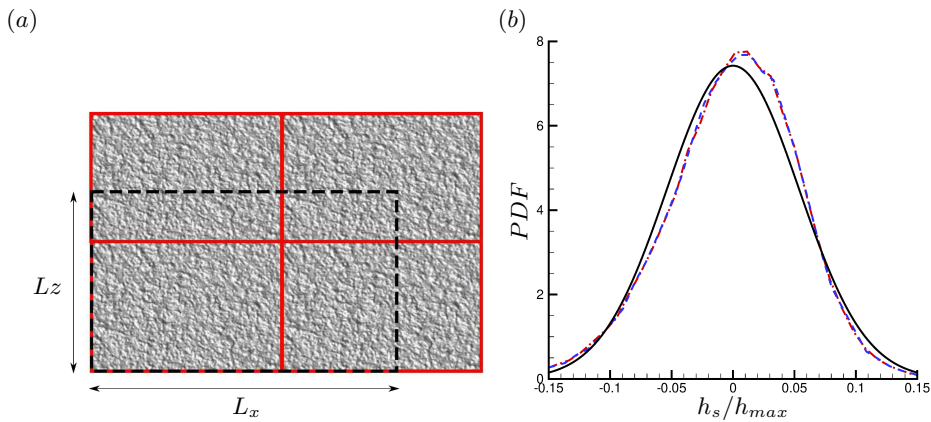
\begin{figure}
\centering{
\includegraphics[width=55mm]{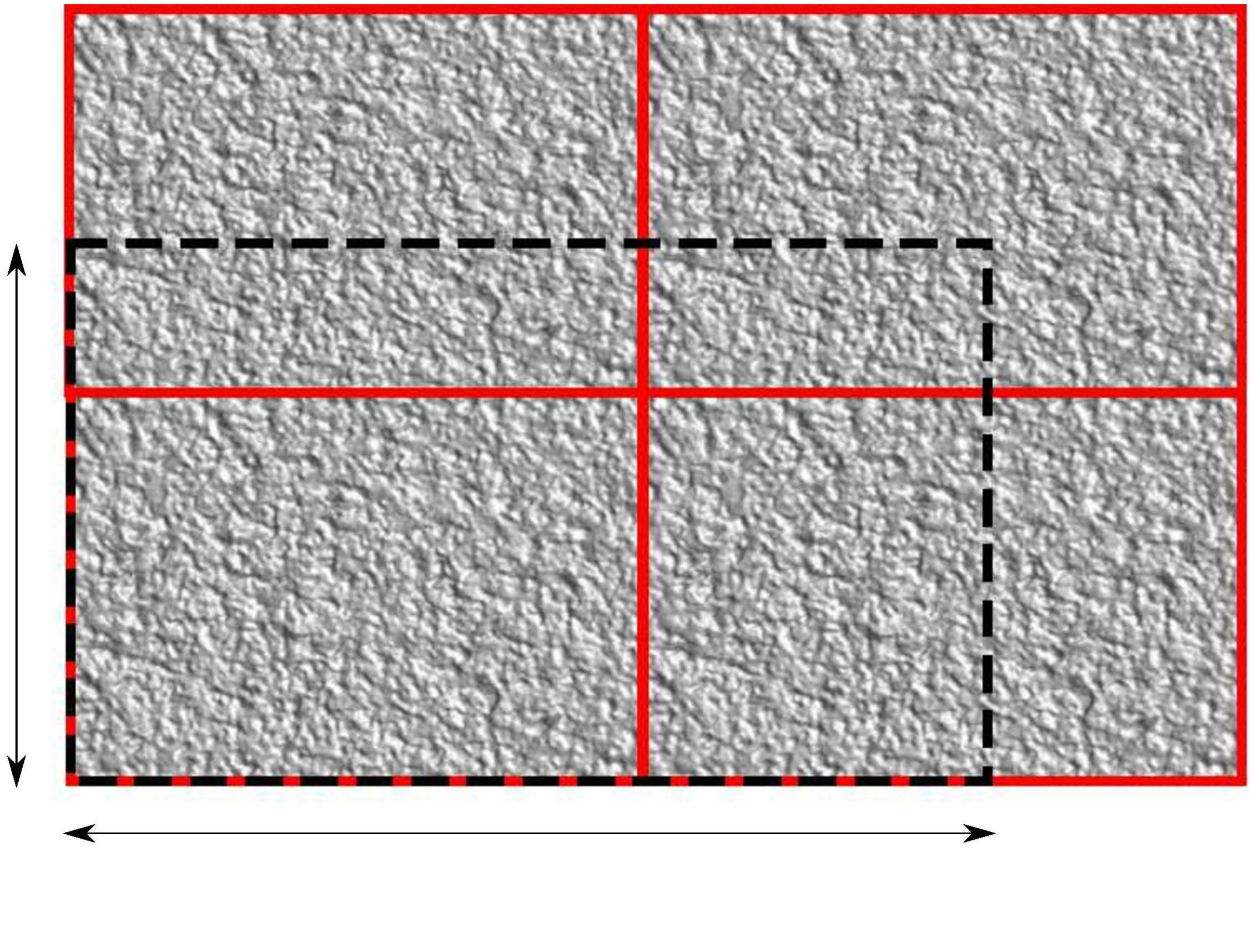}
\put(-95,5){$L_x$}
\put(-170,50){$Lz$}
\put(-180,150){$(a)$}
\includegraphics[width=65mm]{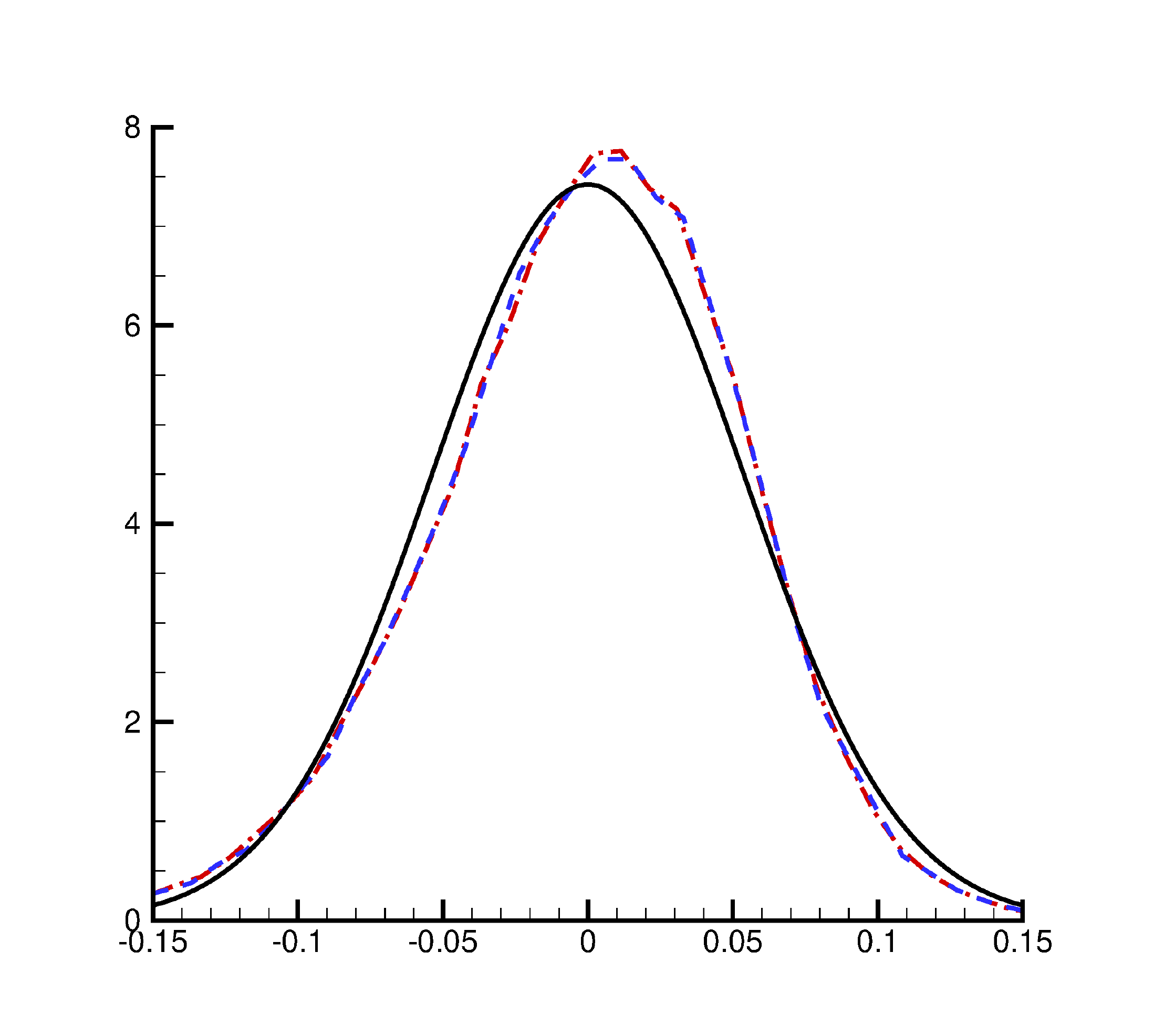}
\put(-105,5){$h_s/h_{max}$}
\put(-175,70){\rotatebox{90}{$PDF$}}
\put(-180,150){$(b)$}
}
\caption{(a) Turbulent channel domain extent (dashed line) and the tiled surface at different orientations with their physical boundaries (solid red) not drawn to scale. (b) The probability density function (PDF) distribution of the real surface height for the laminar Couette (red dash dot) and turbulent channel (blue dashed) compared to a Gaussian (black solid line) of the same root-mean-square height.}
\label{fig:psd_channel_size_comparison}
\end{figure}

A schematic diagram describing the turbulent channel domain is given in figure \ref{fig:turbchan_domain}. No-slip boundary conditions are applied on both the top smooth wall and the bottom rough wall with periodicity in the streamwise ($x$) and spanwise ($z$) directions: non-uniform grids are used in the wall-normal ($y$) direction where the grid is clustered near the rough wall region. The grid details are given in table \ref{tab:gridres_turbchan}. A constant body force in the liquid phase is applied such that the friction Reynolds number is $Re_{\tau}=u_{\tau} \delta / \nu=180$ where $u_{\tau}$ is the wall friction velocity, $\delta=(L_y-y_o)/2$ the channel half-height and $y_o$ the reference bottom plane. Four cases were considered: (i) fully wetted rough channel for Case T-RFW, (ii, iii) two-phase rough channel with $h=0$ for Case T-RI1, $h=S_q$ for Case T-RI2, and (iv) $h=S_p$ for Case T-RI3. The reference plane $y_o$ is taken to be the arithmetic mean elevation of the roughness for Case T-RFW, and the location of the interface for Cases T-RI1, T-RI2 and T-RI3.  The viscosity ratio is that of an air--water interface given by $\mu_{r}=\mu_{a}/\mu_{w}=1/50$. The solver was validated (not shown here) for the flat smooth channel \citep{kim_1987}. The original surface is scaled such that $S_q^+ \approx 1.6$. The original surface scan was not large enough to cover the bottom wall after scaling, and the roughness patch had to be tiled in random orientations to minimise any directional bias and create a larger area. The required domain extents were then extracted from the tiled surface as shown in figure \ref{fig:psd_channel_size_comparison}(a). The computational domain required after scaling was twice as long in the streamwise ($x$) direction and $35\%$ longer in the spanwise ($z$) direction. The ratio of $L_x$ to $L_z$ is $4:3$ in the laminar Couette case compared to the $2:1$ ratio of the turbulent channel case. In order to ensure this does not affect the roughness height distribution, a probability density function (p.d.f) distribution is plotted for both cases in figure \ref{fig:psd_channel_size_comparison}(b) and is compared to a Gaussian distribution. No appreciable difference is observed between the two cases. The p.d.f distribution is negatively skewed when compared to a Gaussian which is also calculated in the surface statistics presented in table \ref{tab:stats} of appendix \ref{appB}. 

\begin{figure}
\centering{
\includegraphics[width=80mm]{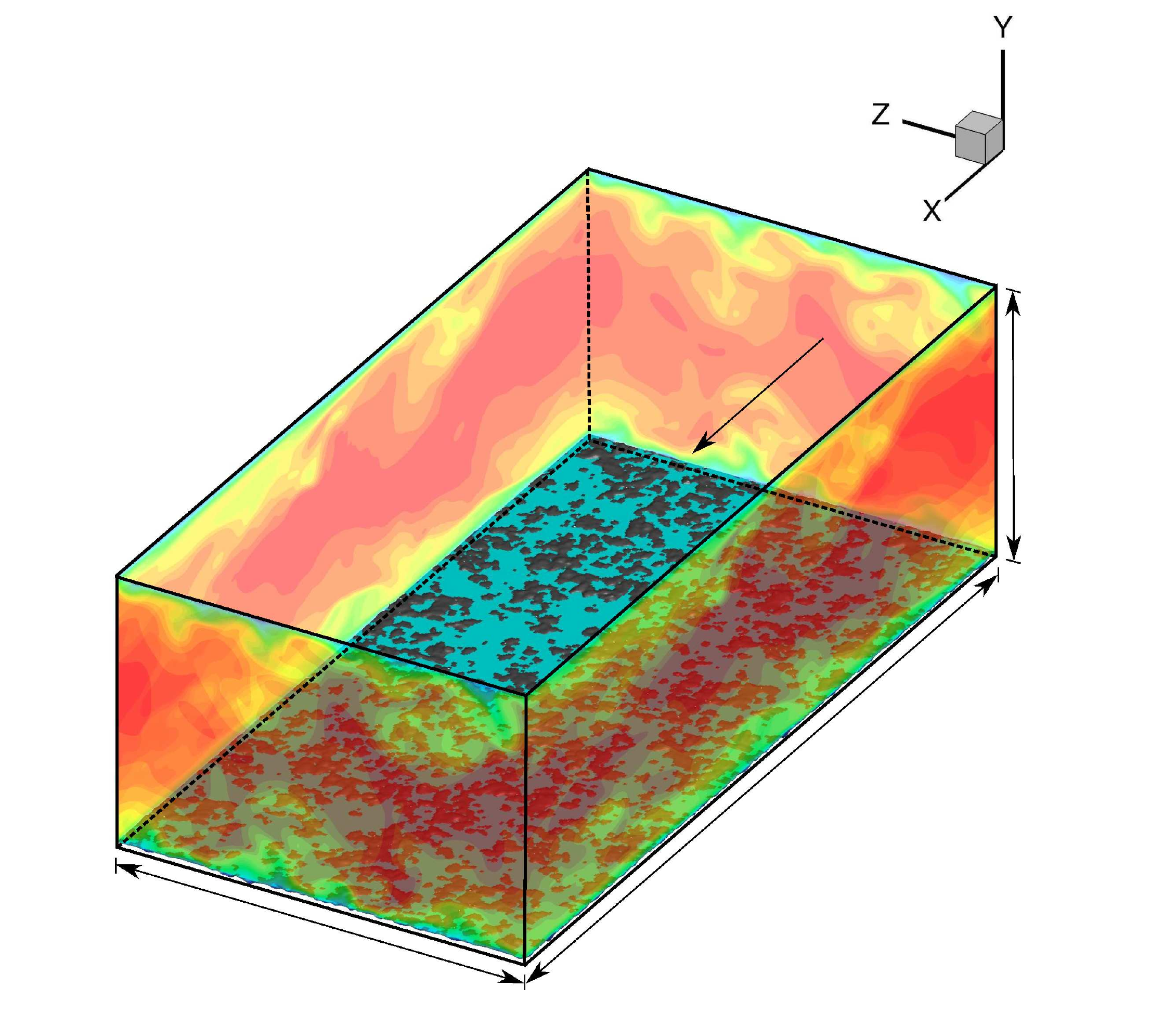}
\put(-170,10){$L_z$}
\put(-75,40){$L_x$}
\put(-30,115){$L_y$}
\put(-90,125){$K$}
}
\caption{Illustration of the computational domain for the turbulent channel flow, interface location embedded within the roughness and the instantaneous velocity field.}
\label{fig:turbchan_domain}
\end{figure}
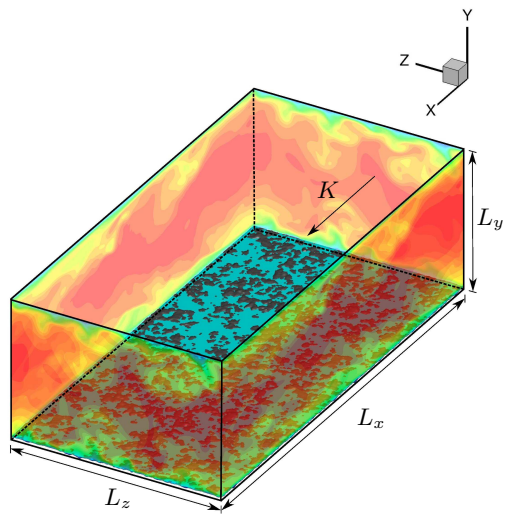
\begin{table}
 \begin{center}
 \begin{tabular}{l|{c}|{c}{c}{c}{c}{c}{c}{r}}
& Case & h & $L_{x}$ & $L_{y}$ & $L_{z}$ & $N_{x}\times N_{y}\times N_{z}$ & $\Delta y^+_{min}$ & $\Delta y^+_{max}$ \\
\hline
Turbulent Channel Flow  & T-S   & - & $2\pi \delta$ & $2.08\delta$  & $\pi \delta$ & $341 \times 128 \times 207$ & $1.8$ & $6.12$ \\
			& T-RFW & $S_v$ & $2\pi \delta$ & $2.08\delta$  & $\pi \delta$ & $448 \times 256 \times 271$ & $0.42$ & $2.4$ \\
			& T-RI1 & $0$ & $2\pi \delta$ & $2.08\delta$  & $\pi \delta$ & $448 \times 256 \times 271$ & $0.42$ & $2.4$ \\
			& T-RI2 & $S_q$ & $2\pi \delta$ & $2.08\delta$  & $\pi \delta$ & $448 \times 256 \times 271$ & $0.42$ & $2.4$ \\
			& T-RI3 & $S_p$ & $2\pi \delta$ & $2.08\delta$  & $\pi \delta$ & $448 \times 256 \times 271$ & $0.42$ & $2.4$ \\
\hline
\end{tabular}
\caption{Case names, interface location, domain extents and the grid resolution in wall units for the turbulent channel flow problem. T denotes the turbulent cases. S and R denote a smooth and rough wall respectively. Fully wetted roughness is denoted by FW. I1, I2 and I3 represent the interface height at three locations: $0$, $S_q$ and $S_p$ respectively.}
\label{tab:gridres_turbchan}
\end{center}
\end{table}

For the sake of brevity, the streamwise mean velocity is denoted by $U$ where the overline (denoting temporal averaging) and angle brackets (denoting spatial averaging) are dropped e.g.
\begin{equation}
 U(y) = \overline{\left< U \right>} = (1/L_x L_z) \int_{0}^{L_x} \int_{0}^{L_z} \overline{u} \mathrm{d}x \mathrm{d}z.
\end{equation}
The bulk velocity is defined as follows:
\begin{equation}
 U_b = (1/L_y) \int_{y_o}^{L_y} U(y) \mathrm{d}y.
\end{equation}
Similarly, the Reynolds stresses are denoted by $u'_i u'_j$ dropping the angle brackets and overline. 
Given that the channel is under a constant pressure gradient, at a statistically stationary state, the average friction wall velocity is given by $u_{\tau}=(\delta K_1)^{1/2}$ and the average shear stress by $\tau_w=\delta K_1$. It also holds that $\tau_w=(\tau^T_w+\tau^B_w)/2$ where $\tau^T_w$ is the top wall shear stress and $\tau^B_w$ the bottom wall shear stress. The top wall is flat therefore $\tau^T_w$ is calculated directly by averaging $\mu (\partial{U}/\partial{y})_{y=L_y/2}$ and $\tau^B_w$ is calculated indirectly to avoid averaging over the masks using $\tau^B_w=2\tau_w-\tau^T_w$. The bottom wall friction velocity is then calculated using $u^B_{\tau}=(\tau^B_w)^{1/2}$. Results are plotted against the channel height in wall units $y^+=u_{\tau}y/\nu$. If the bottom wall friction velocity is used as a reference, then a distinction is made explicitly. For example $y^+(u^B_{\tau})$ denotes the channel height in wall units based on the bottom wall friction velocity.

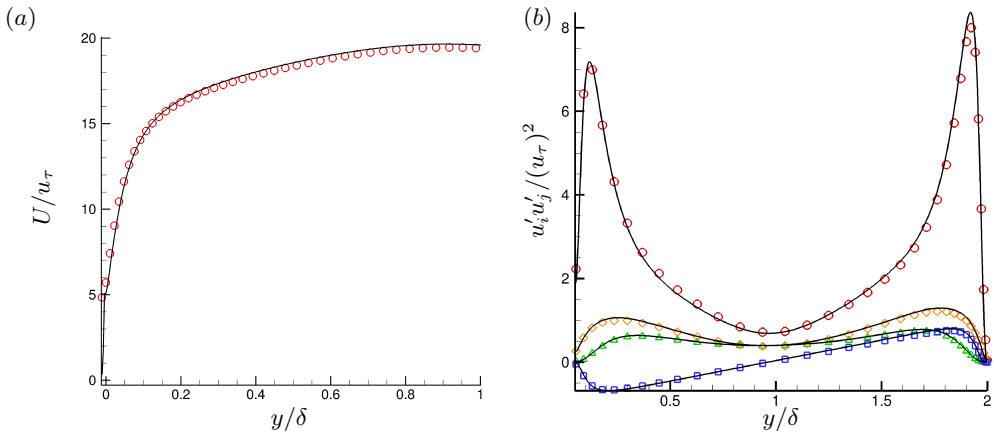
\begin{figure}
\centering{
\includegraphics[width=65mm]{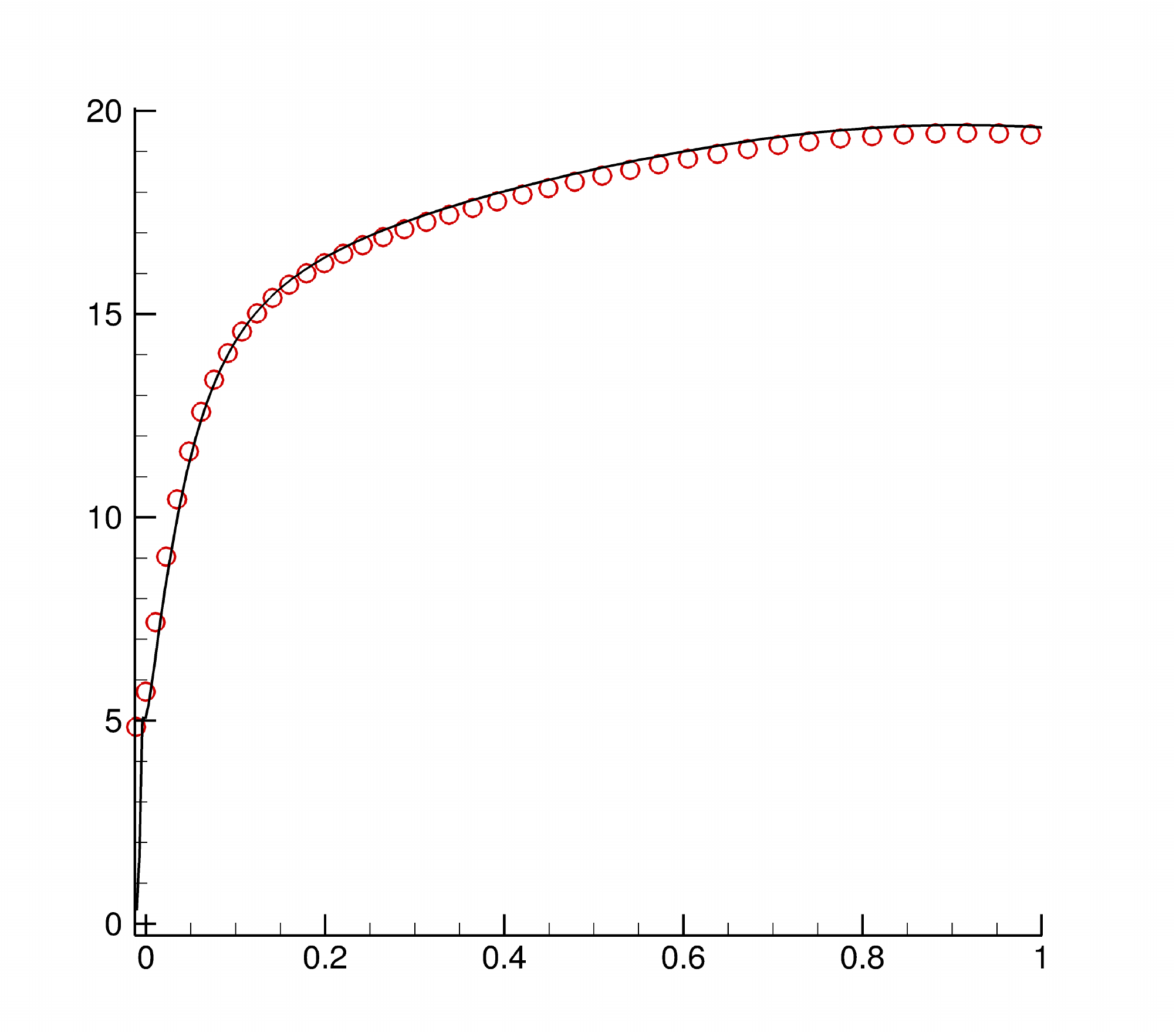}
\put(-100,0){$y/\delta$}
\put(-190,75){\rotatebox{90}{$U/u_{\tau}$}}
\put(-200,150){$(a)$}
\includegraphics[width=65mm]{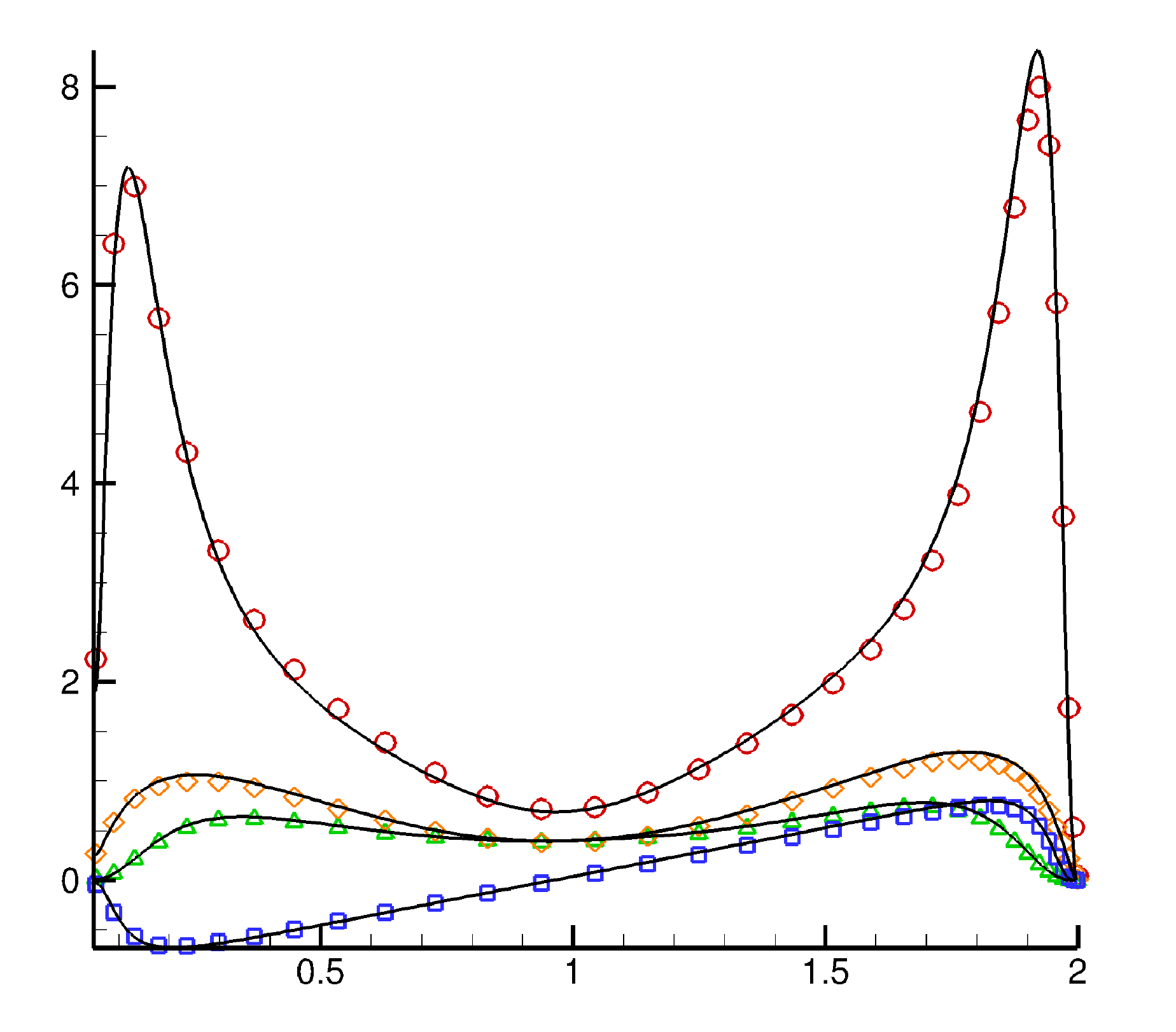}
\put(-100,0){$y/\delta$}
\put(-190,70){\rotatebox{90}{$u'_iu'_j/(u_{\tau})^2$}}
\put(-190,150){$(b)$}
}
\caption{Grid refinement comparison between (a) mean velocity profiles and (b) the Reynolds stresses. Black solid lines represent the fine grid and symbols the coarse grid.}
\label{fig:grid_refinement}
\end{figure}

Two simulations were performed at different resolutions to quantify the effect of grid size. The refined grid is $\thicksim 3.5$ as fine as the previous grid. Case T-RI2 was used as a baseline for the grid refinement comparison. No appreciable difference (less than $1\%$) in the mean velocity profiles, bulk velocity and Reynolds stresses is observed in figure \ref{fig:grid_refinement}. The slip velocity increased by $3.76\%$ and the bottom wall shear stress $\tau^B_w$ decreased by $2.3\%$. We report results from the finer grid in this paper.

\section{Results}
\label{sec:res}

\subsection{Laminar Couette flow}

\subsubsection{Steady-state flow field} 

\begin{figure}
\centering{
\includegraphics[width=70mm]{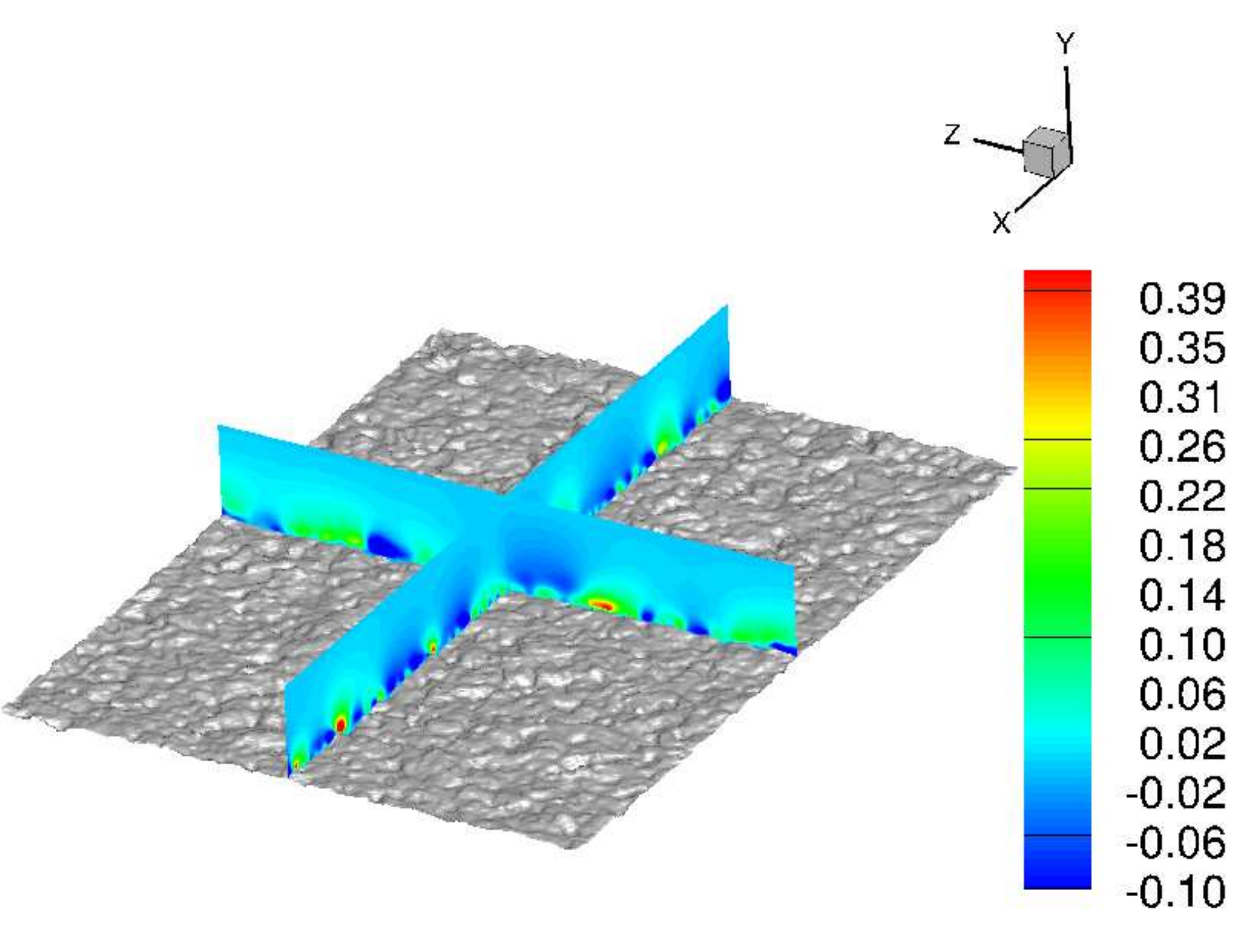}
\put(-200,150){$(a)$}
\includegraphics[width=70mm]{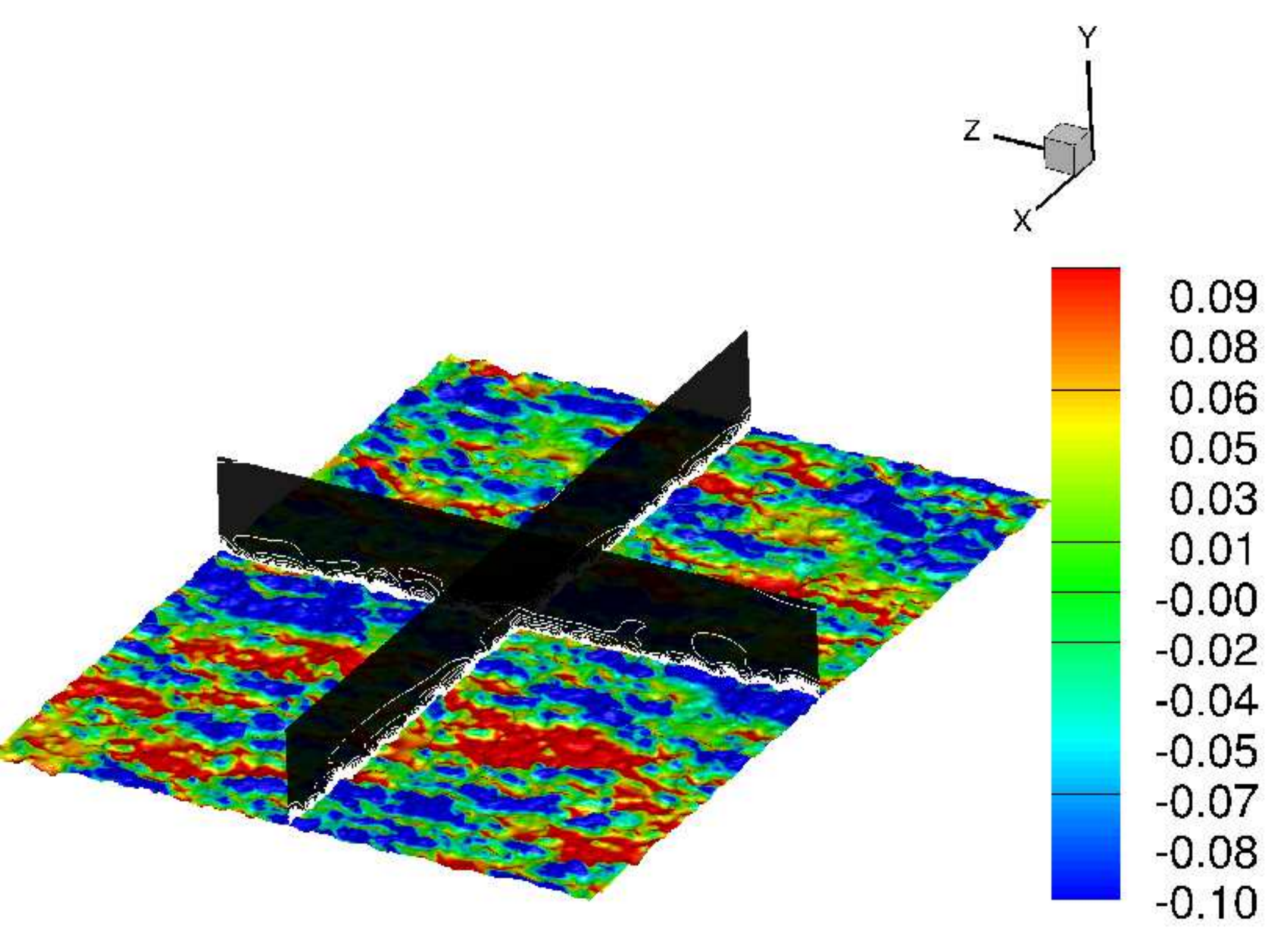}
\put(-200,150){$(b)$}
}
\caption{Laminar Couette flow (Case L-RFW) with (a) Wall-normal velocity contours normalised by the maximum wall-normal velocity $v_{max}$ and (b) vorticity magnitude line contours normalised by the maximum vorticity $\omega_{max}$ with surface pressure (normalised by $p_{max}$) on the roughness for the range shown in the colour bar.}
\label{fig:lamcouette_vcont}
\end{figure}

\begin{figure}
\centering{
\includegraphics[width=70mm]{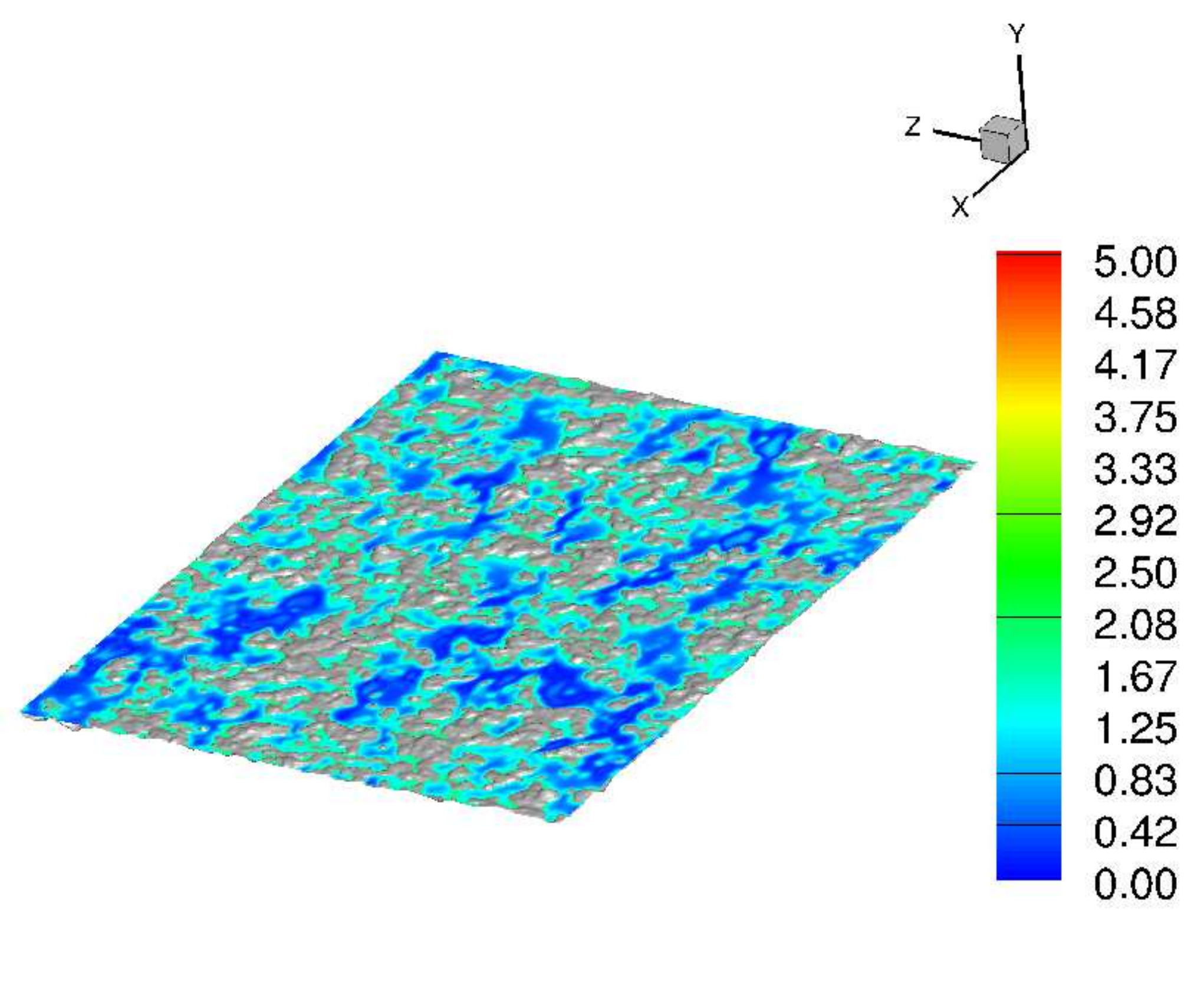}
\put(-200,150){$(a)$}
\includegraphics[width=70mm]{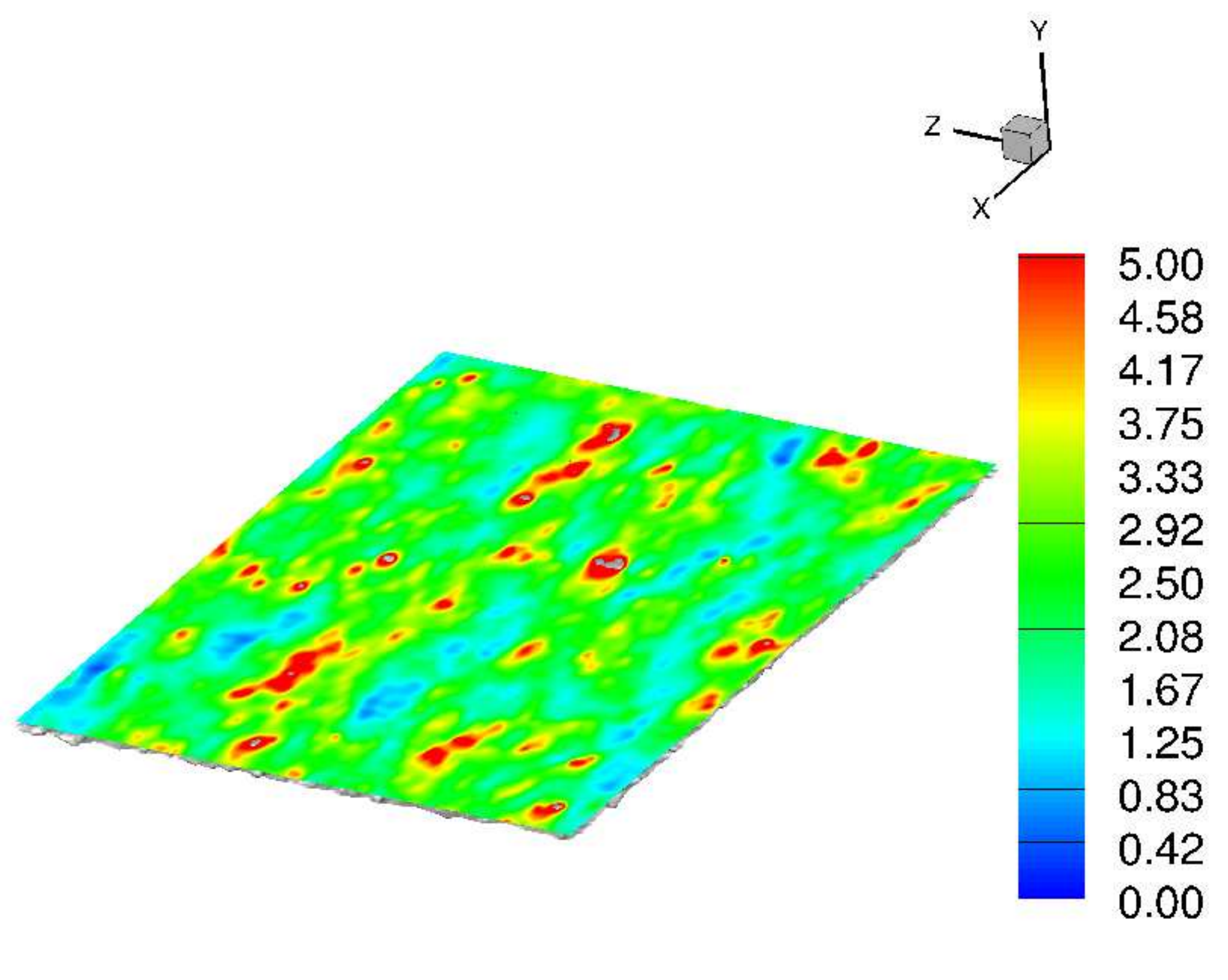}
\put(-200,150){$(b)$}
}
\\
\centering{
\includegraphics[width=70mm]{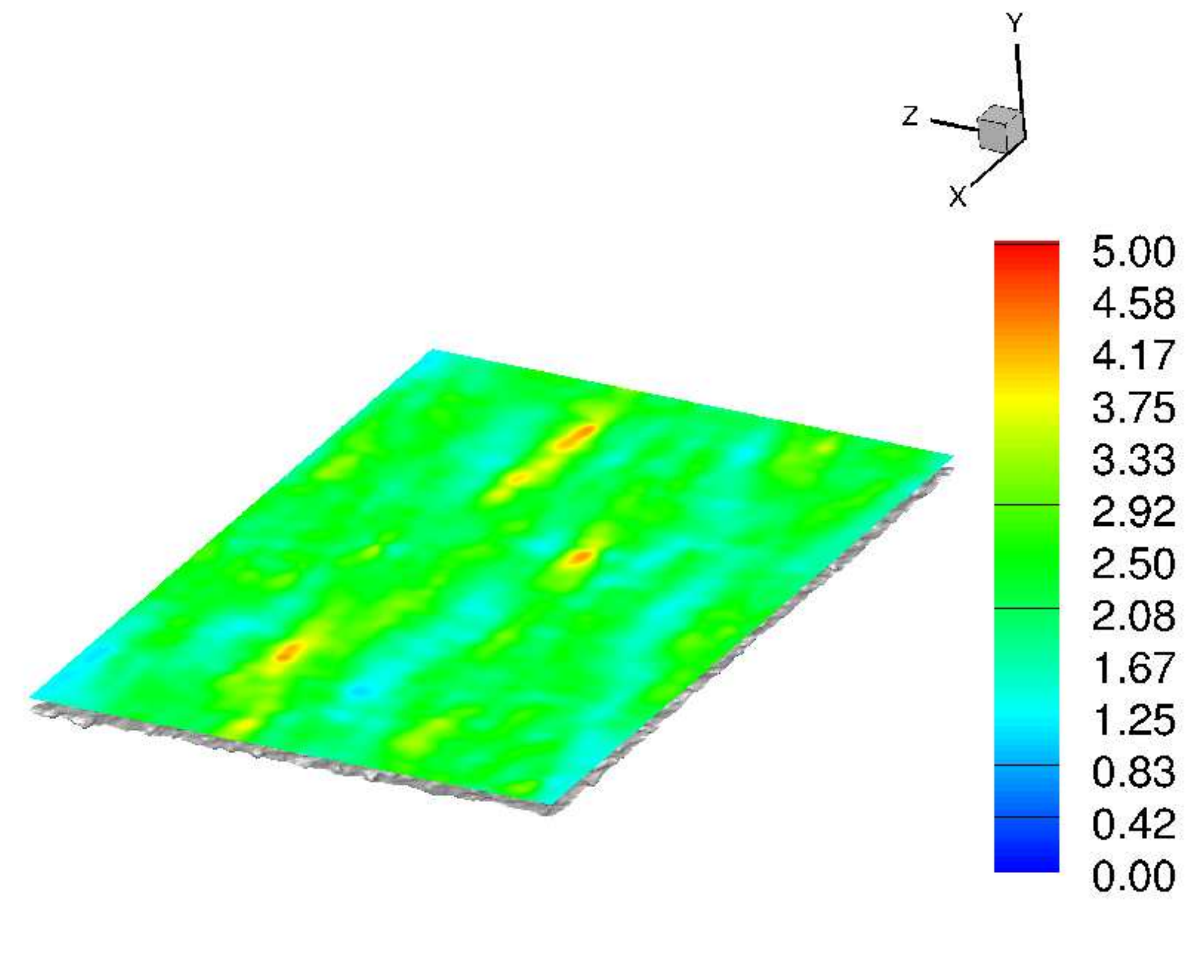}
\put(-200,150){$(c)$}
\includegraphics[width=70mm]{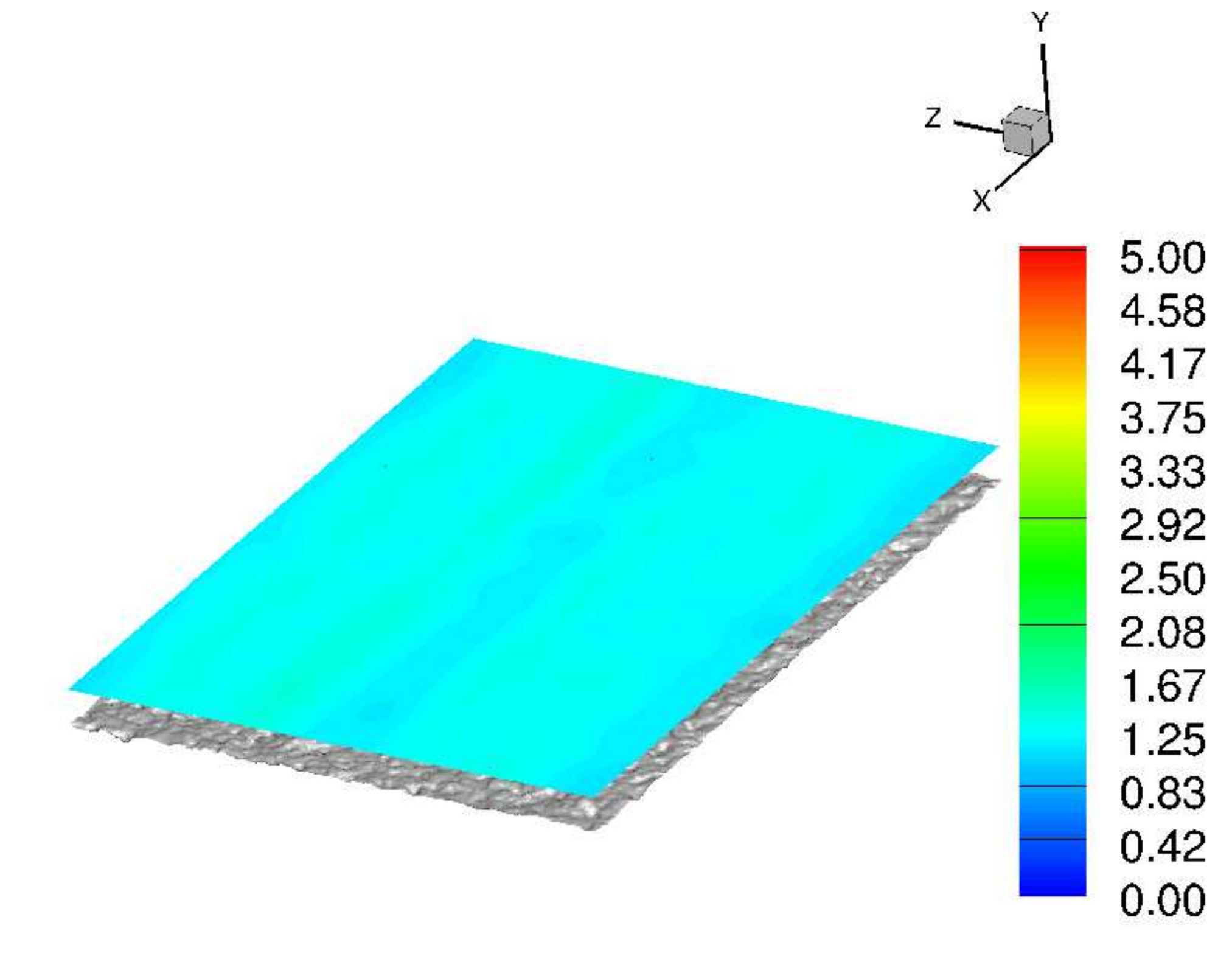}
\put(-200,150){$(d)$}
}
\caption{Percent change in the streamwise velocity field for the fully wetted laminar Couette flow (Case L-RFW) compared to the baseline smooth wall (Case L-S) at wall parallel planes: (a) $0.02H$, (b) $0.08H$, (c) $0.16H$ and (d) $0.40H$ from the bottom surface. }
\label{fig:lamcouette_upercent}
\end{figure}

Initially, the fully wetted Case L-RFW is considered. Figure \ref{fig:lamcouette_vcont} shows the flow field after it is fully developed.  The wall-normal velocity (figure \ref{fig:lamcouette_vcont}a) and the vorticity magnitude along with surface pressure (figure \ref{fig:lamcouette_vcont}b) are shown. A wall-normal velocity component into the flow is induced due to the surface asperities. Additional vorticity is generated due to the surface roughness, and large variations of pressure on the surface are evident due to the presence of peaks and valleys. 
The penetration effect of the surface roughness is illustrated in figure \ref{fig:lamcouette_upercent}, where the percent change in instantaneous streamwise velocity $(u(y)-u_o(y))/u_o(y)$ is shown for the fully wetted rough case (Case L-RFW) compared to the smooth channel case (Case L-S) at four wall-parallel planes varying from $y=0.02H$ to $y=0.4H$. The baseline streamwise velocity $u_o(y)$ represents Case L-S and $u(y)$ represents Case L-RFW. Notice that it is not until the location $y=0.4H$ that the change in velocity is less than $1\%$ suggesting that the surface roughness effects can penetrate up to that distance. 

\subsubsection{Mean flow field properties} 
\label{sec:mff}

\begin{figure}
 \centering{
 %[width=140mm]{laminar_couette_dr_3}
 \includegraphics[width=140mm]{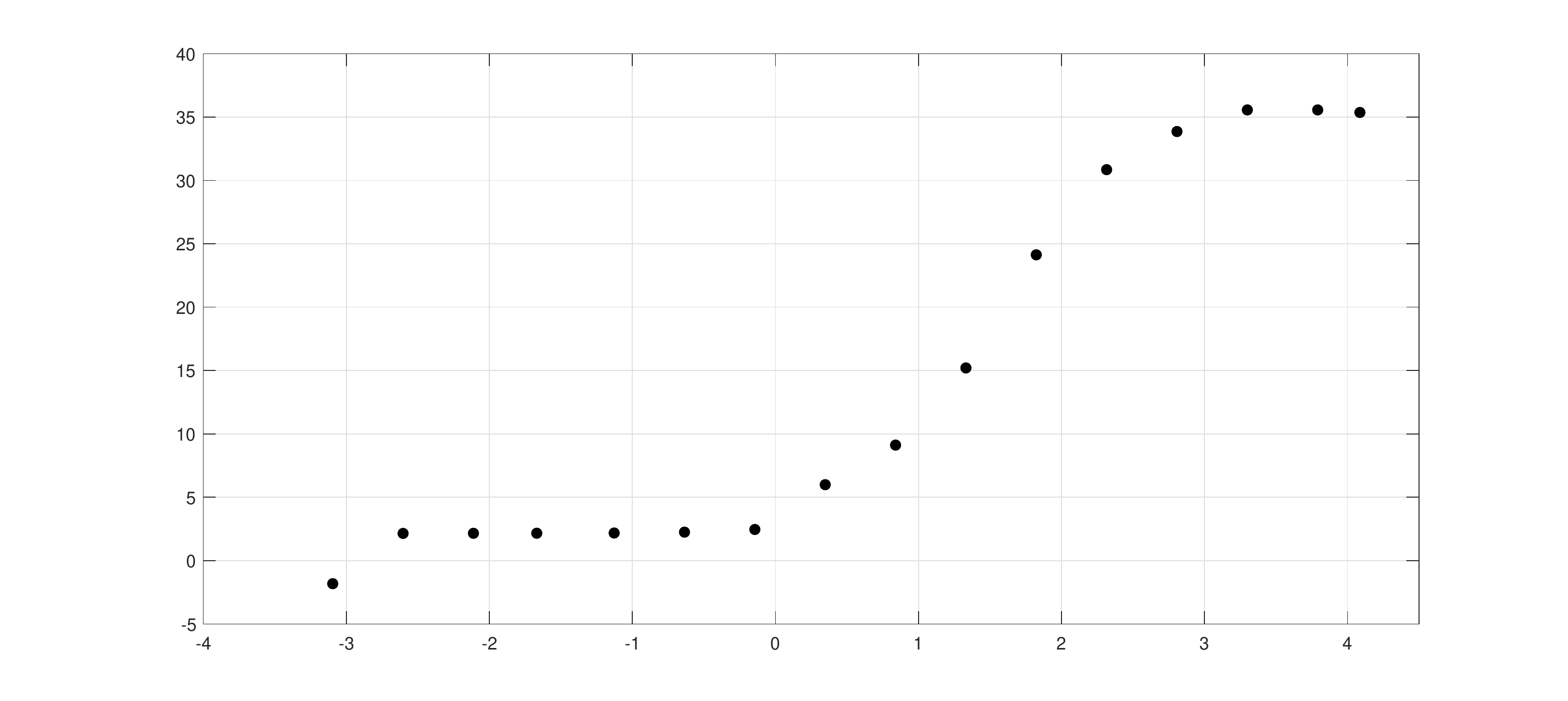}
 \put(-370,75){\rotatebox{90}{$DR(\%)$}}
 \put(-205,-5){$h/S_q$}
 \put(-250,50){Region $I$}
 \put(-140,70){Region $II$}
 \put(-90,125){Region $III$}
 }
 \caption{Laminar Couette flow: drag reduction as a function of interface height normalised by the RMS roughness height $S_q$.}
 \label{fig:dr_lam_couette}
\end{figure}

\begin{figure}
 \centering{
 \includegraphics[width=110mm]{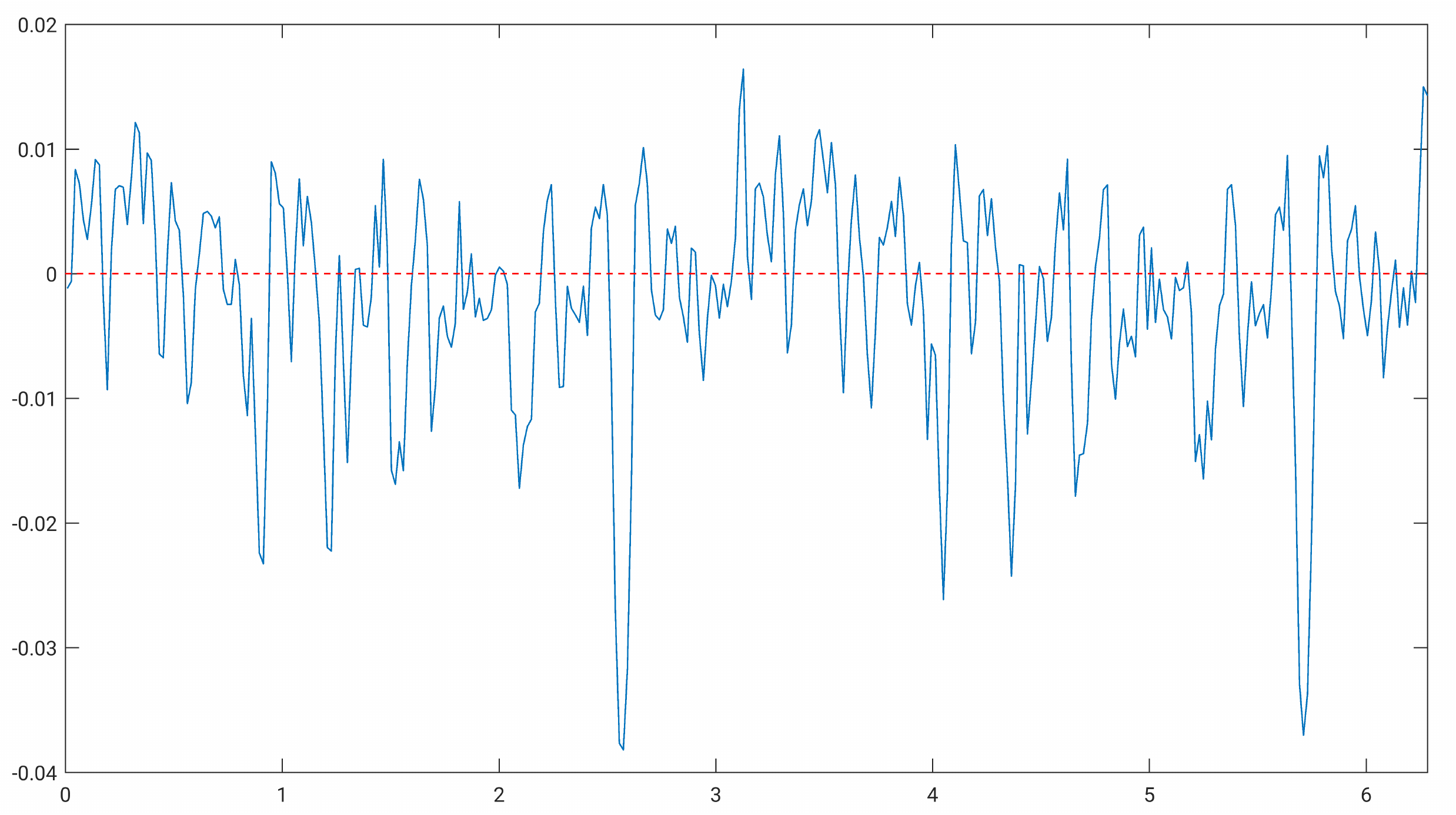}
 \put(-325,87){$L_y$}
 \put(-150,-10){$L_x$}
 }
  \caption{A 2-D slice of the surface roughness (solid blue line) to highlight negative skewness about the reference line (dashed red line).}
 \label{fig:2d_negative_skewness}
\end{figure}

\begin{figure}
\centering{
\includegraphics[width=70mm]{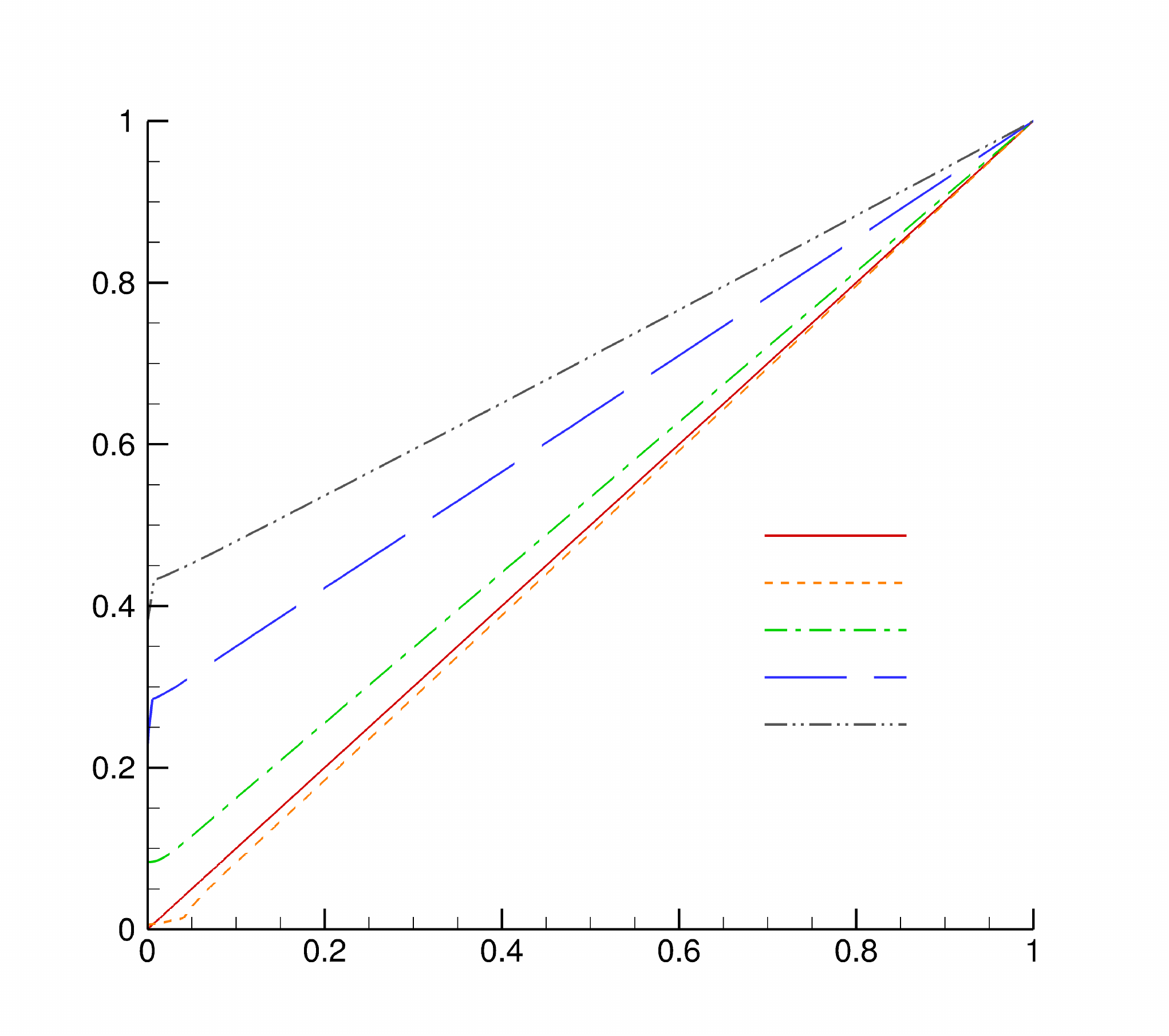}
\put(-108,0){$y/H$}
\put(-205,78){\rotatebox{90}{$u/U_\infty$}}
\put(-45,84){\scriptsize L-S}
\put(-45,76){\scriptsize L-RFW}
\put(-45,68){\scriptsize L-RI1}
\put(-45,60){\scriptsize L-RI2}
\put(-45,52){\scriptsize L-RI3}
}
\caption{Mean streamwise velocity $u$ as a function of the wall-normal distance $y$, where $u$ and $h$ are normalised with $U_\infty$ and $H$ respectively.}
\label{fig:lamcouette_uvsy}
\end{figure}

\begin{figure}
\centering{
\includegraphics[width=90mm]{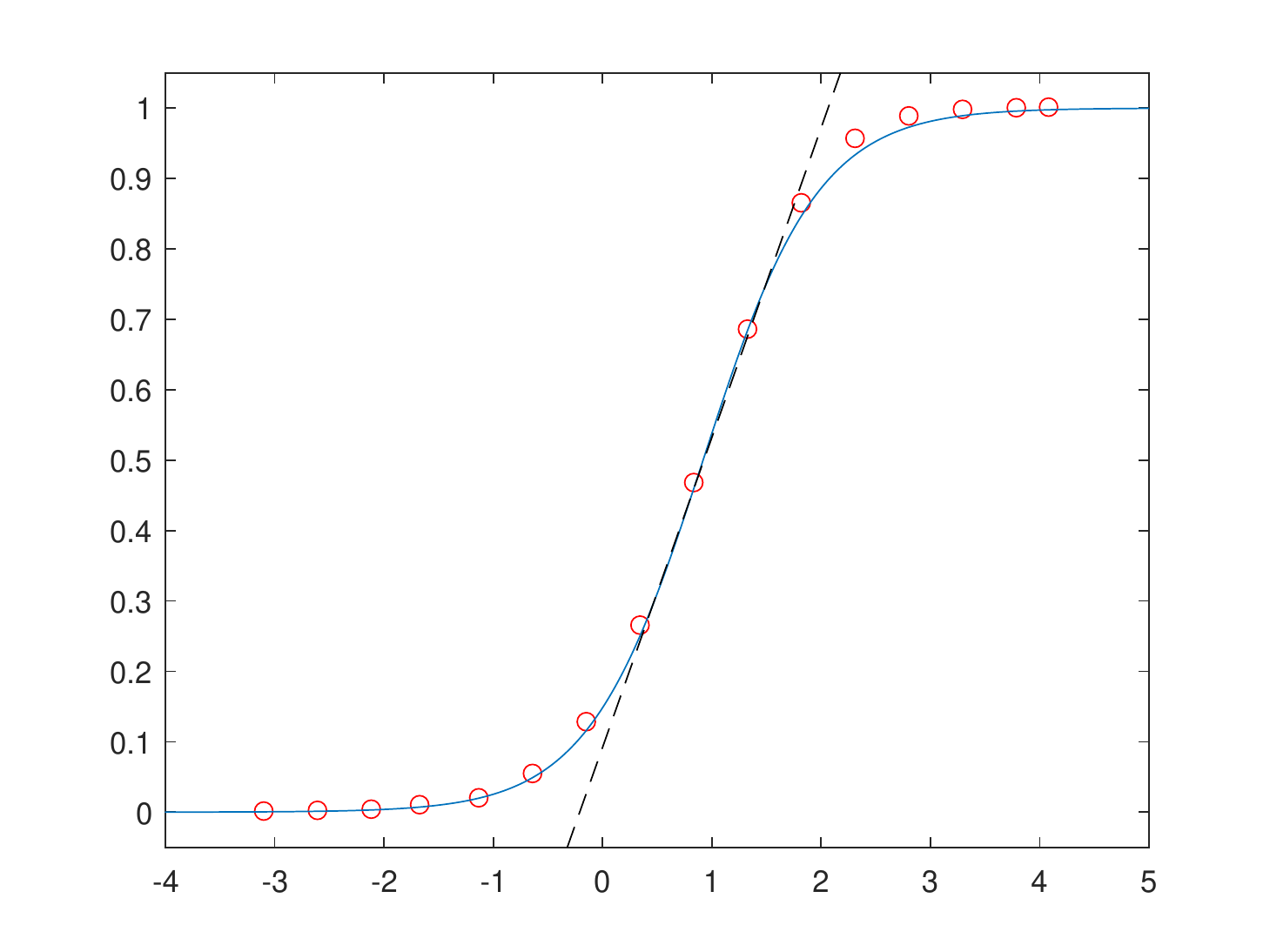}
\put(-125,0){$h/S_q$}
\put(-250,97){$\phi_g$}
\put(-130,30){$S_{sk}=-0.32$}
}
\caption{Gas fraction $\phi_g$ as a function of interface height $h$ normalised with the RMS roughness height $S_q$. The red symbols represent the data and the solid blue line represents the non-linear fit. }
\label{fig:h_phig}
\end{figure}

\begin{figure}
 \centering{
 \includegraphics[width=140mm]{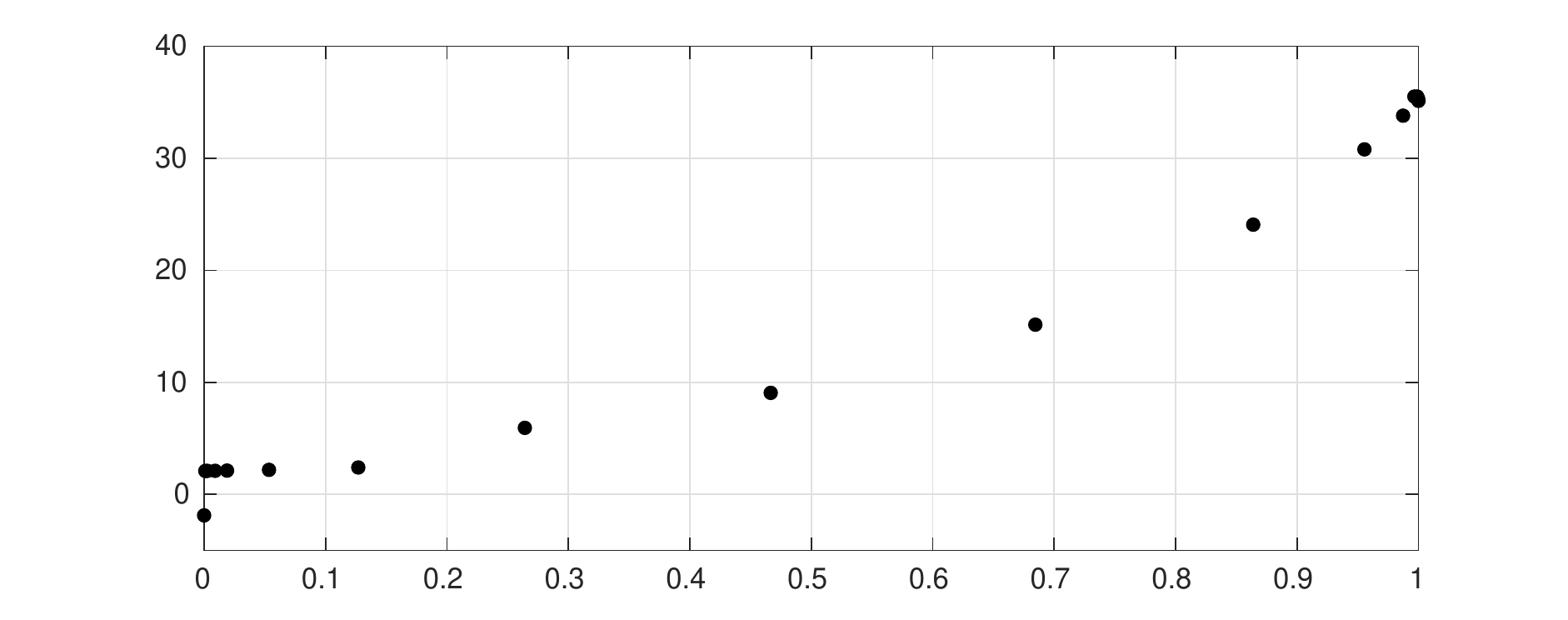}
 \put(-375,70){\rotatebox{90}{$DR(\%)$}}
 \put(-198,-5){$\phi_g$}
 }
 \caption{Laminar Couette flow: drag reduction $DR$ as a function of gas fraction $\phi_g$.}
 \label{fig:dr_lam_couette_phig}
\end{figure}

\begin{figure}
\centering{
\includegraphics[width=70mm]{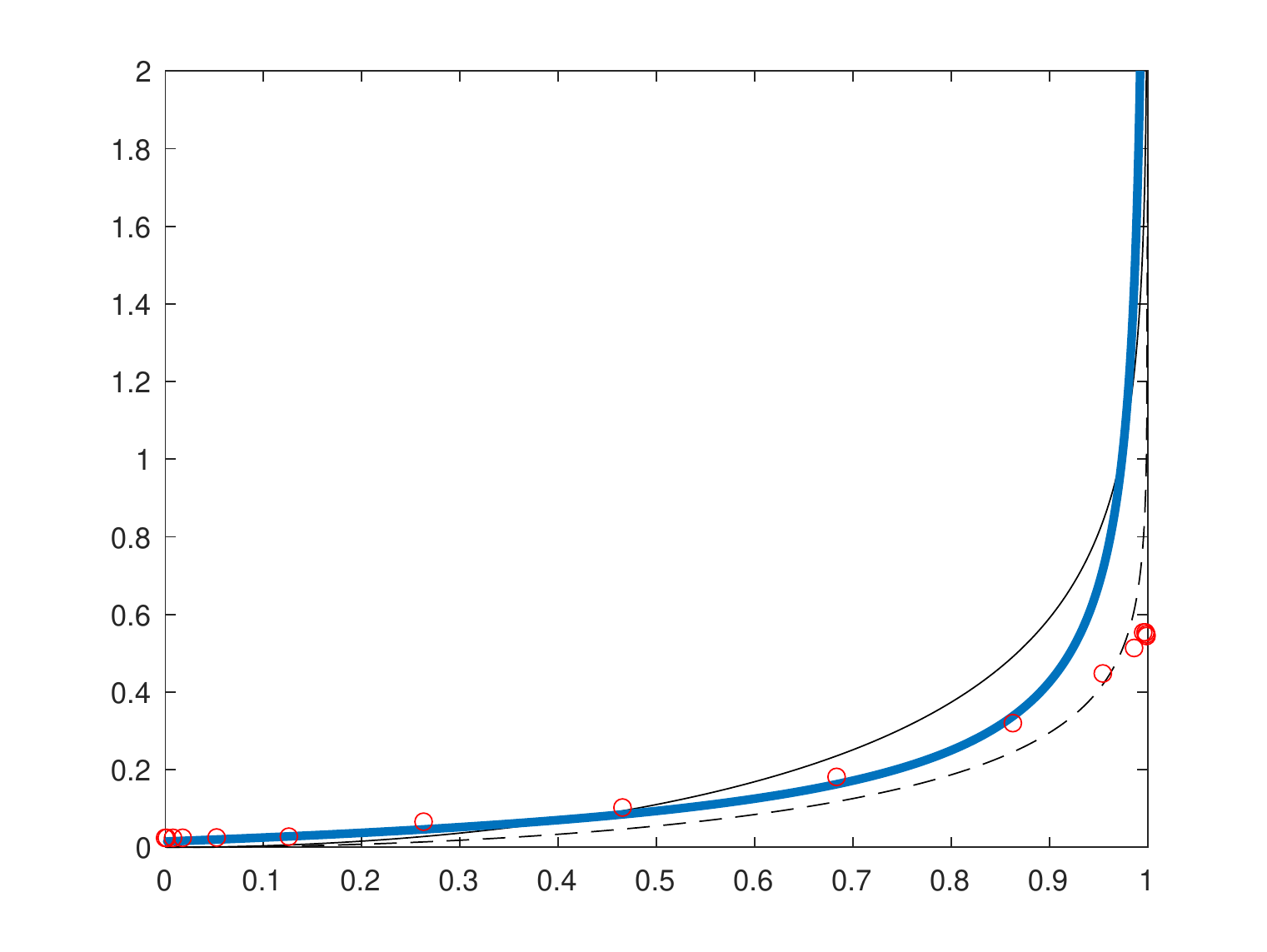}
\put(-100,2){$\phi_g$}
\put(-200,65){\rotatebox{90}{$b_{eff}/H$}}
\put(-200,150){$(a)$}
\includegraphics[width=70mm]{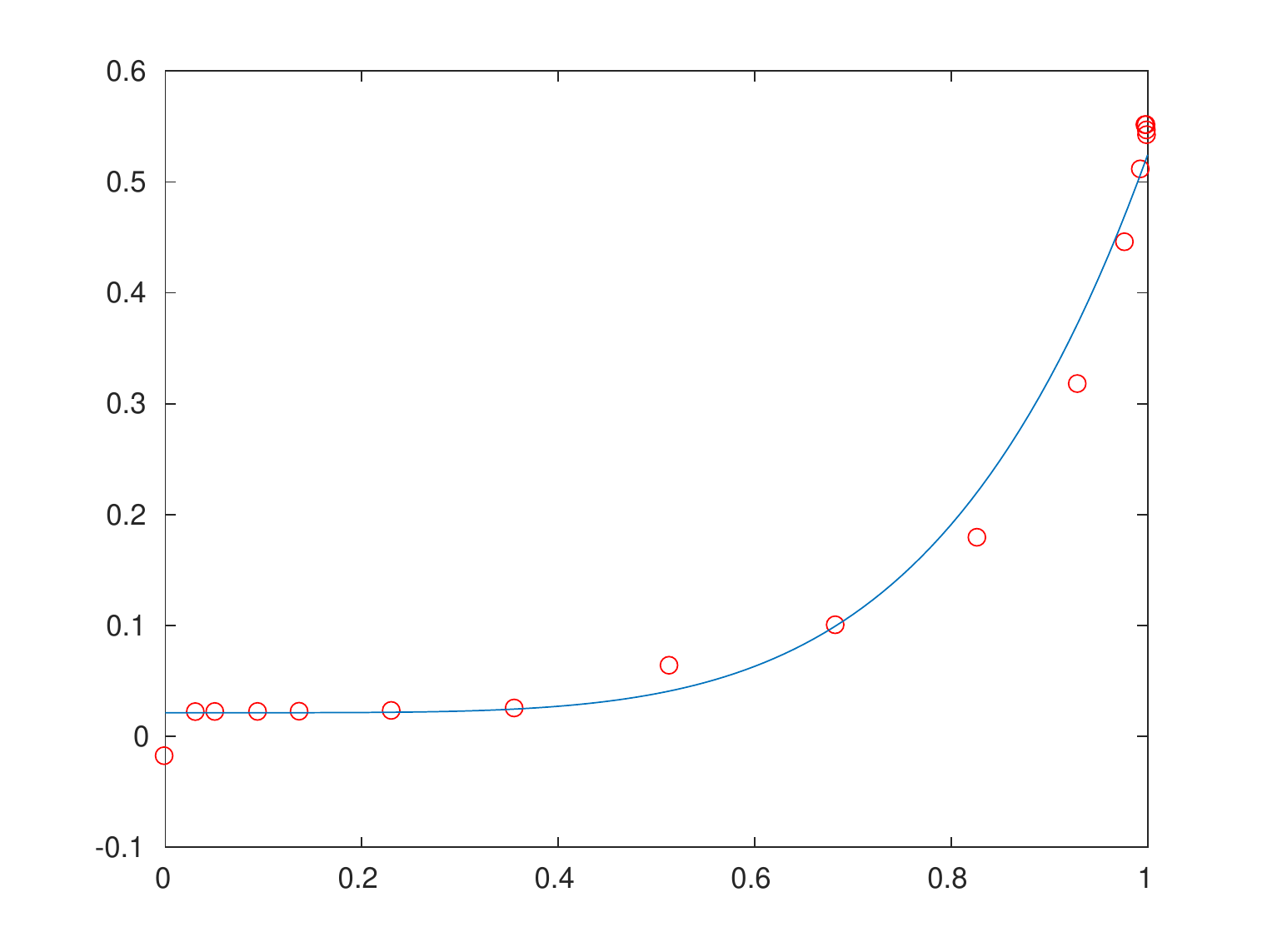}
\put(-105,2){$\sqrt{\phi_g}$}
\put(-200,65){\rotatebox{90}{$b_{eff}/H$}}
\put(-200,150){$(b)$}
}
\caption{Effective slip length $b_{eff}$ normalised with the channel height $H$ as a function of (a) the gas fraction $\phi_g$ and (b) the square root of gas fraction area $\phi_g$. In (a), the solid black line is for longitudinal grooves, solid dashed line for transverse grooves, thick solid blue line for post geometry and symbols for the random roughness. In (b), the solid blue line is the power-law fit and the symbols represent the random roughness.}
\label{fig:beff_comparison}
\end{figure}

Simulations are performed for each of the interface heights $h$ varying from $S_{v}$ to $S_{p}$. $DR$ is shown for all the interface heights in figure \ref{fig:dr_lam_couette}. The increase in $DR$ is not linear when $h$ is varied from $S_{v}$ to $S_{p}$. Note that the fully wetted case has negative $DR$ indicating that the absence of the interface has increased drag due the exposed asperities.
The presence of the interface produces nearly the same drag reduction for $h/S_q \leq -0.32$  which we will refer to as region $I$. This suggests that the value of $DR$ is insensitive to $h$ in region $I$. This holds in general for any surface with negative skewness ($S_{sk}=-0.32$) since that region $I$ holds most of the valleys. Figure \ref{fig:2d_negative_skewness} shows a 2-D slice of the surface roughness; note that the valleys dominate over the peaks about the reference line. As the interface fills up more of the cavities, the slip area becomes significant enough to cause drag reduction. In region $II$, for $-0.32 < h/S_q < 2.15 $, the increase in drag reduction is rapid since the increase in interface height exposes fewer asperities to the outer flow. The slip is enhanced due to a much larger area of air--water interface. In region $III$ beyond $h/S_q \geq 2.15$, $DR$ hits a plateau and becomes insensitive to the interface height since it covers most of the asperities. $DR$ is therefore sensitive to $h$ in the vicinity of mean roughness $S_q$. It is therefore evident from figure \ref{fig:dr_lam_couette} that we can classify the interface cases into three distinct regions: 

\begin{equation}
Region = \begin{cases}
              I ,   &\text{ if  $h/S_q \leq -0.32$};\\
              II ,  &\text{ if  $-0.32 < h/S_q < 2.15$};\\
              III , &\text{ if  $h/S_q \geq 2.15$}.
             \end{cases}
\end{equation}

One can extract a representative case from each of the three regions shown above. The baseline smooth wall is denoted as Case L-S, the fully wetted case in region $I$ is denoted as Case L-RFW, the interface at the mean elevation height of the roughness $h=0$ denoted by Case L-RI1, the interface at $h=S_q$ in region $II$ is represented by Case L-RI2 and the interface at $h=S_p$ in region $III$ is represented by Case L-RI3. The velocity profiles are extracted and compared in figure \ref{fig:lamcouette_uvsy}. Case L-RFW exhibits an increase in velocity gradient when compared to Case L-S indicating an increase in drag. Once the interface is introduced, the effect is reversed and the velocity gradient decreases for Cases L-RI1, L-RI2 and L-RI3. The effect is more pronounced in Case L-RI3 since it corresponds to the interface being at the highest peak where most of the asperities are covered and the largest slip effect is achieved.

The increase in interface height reduces the amount of rough surface area exposed to the flow. As a result, the flow is subjected to an increase in slip area. The asperities exposed to the outer flow can be represented by a solid fraction $\phi_s$, which is found by calculating the area of the rough surface above the interface normalised by the projected area of the bottom wall. It is evident that there exists a relationship between the interface height $h$ and gas fraction $\phi_g$ defined by $\phi_g = 1 - \phi_s$. This is useful since $\phi_g$ is not known \textit{a priori} and $h$ is prescribed as an initial condition. A simple nonlinear fit relates $\phi_g$ to $h/S_q$ of the form of $c[1+\tanh(ax+b)]$ and is described by the following equation:  

\begin{equation}
 %\phi_g = 0.52\tanh\left[1.23\left(\frac{h}{S_q}-S_{sk}\right)\right]-0.007\frac{h}{S_q}+0.5.
 \phi_g = 0.5\left[1+\tanh\left(0.95\frac{h}{S_q}-0.875\right)\right].
 \label{eq:phi_h}
\end{equation}
This equation can be applied to any general rough surface: the coefficients do change for different surfaces, but the overall fit is general since any surface roughness can be represented by a bearing area curve (BAC). Region $I$ represents the index of the deepest valleys where the interface is retained, region $II$ represents the core interface retention index where the maximum amount of air is trapped within the cavities, and region $III$ the upper zone index related to the largest asperities that contribute to drag.  Figure \ref{fig:h_phig} shows a comparison between the actual data and (\ref{eq:phi_h}) for $\phi_g$ as a function of $h/S_q$. The negative skewness $S_{sk}=-0.32$ coincides with the transition between regions $I$ and $II$. 

Alternatively, $DR$ can be represented as a function of slip area instead of interface height by using $\phi_g$ as shown in figure \ref{fig:dr_lam_couette_phig}. Based on the definition of $DR$ in (\ref{eq:dr}), it can be shown that $DR$ is related to the slip length $b_{eff}$ by the following equation:

\begin{equation}
DR = \frac{1}{1+\frac{H}{b_{eff}}},
\end{equation}
therefore,
\begin{equation}
\frac{b_{eff}}{H} = \frac{DR}{1-DR}.
\end{equation}

\cite{philip:1,philip:2} obtained an analytic solution for the normalised slip lengths $b_{eff}/H$ for periodic grooves oriented parallel and perpendicular to the flow respectively:
\begin{equation}
 \frac{b_{eff}}{H} = -\frac{1}{\pi}\log \bigg[ \cos \bigg(\frac{\pi}{2}\phi_g \bigg) \bigg],
\end{equation}
\begin{equation}
 \frac{b_{eff}}{H} = -\frac{1}{2\pi}\log \bigg[ \cos \bigg(\frac{\pi}{2}\phi_g \bigg) \bigg].
\end{equation}
\cite{ybertetal_2007} showed that for a post geometry, $b_{eff}$ scales with the solid fraction as
\begin{equation}
 \frac{b_{eff}}{H} \sim \frac{\alpha}{\sqrt{\phi_s}},
\label{eq:ybert_post}
\end{equation}
where $\alpha$ is a prefactor that depends on the geometry. \cite{davis_lauga_2010} were able to use superposition of point sources, where the infinite series is interpreted as a Riemann sum to obtain an analytical solution that agrees with \cite{ybertetal_2007} in the asymptotic limit of small surface coverage. A linear regression is performed on the numerically obtained data using the scaling given by eq.(\ref{eq:ybert_post}) to obtain the following expression: 
\begin{equation}
 \frac{b_{eff}}{H} = \frac{0.19}{\sqrt{\phi_s}} - 0.175.
\end{equation}

Figure \ref{fig:beff_comparison}(a) shows a comparison between the different solutions obtained for the longitudinal and transverse grooves and posts to that of random roughness. The solution for post geometry best approximates the data for a random rough geometry for $\phi_g<0.85$ but starts to diverge as gas fraction increases. In the limit of large $\phi_g$, the transverse groove solution captures the slip effect in random roughness more accurately. We present a power-law fit to the current data to obtain an expression for $b_{eff}$ as a function of $\phi_g$: 
%in figure \ref{fig:beff_phig} against $\sqrt{\phi_g}$. A linear regression is performed on the calculated data when $\phi_g > 0$ and the scaling coefficients of the power law are rounded to the nearest number and given by the following equation:
\begin{equation}
 \frac{b_{eff}}{H} \bigg|_{\phi_g>0} = 0.5\left({\phi_g}\right)^{5/2} + 0.02.
\end{equation}
This formula provides a simple expression for slip length over the rough surface given the amount of gas fraction present when $\phi_g>0$ and is shown in figure \ref{fig:beff_comparison}(b). The fully wetted roughness $b_{eff}$ remains an outlier. 

\subsection{Turbulent channel flow}
\subsubsection{Mean velocity profiles}
 
\begin{figure}
\centering{
\includegraphics[width=70mm]{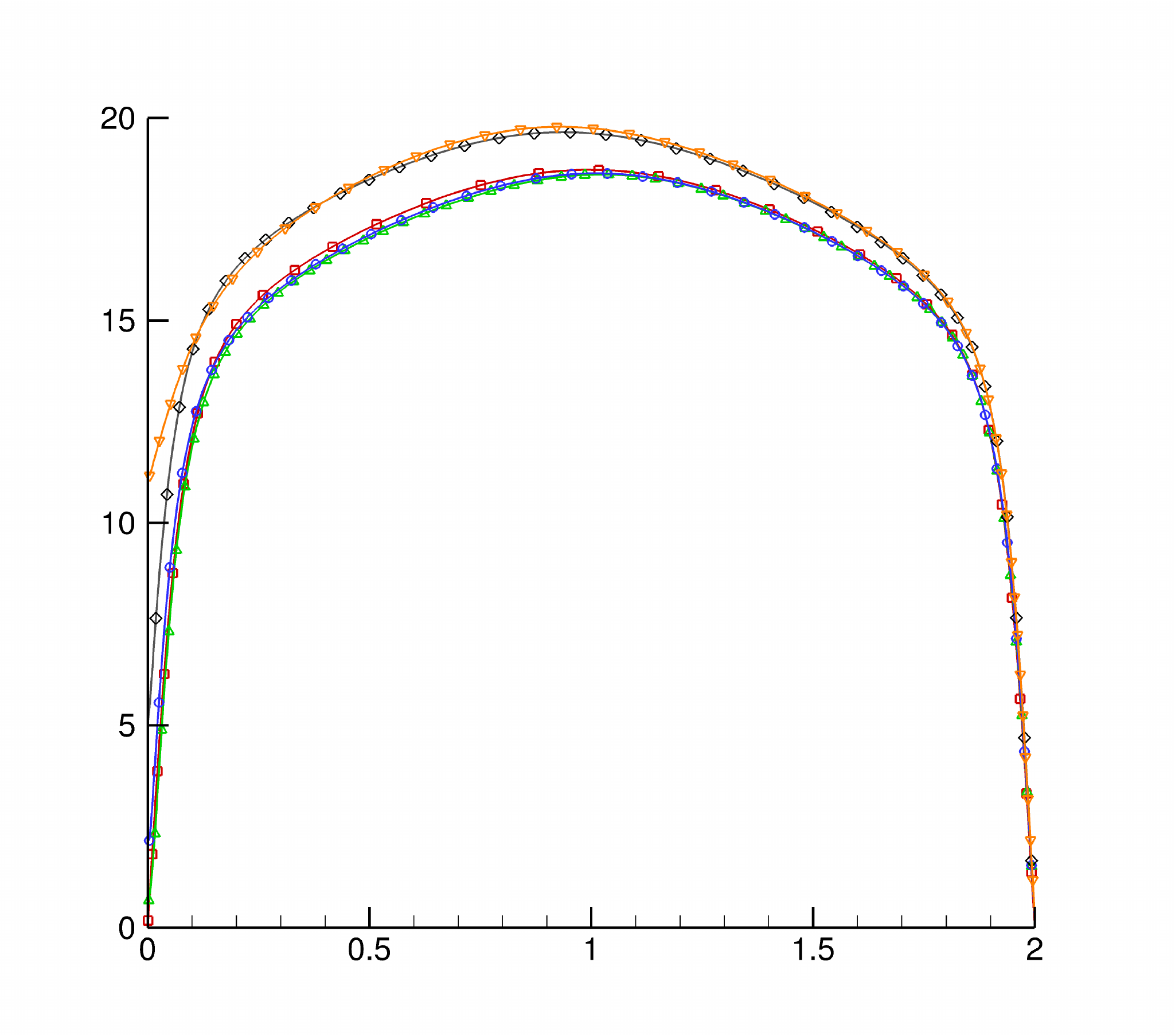}
\put(-105,5){$y/\delta$}
\put(-195,80){\rotatebox{90}{$U/u_{\tau}$}}
%\put(-120,75){\scriptsize T-S}
%\put(-120,64){\scriptsize T-RFW}
%\put(-120,53){\scriptsize T-RI1}
%\put(-120,43){\scriptsize T-RI2}
\put(-200,150){$(a)$}
\includegraphics[width=70mm]{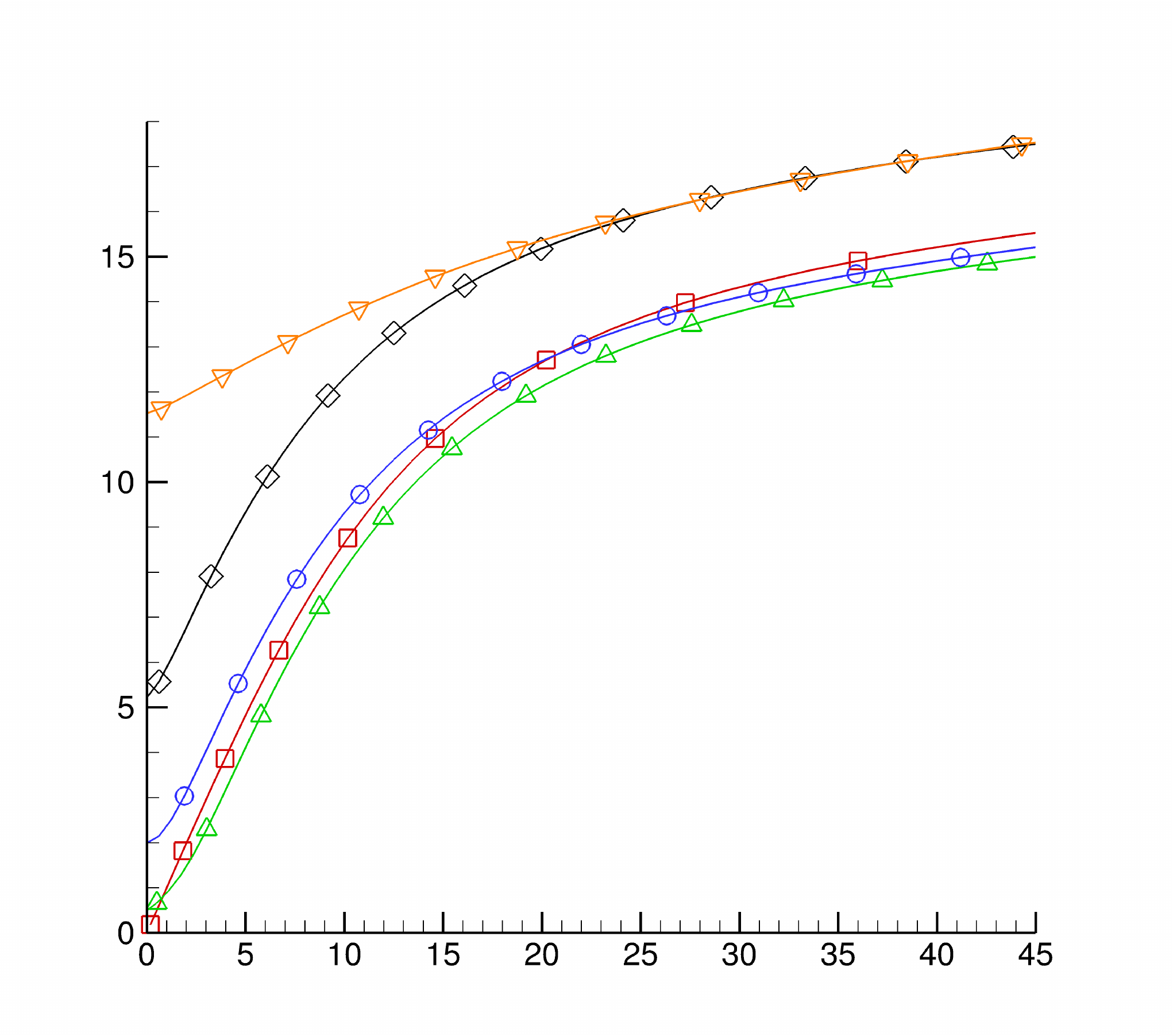}
\put(-110,5){$y^+(u^B_{\tau})$}
\put(-195,80){\rotatebox{90}{$U/u^B_{\tau}$}}
%\put(-135,150){\scriptsize T-S}
%\put(-135,140){\scriptsize T-RFW}
%\put(-135,129){\scriptsize T-RI1}
%\put(-135,119){\scriptsize T-RI2}
\put(-200,150){$(b)$}
}
\caption{Mean profile of (a) velocity normalised with the average friction velocity $u_{\tau}$ as a function of wall-normal distance normalised by the channel half-height $\delta$ and (b) close-up near the wall region of the velocity normalised by the bottom wall friction velocity $u^B_{\tau}$ as a function of wall-normal distance $y^+(u^B_{\tau})$. Symbols for each case are: Case T-S ($\Box$), Case T-RFW ($\vartriangle$), Case T-RI1 ($\ocircle$), Case T-RI2 ($\Diamond$), Case T-RI3($\triangledown$). The symbols are not representative of the grid resolution.}
\label{fig:mean_profile}
\end{figure}

\begin{figure}
\centering{
\includegraphics[width=70mm]{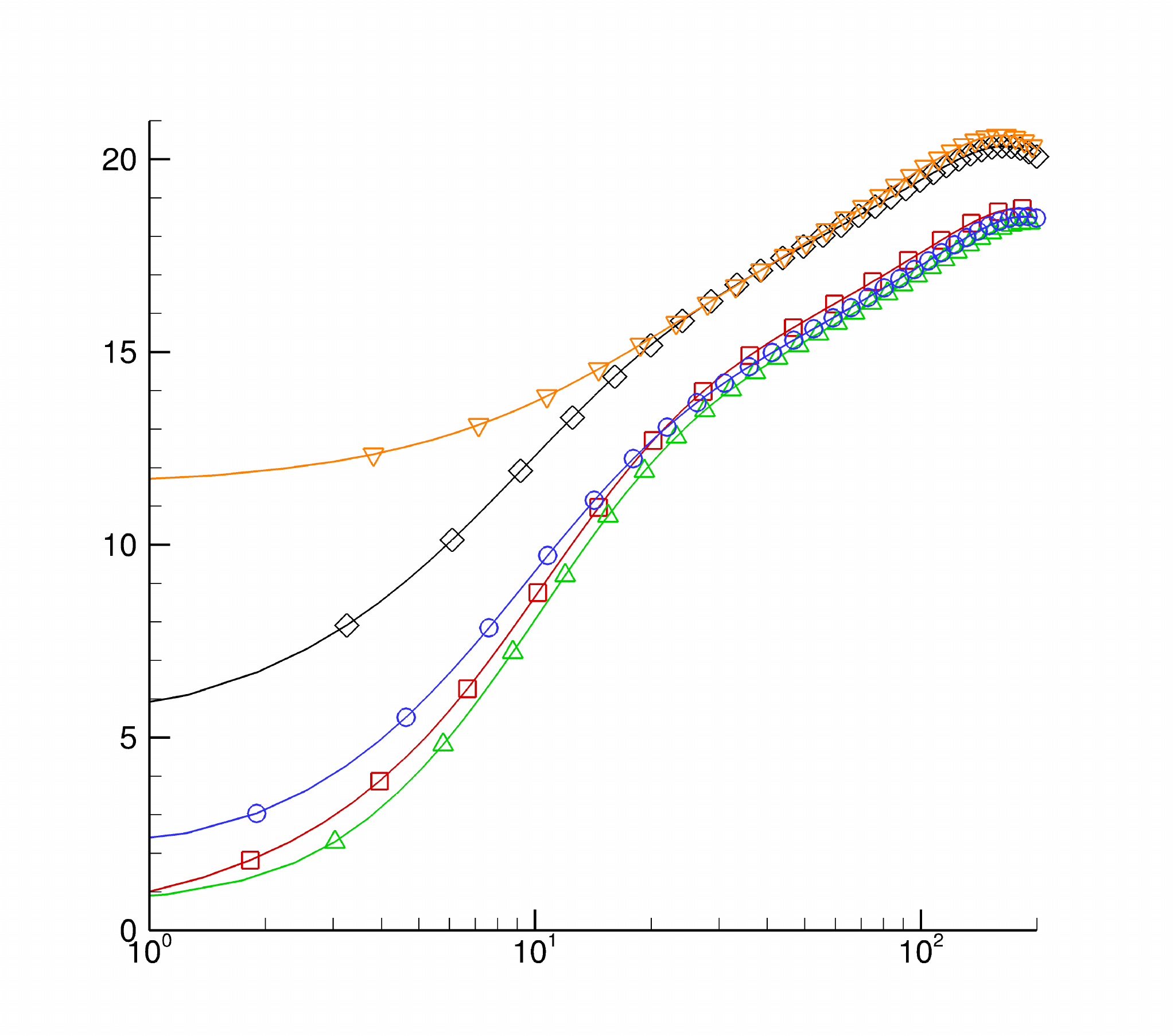}
\put(-100,5){$y^+(u^B_{\tau})$}
\put(-195,80){\rotatebox{90}{$U/u^B_{\tau}$}}
% \put(-120,75){\scriptsize T-S}
% \put(-120,64){\scriptsize T-RFW}
% \put(-120,53){\scriptsize T-RI1}
% \put(-120,43){\scriptsize T-RI2}
\put(-200,150){$(a)$}
\includegraphics[width=70mm]{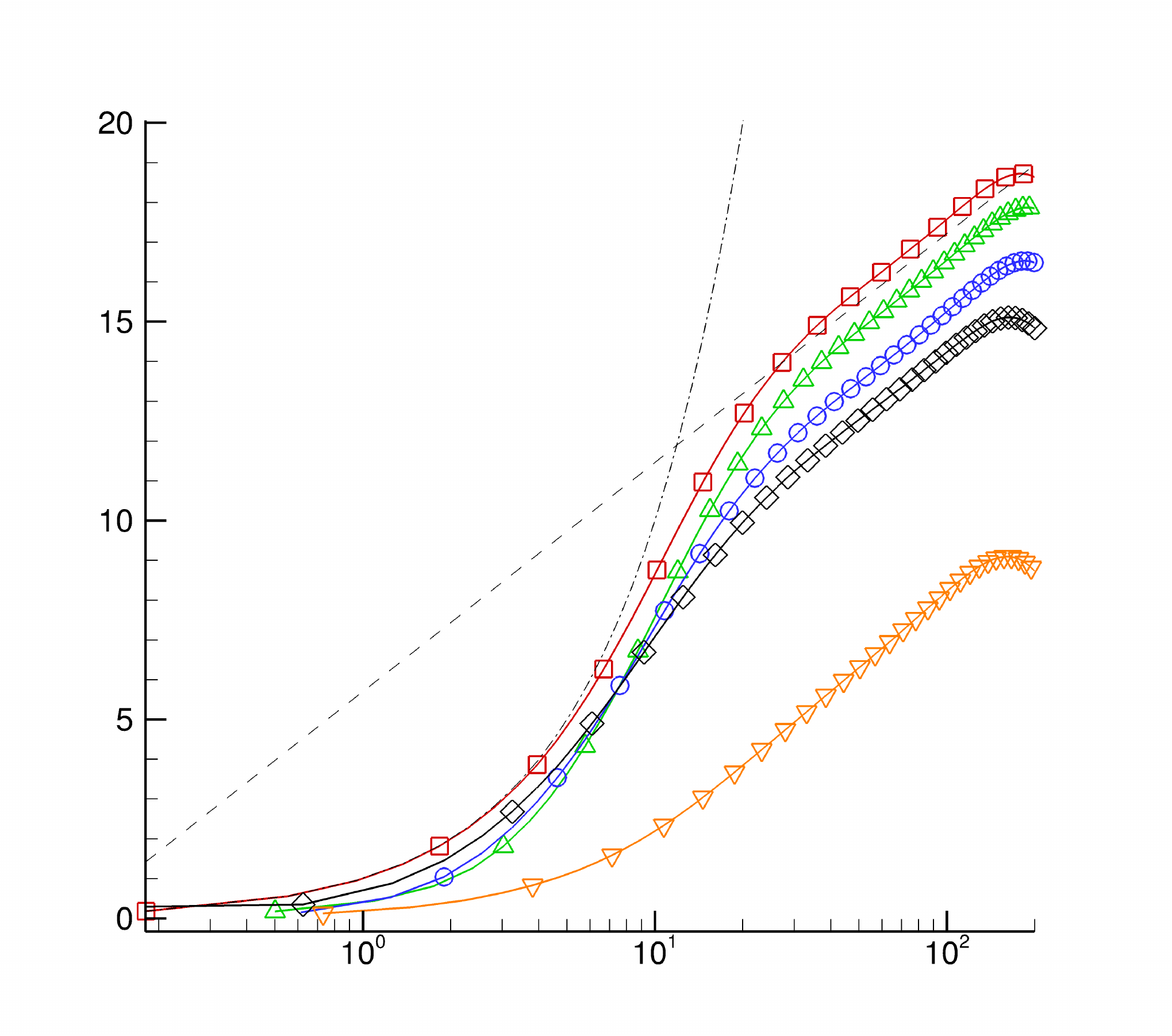}
\put(-105,5){$y^+(u^B_{\tau})$}
\put(-195,65){\rotatebox{90}{$(U-u_s)/u^B_{\tau}$}}
% \put(-135,150){\scriptsize T-S}
% \put(-135,140){\scriptsize T-RFW}
% \put(-135,129){\scriptsize T-RI1}
% \put(-135,119){\scriptsize T-RI2}
\put(-200,150){$(b)$}
}
\caption{Semi-log plot of the mean profile for (a) velocity normalised with the bottom wall friction velocity $u^B_{\tau}$ of each respective case and (b) velocity shifted by the corresponding slip velocity all normalised by the bottom wall friction velocity $u^B_{\tau}$ as a function of the wall-normal distance $y^+(u^B_{\tau})$. Symbols for each case are: Case T-S ($\Box$), Case T-RFW ($\vartriangle$), Case T-RI1 ($\ocircle$), Case T-RI2 ($\Diamond$), Case T-RI3($\triangledown$). The symbols are not representative of the grid resolution.}
\label{fig:mean_profile_log}
\end{figure}
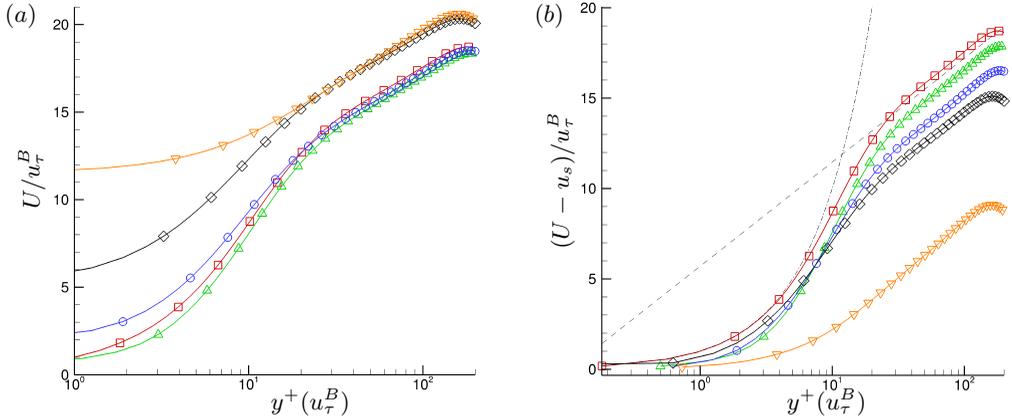

Mean velocity profiles and Reynolds stresses are computed for five cases (T-S, T-RFW, T-RI1, T-RI2 and T-RI3). Figure \ref{fig:mean_profile}(a) shows the mean velocity profile $U$ normalised by the average friction velocity $u_{\tau}$ of each corresponding case, as a function of the wall-normal distance $y$ normalised by the channel half-height $\delta=L_y/2$. The smooth channel (Case T-S) is also shown for reference. The presence of roughness (Case T-RFW) causes a small slip effect: it shows an insignificant decrease of $0.6$\% in peak value of $U/u_{\tau}$ and about a $3$ \% shift in its centreline location away from the rough wall. The presence of an interface (Case T-RI1) shows a further increase in slip, a slight increase in the centreline peak value of $U/u_{\tau}$ and a $1$\% shift in its location towards the SHS wall. The presence of an interface for Case T-RI1 adds just enough slip to offset the effect of roughness. The slip effect is more pronounced for Case T-RI2 and is largest for Case T-RI3 when the interface location covers all the roughness. The mean peak velocity $U/u_{\tau}$ increases and the mean profile shifts towards the SHS wall by $5$\% and $6$\% for Cases T-RI2 and T-RI3 respectively when compared to Case T-S. 

Figure \ref{fig:mean_profile}(b) shows a close-up view near the SHS wall of the mean velocity profile $U$ normalised by the bottom wall friction velocity $u^B_{\tau}$ as a function of the wall-normal distance $y^+(u^B_{\tau})$. The scaling with $u^B_{\tau}$ describes a more accurate picture in terms of slip and drag reduction. Case T-RFW shows a slip velocity at the wall due to the presence of roughness. The mean velocity profile is $5.8$\% lower than the baseline case in the viscous wall region ($y^+<50$), indicating an overall increase in drag. This effect is not apparent when the mean velocity profile is scaled with the average $u_{\tau}$. Case T-RI1 sees a further increase in slip at the wall due to the presence of an interface: the mean velocity profile is around $14$\% higher within the viscous sublayer and extends into the buffer layer where the two velocity profiles of Case T-S and Case T-RI1 intersect at $y^+ \thicksim 22$. Cases T-RI2 and T-RI3 exhibit the largest slip at the wall: the two profiles intersect at $y^+ \thicksim 28$. The intersection of the velocity profiles is due to asymmetry caused by the slip effect which shifts the profile towards the SHS. This trend has been observed in \cite{Martell2009,Martell2010} for longitudinal grooves and post geometries. In terms of scaling laws, the law of the wall $u^+=y^+ + C$ still holds ($C$ is the constant shift that represents the normalised slip velocity), with the exception of Case T-RFW and Case T-RI3. Case T-RFW shows a deviation from the law of the wall at $y^+<3$, where the Reynolds stress is negligible compared to the viscous stress, but follows it for $3<y^+<10$. This implies that shifting the profile by the normalised slip velocity $u^+_s$ will cause the profile to move further below the baseline case. This behaviour in the profile for the fully wetted case has been demonstrated in the literature \citep{yuan2014}. Case T-RI3 shows a complete deviation from the law of the wall due to the large amount of slip which modifies the slope of the velocity profile such that $u^+=\alpha y^+ + C$ where $\alpha<1$. 

A semi-log plot is shown in figure \ref{fig:mean_profile_log}(a) where unlike figure \ref{fig:mean_profile}(a), the mean velocity profile $U$ is normalised with the bottom wall friction velocity $u^B_{\tau}$. The slip effect is more pronounced in the near-wall region and the difference in peaks are more apparent. The mean velocity profile of Case T-RFW shows a positive slip effect as mentioned earlier: the mean velocity profile is lower than Case T-S over all regions ranging from the viscous sublayer through the log law. The roughness reduces the overall mass flux indicating an increase in drag. The profile shifts down and away from the wall by roughly $6$\% from the centreline location of the baseline. The peak velocity at the centreline is lower than Case T-S by $2$\%. Case T-RI1 shows a further increase in slip and the velocity profile is shifted further up indicating a drag reduction. The presence of an interface at that specific height is not enough to overcome the effect of roughness. This is evident from the velocity profile which is still $1$\% below the baseline case. The profile shifts back towards the SHS by around $2$\% from the centreline location of Case T-RFW. The near-wall slip seems to affect the velocity profile only within the viscous wall region for Cases T-RFW and T-RI1 when compared to Case T-S; however the log law region shows a collapse in the data where the difference is within $2$\%-$3$\%. Cases T-RI2 and T-RI3 show a large increase in slip near the wall with Case T-RI3 having the largest slip given that the roughness is fully covered by an interface. The peaks for those cases are $8$\% and $10$\% higher than Case T-S respectively. The profiles for both Cases T-RI2 and T-RI3 shift closer to the SHS wall such that their centrelines are $8.75$\% and $11$\% away from Case T-S respectively. Overall, the largest difference is clearly seen in the viscous wall region but the log law region shows a collapse for Cases T-RI2 and T-RI3. This collapse holds until $y^+=40$ and the deviation in the slope of the log law region in Case T-RI3 becomes more apparent for $y+>50$ when compared to Case T-RI2. Overall, fully wetted roughness exhibits a decrease in mass flux whereas the presence of an interface increases mass flux. This is evident by the downward shift in the log-law region for Case T-RFW and an upward shift for Cases T-RI1, T-RI2 and T-RI3. We can conclude that the trend for an increase in mass flux (more fluid mass moving) directly correlates with drag reduction and vice versa. This is due to the fact that for all the cases, $Re_{\tau}$ and pressure gradient are held constant. It has been shown that for structured geometries (grooves and posts) not only does the gas fraction $\phi_g$ matter, but also the gap spacing \citep{ourothstein_2005,Daniello2009,Martell2009}. For a random rough geometry with varying interface heights, the gas fraction is indeed increasing, but the gap spacing is also altered since more pockets are being filled with air. Notice that between Cases T-RI2 and T-RI3 there was not much increase in peak centreline velocity aside from the large increase in slip near the wall which does not substantially alter the total mass flux. However a significant increase in mass flux is observed between Cases T-RI1 and T-RI2. This can be attributed to the change in gap spacing. Although the gas fraction increases with increasing interface height, more gaps and surface valleys are covered up, since the surface roughness is dominated by valleys due to negative skewness. As the interface covers nearly all the valleys then a sharp increase in drag reduction is observed. Case T-RI1 is dominated by small gap size features which may be ineffective in reducing drag; this is in agreement with the literature on longitudinal grooves and posts. 

The velocity profiles can be corrected by offsetting them with the slip velocity as shown in figure \ref{fig:mean_profile_log}(b). Close to the wall we see a good collapse in the viscous sublayer with an early departure $y^+>1$ from the $u^+=y^+$. With the roughness fully covered by the interface in Case T-RI3, a large deviation from the law of the wall is observed when compared to other cases. This implies that we should expect the structures of the wall-normal turbulence to remain intact for Cases T-RFW, T-RI1 and T-RI2, and Case T-RI3 to be fundamentally different. Away from the near-wall, the relative velocity $(U-u_s)/u^B_{\tau}$ decreases with roughness and increasing interface height (equivalently with increasing $\phi_g$). This trend has also been observed in \cite{Turk2014} for structured geometries. The profile in the log-law region is given by
\begin{equation}
\frac{(U-u_s)}{u^B_{\tau}}=\frac{1}{\kappa}\log(y^+(u^B_{\tau})) + B,
\end{equation}
where $\kappa$ is the von K{\'a}rm{\'a}n constant and $B$ is the intercept. The value of $\kappa$ decreases from $0.41$ to $0.38$ for Case T-RI3 but remains the same for the other cases. The value of $B$ decreases with increasing $\phi_g$ going from $B \approx 5.5$ to $B \approx 2$ for Case T-RI2 and $B \approx -4$ for Case T-RI3. Similar trends have been observed in the literature \citep{Busse2012,yuan2014,Busse2017}. The decrease in $B$ implies an increase in friction as discussed in the literature through surface manipulation \citep{Luchini1991,jimenez1994,mayoraljimenez_2011}. Therefore a decrease in $B$ is associated with an increase in friction due to roughness and a decrease in friction due to SHS, and in order to differentiate the two, we consider the following argument. This increase in friction that is typically associated with surface roughness is offset by the drag reducing slip velocity $u_s$ due to the presence of an interface in SHS. Roughness induces slip, but the velocity profile in the viscous wall region does not necessarily follow the law of the wall. If we were to compare it to a smooth channel, the profile of the rough channel would intersect the profile of the smooth channel somewhere in the viscous sublayer ($y^+<5$) to merge into the log region which is shifted below the baseline indicating an increase in drag. If we offset the profile by the slip velocity, the near-wall region does not collapse but moves further below. If we take the SHS with structured geometries, the interface is flush with the top location of the roughness. The slip effect is also present but the velocity profile in the viscous wall region obeys the law of the wall with some offset $u^+=y^+ + C$; therefore if the velocity profile is shifted by the slip velocity, then a collapse in the near-wall region is observed. What we see in our simulation is somewhere in between. This is simply due to the fact that even when we have an interface, some roughness protrudes. Our analysis shows a combination of both behaviours where the law of the wall holds to a certain extent in the viscous sublayer before any appreciable deviation is observed. Also with increasing interface height, the profile in the viscous wall region tends to move closer to the baseline case and away from the fully wetted roughness, with the exception of Case T-RI3 in which the log-law region seems to extend to the vicinity of the wall, disrupting the near-wall cycle. In general, rough surfaces tend to shift both the law of the wall and the logarithmic region away from baseline while SHS tend to shift the law of the wall closer and simultaneously moving the logarithmic region away from baseline. 

\subsubsection{Scaling laws}
\begin{figure}
 \centering{
 \includegraphics[width=65mm]{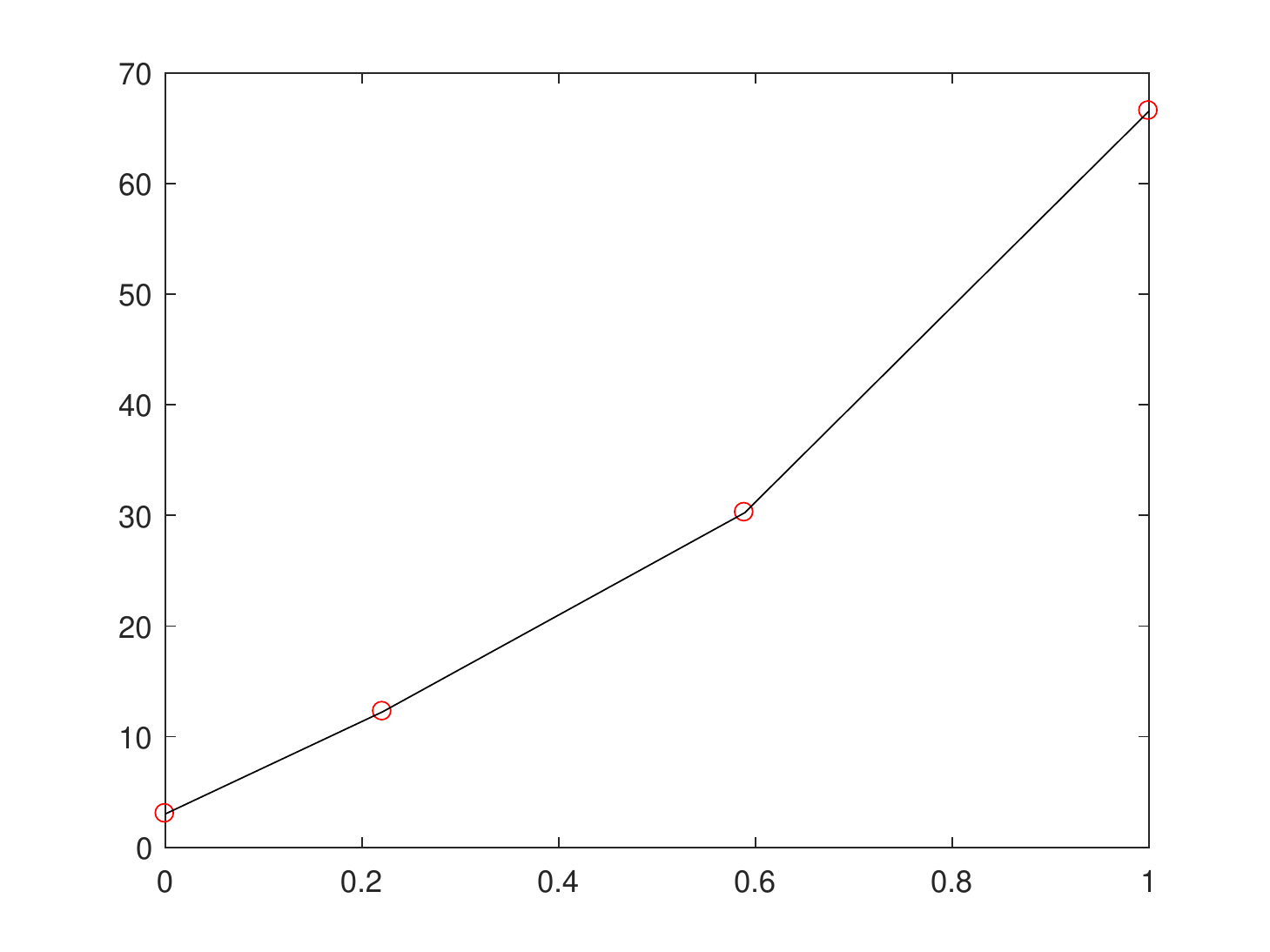}
 \put(-92,0){$\phi_g$}
 \put(-185,55){\rotatebox{90}{$u_{s}/U_{b}$(\%)}}
 \put(-180,140){$(a)$}
 \includegraphics[width=65mm]{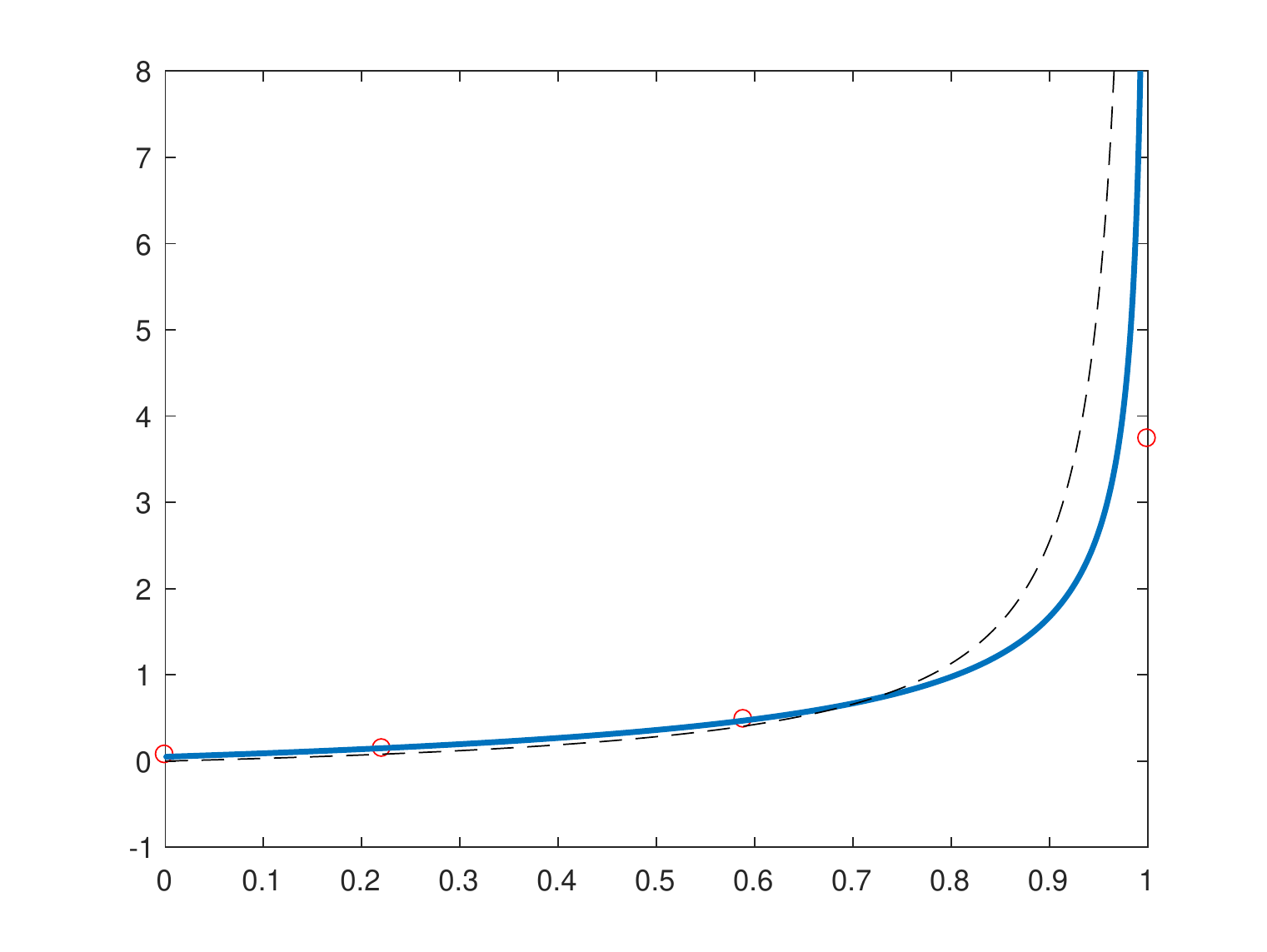}
 \put(-92,0){$\phi_g$}
 \put(-180,60){\rotatebox{90}{$b_s/L$}}
 \put(-180,140){$(b)$}
 }
 \centering{
 \includegraphics[width=65mm]{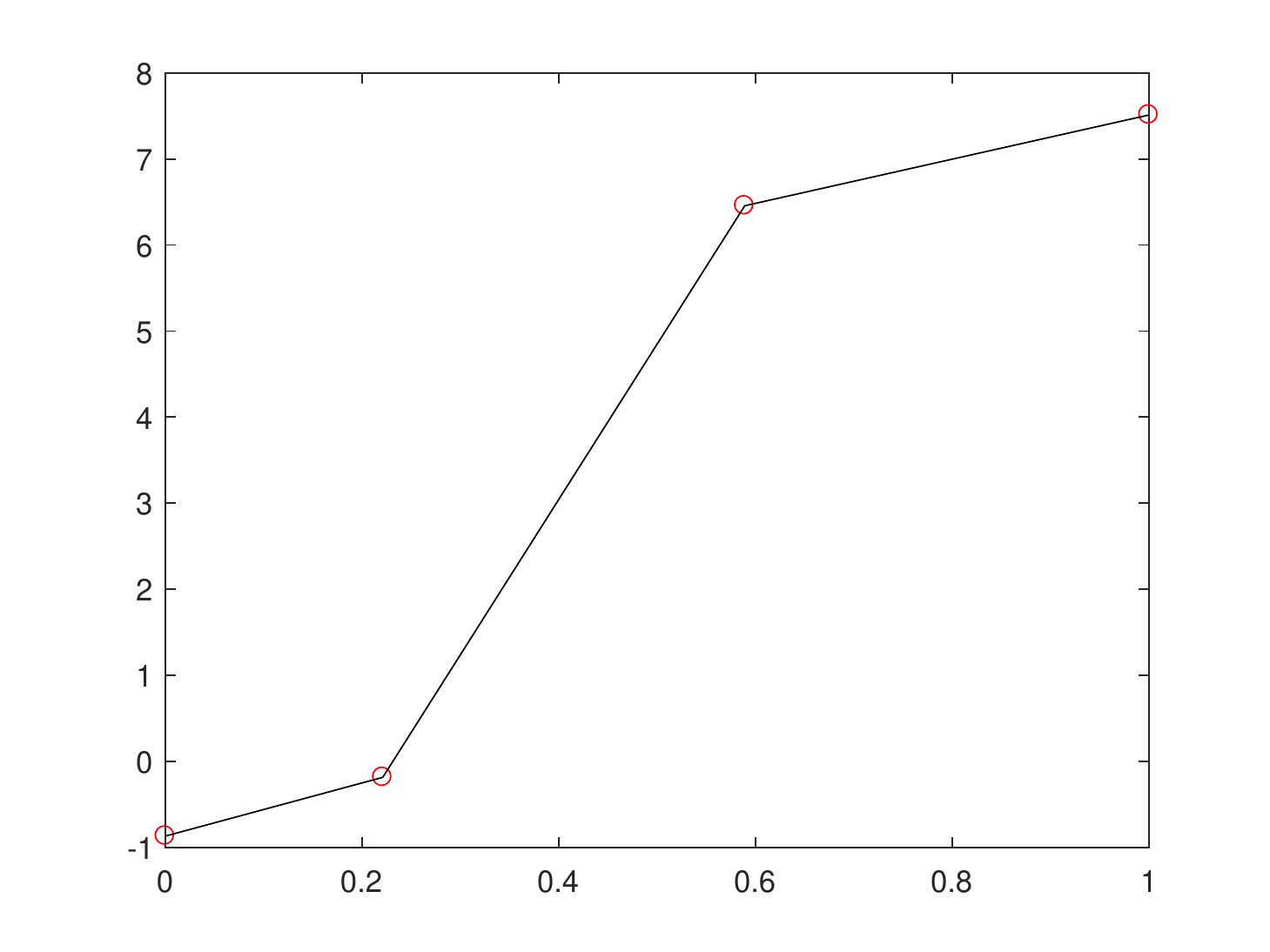}
 \put(-92,0){$\phi_g$}
 \put(-180,45){\rotatebox{90}{$\Delta U_{b}/U_{b,0}$(\%)}}
 \put(-180,140){$(c)$}
 \includegraphics[width=65mm]{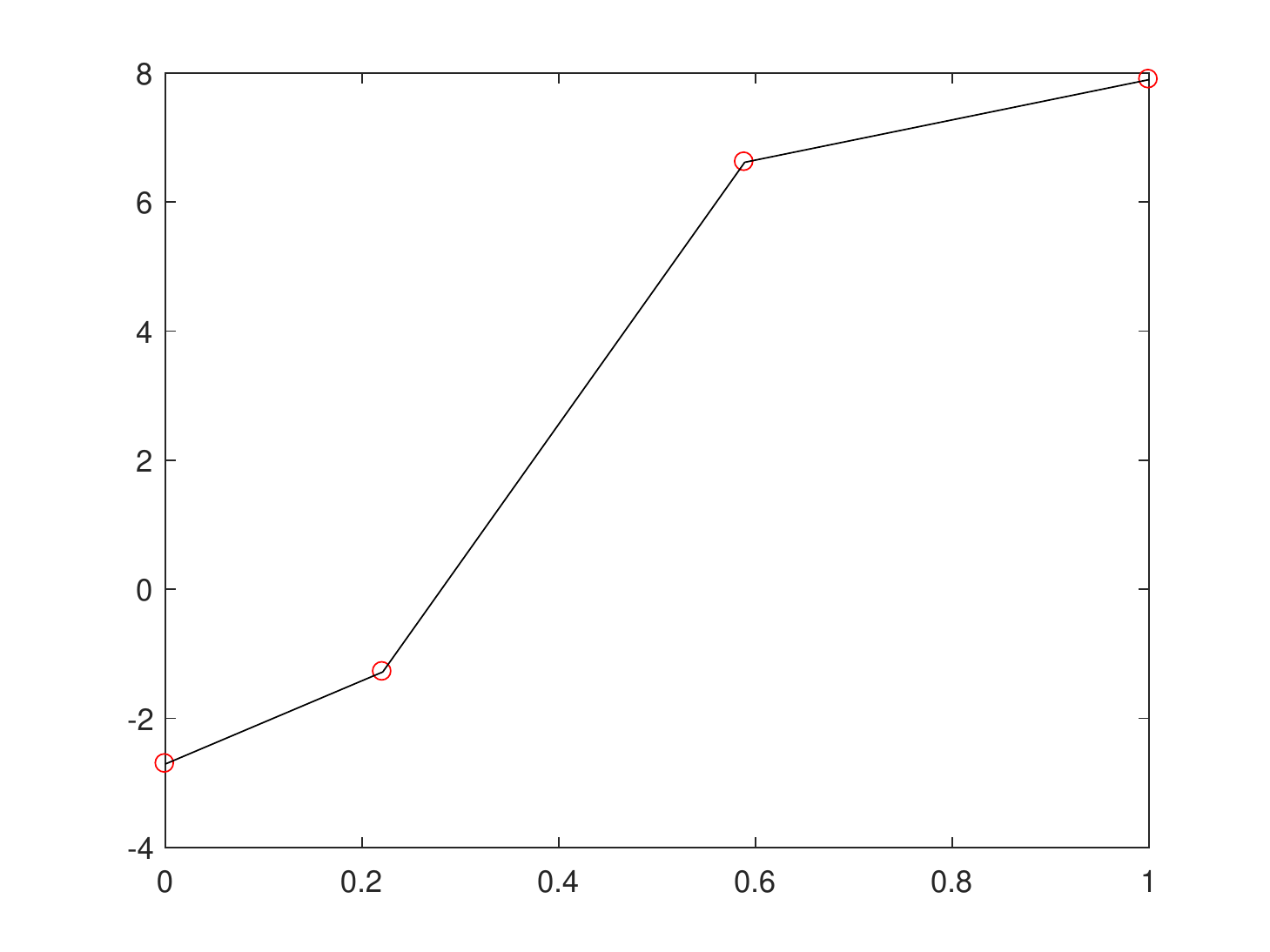}
 \put(-92,0){$\phi_g$}
 \put(-180,25){\rotatebox{90}{$\Delta \tau_w = 1 - (\tau^B_w/\tau_w)$(\%)}}
 \put(-180,140){$(d)$}
}
\centering{
\includegraphics[width=70mm]{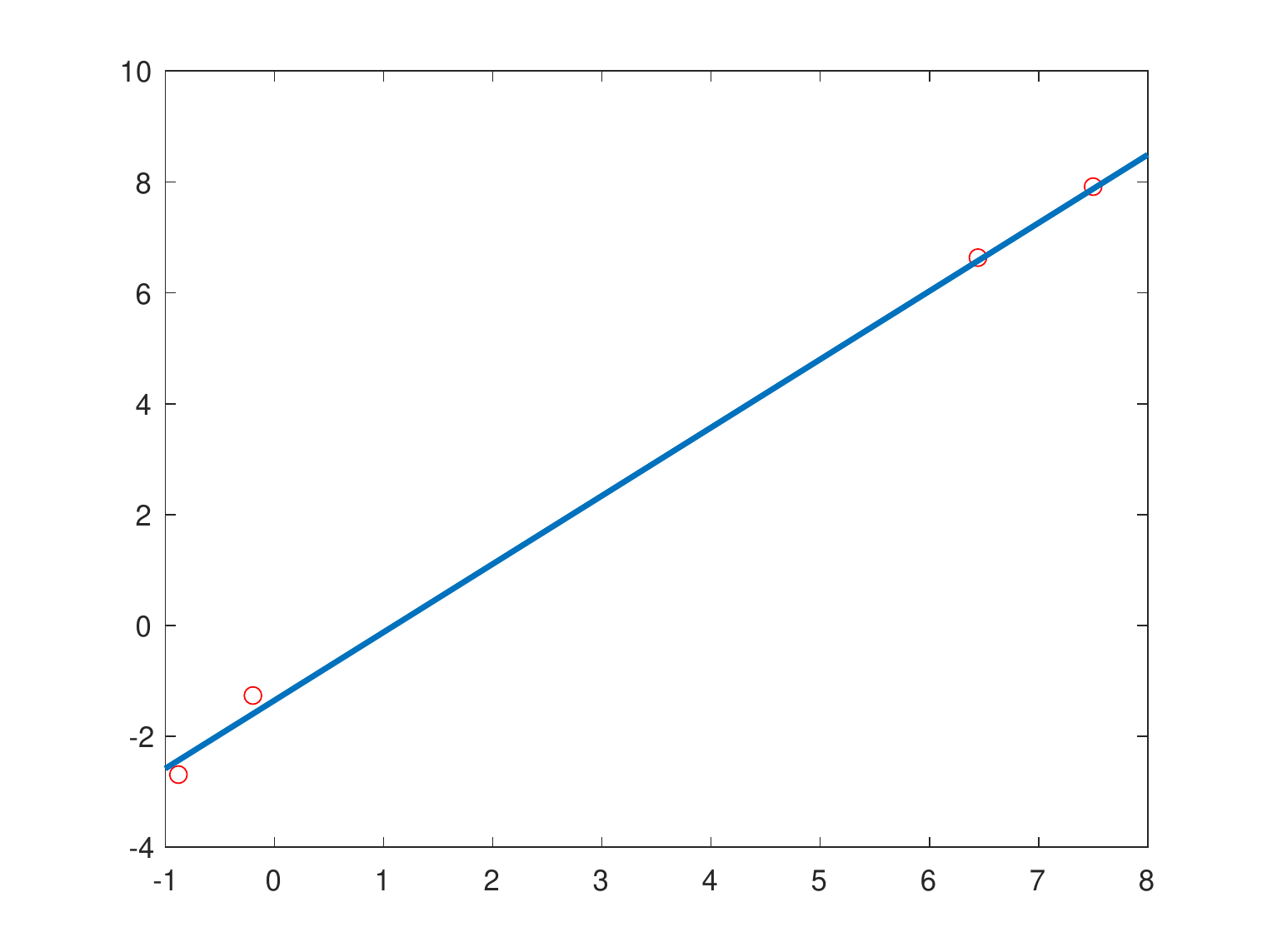}
\put(-115,0){$\Delta U_{b}/U_{b,0}$(\%)}
\put(-193,25){\rotatebox{90}{$\Delta \tau_w = 1 - (\tau^B_w/\tau_w)$(\%)}}
\put(-200,140){$(e)$}
}
\caption{Mean flow field properties as a function of $\phi_g$. (a) Percentage slip velocity $u_s$ normalised by the mean bulk velocity $U_b$ for each corresponding case, (b) slip length $b_s$ normalised by the average gap width of the roughness $L$ where the solid blue line represents the scaling law of \cite{ybertetal_2007} and the dashed black line of \cite{sbragaglia_prosperetti_2007}, (c) percentage change in bulk velocity $\Delta U_b$ normalised by the mean bulk velocity $U_{b,0}$ of the baseline case, (d) percentage shear stress reduction $\Delta \tau_{w}$ based on the ratio of the bottom wall $\tau^B_w$ to the average channel $\tau_w$ and (e) correlation between the percentage shear stress reduction and percent change in bulk velocity compared to a linear fit (solid blue line). }
\label{fig:slipeffect_phig}
\end{figure}
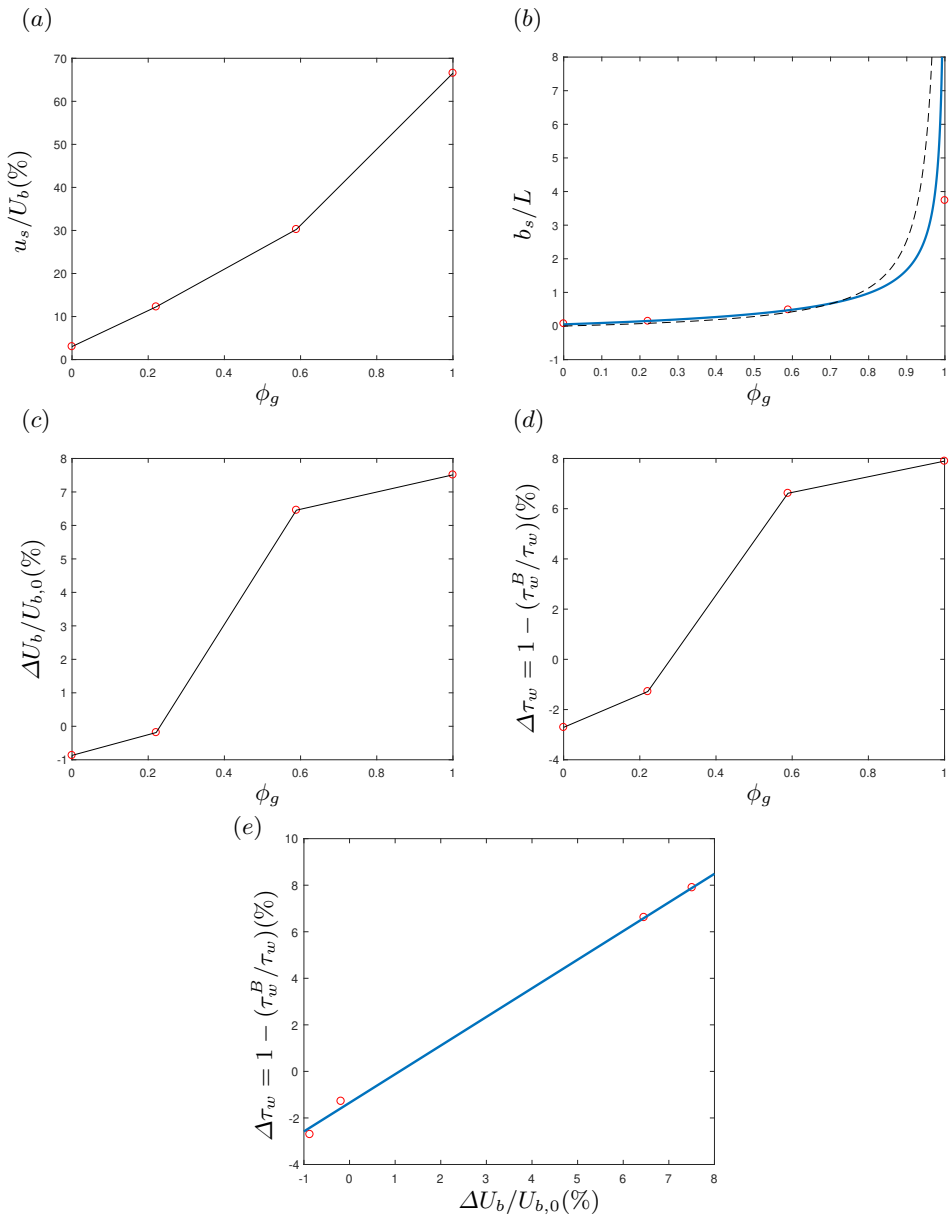

Figure \ref{fig:slipeffect_phig}(a) shows the slip effect as a function of $\phi_g$. The slip velocity $u_s$ is normalised by the mean bulk velocity $U_b$ for each corresponding case. As $\phi_g$ increases, the slip effect is more pronounced: $u_s$ exhibits a steady increase. Slip velocities can reach as much as $68$\% of the bulk velocities which can obscure some of the effects due to roughness. This is another reason why it would be important to plot the mean profile offset by the slip as shown in figure \ref{fig:mean_profile_log}(b). $DR$(\%) as a function of $u_s$ is not an accurate predictor of drag reduction since fully wetted roughness increases drag while achieving positive slip velocities as shown in figure \ref{fig:slipeffect_phig}(a).  The slip length $b_s$ normalised by the average roughness gap width $L$ is reported in figure \ref{fig:slipeffect_phig}(b) as a function of $\phi_g$. The average roughness gap $L$ is not a straightforward property to obtain since there is no periodicity (typically associated with structured geometry such as grooves and posts). We use a two-step process to calculate $L$. First we find the profile peak count $HSC$, the number of profile peaks that exceed a pre-selected threshold (e.g. arithmetic mean elevation of the roughness), calculated over the entire streamwise length for all the spanwise slices. Second we calculate the mean peak spacing $S_m$, the mean spacing between profile peaks, averaged over all the spanwise slices to obtain $L$. We refer the reader to figure \ref{fig:peak_count} in appendix \ref{appB} for an example of profile peak count. There is a good agreement between our data and a fit based on the scaling presented by \cite{ybertetal_2007} for posts given by the following equation:
\begin{equation}
  \frac{b_{s}}{L} = \frac{0.75}{\sqrt{\phi_s}} - 0.7.
\end{equation}
The solution by \cite{sbragaglia_prosperetti_2007} which does not require a fit gives good agreement for $\phi_g<0.7$. Slip length is not a good indicator of $DR$ since both fully wetted roughness and SHS produce a positive $b_s$. As mentioned earlier during the discussion of the mean velocity profiles, overall mass flux appears to be a good indicator for drag reduction given that our channels are run at a constant $Re_{\tau}$ and pressure gradient. Drag reduction would imply that more fluid mass is moving for the same conditions i.e. a larger mass flux and a larger change in bulk velocity; the opposite is also true. Figure \ref{fig:slipeffect_phig}(c) shows the percent change in mean bulk velocity $\Delta U_b$ normalised by $U_{b,0}$ of the baseline Case T-S. A negative change in $\Delta U_b$ indicates an increase in drag i.e. a lower mass flux. Once an interface is present (Case T-RI1) drag is reduced but is not enough to offset the effect roughness since $\Delta U_b$ is still negative. As the interface height increases, $\Delta U_b$ increases to become a positive value indicating an increase in mass flux and a larger $DR$ effect. This description is exactly what we see when compared to the analysis of the mean velocity profile $U/u^B_{\tau}$ done earlier. $\Delta U_b$ does not exhibit a steady increase and plateaus for large values of $\phi_g$: this was also observed by \cite{Turk2014}. The reduction in the shear stress on the bottom wall $\tau^B_w$ compared to the average shear stress of the channel $\tau_w$ provides a straightforward result for $DR$ as shown in figure \ref{fig:slipeffect_phig}(d). Note the increase in drag by around $2.5$\% for Case T-RFW. The roughness has $S^+_q \approx 1.6$ which is in the hydrodynamically smooth regime. \cite{Busse2017} reported that the surface property $S^+_{z5\times5}$ is a more suitable measure of the sand-grain roughness $k^+$ where $S^+_{z5\times5}\approx 11$ for our surface. This explains why we see a drag increase for the current configuration. A reduction of $1$\% in drag is obtained once an interface is introduced in Case T-RI1. A $7$\% $DR$ is achieved for Case T-RI2 and peaks at $8$\% for Case T-RI3. The results are in agreement with the experimental results of \cite{ling_katz_2016} obtained for turbulent boundary layers over SHS at higher $Re_{\tau}$, where $k^+_{rms} \approx 0.45-0.75$ and $DR$ ranged from $9$\% to $12$\%. 
One can therefore reasonably assume that the change in shear stress and the change in bulk velocity correlate well with each other. Figure \ref{fig:slipeffect_phig}(e) shows $\Delta \tau_w$ as a function of $\Delta U_b$ compared to a linear fit given by
\begin{equation}
\Delta \tau_w  = 1.1 \Delta U_{b}/U_{b,0} -0.014. 
\end{equation}

\begin{figure}
 \centering{
 \includegraphics[width=65mm]{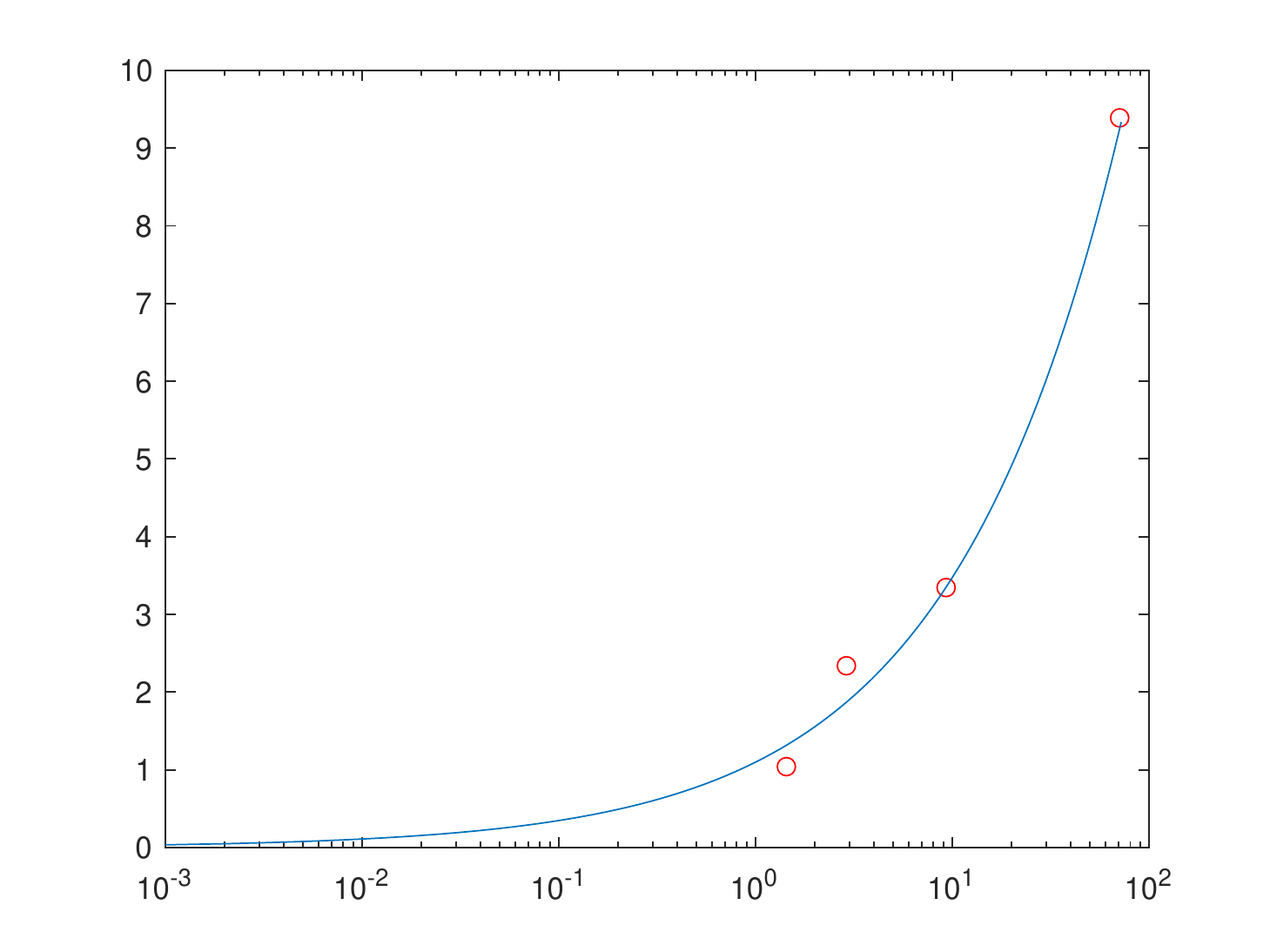}
  \put(-92,-3){$b^+_s$}
 \put(-190,70){$\Delta U^+$}
 \put(-180,140){$(a)$}
 \includegraphics[width=65mm]{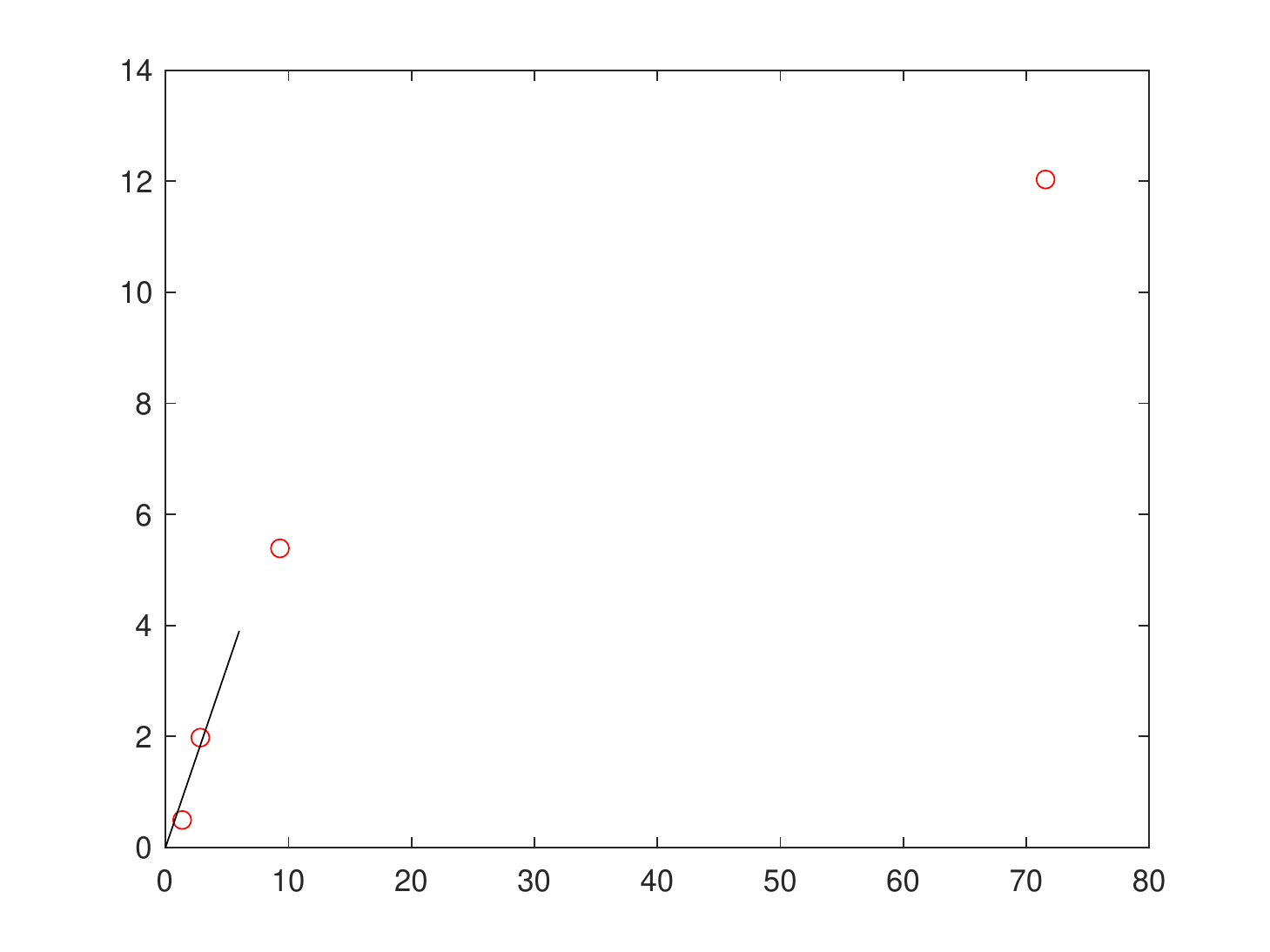}
 \put(-92,-3){$b^+_s$}
 \put(-180,68){$u^+_s$}
 \put(-180,140){$(b)$}
 }
\caption{Roughness function $\Delta U^+$ as a function of slip length $b^+_s$ is shown in (a) and is compared to a power law fit (solid blue line) and (b) shows the slip velocity scaling $u^+_s$ as a function of $b^+_s$ with a comparison to a linear approximation (solid black line).}
\label{fig:slipvel_bslip}
\end{figure}
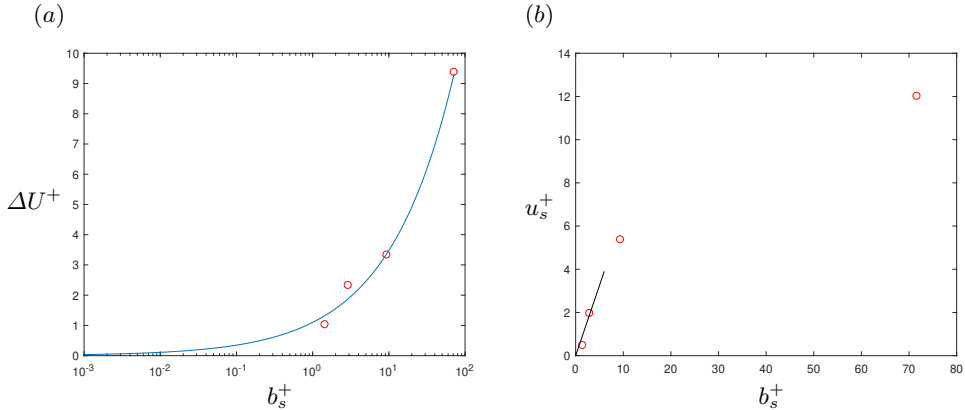

The roughness function $\Delta U^+$ depends on the three-dimensional topography of a surface where \cite{bradshaw2000} suggested a scaling given by $\Delta U^+ \thicksim (k^+_{rms})^\alpha$. Since the interface height essentially modifies $k^+_{rms}$, then a similar scaling argument can be made for $\Delta U^+$ as a function of $b^+_s$. Figure \ref{fig:slipvel_bslip}(a) shows a good agreement with a power-law behaviour where the roughness function goes to zero as the slip length diminishes. The power-law formula obtained from fitting the data is given by
\begin{equation}
 \Delta U^+ = 1.1(b^+_s)^{0.5}.
\end{equation}

A semi-analytical formula by \cite{ybertetal_2007} relates the slip velocity to the cavity width, in terms of wall units $u^+_s=C_YL^+$. We simply use the slip length $b^+_s$ instead since slip is a direct consequence of $L^+$. Figure \ref{fig:slipvel_bslip}(b) shows the linear approximation near the wall for $C_Y=0.65$ where
\begin{equation}
 u^+_s=0.65b^+_s.
\end{equation}
The linear scaling deviates for $b^+_s>10$ which was also shown for $L^+>10$ in \cite{Seo2015} for a value of $C_Y=0.535$. This is expected since the linear relationship is based on Stokes' flow which becomes less accurate as slip increases. It is worth noting that by definition $u^+_s=b^+_s$ for the cases where the air--water interface is aligned with the top of the roughness and where the total stress is equal to the viscous stress. However, as discussed in \cite{ling_katz_2016}, $u^+_s<b^+_s$ for the cases when $-(u^{\prime}v^{\prime})^+ > 0$ or when the total stress $\tau^+_t > \mu(dU/dy)^+$ due the random nature of the roughness and the variable interface heights.

\subsubsection{Reynolds stresses}
\begin{figure}
\centering{
\includegraphics[scale=0.6]{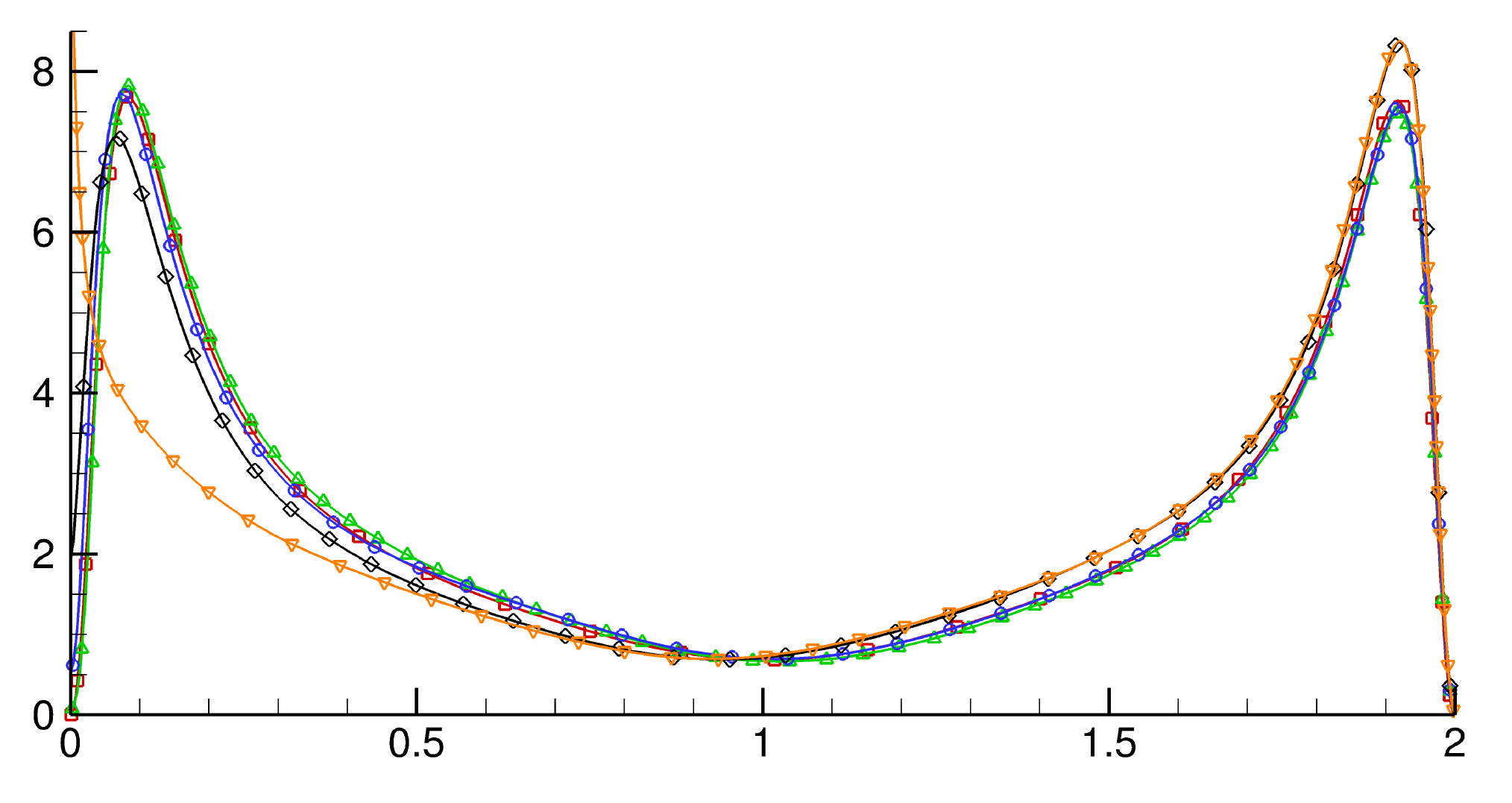}
\put(-355,190){$(a)$}
\put(-180,0){$y/\delta$}
%\put(-355,80){\rotatebox{90}{$\left<u^{\prime}u^{\prime}\right>/u_{\tau}^2$}}
\put(-360,80){\rotatebox{90}{$u^{\prime}u^{\prime}/u_{\tau}^2$}}
}
\centering{
\includegraphics[scale=0.6]{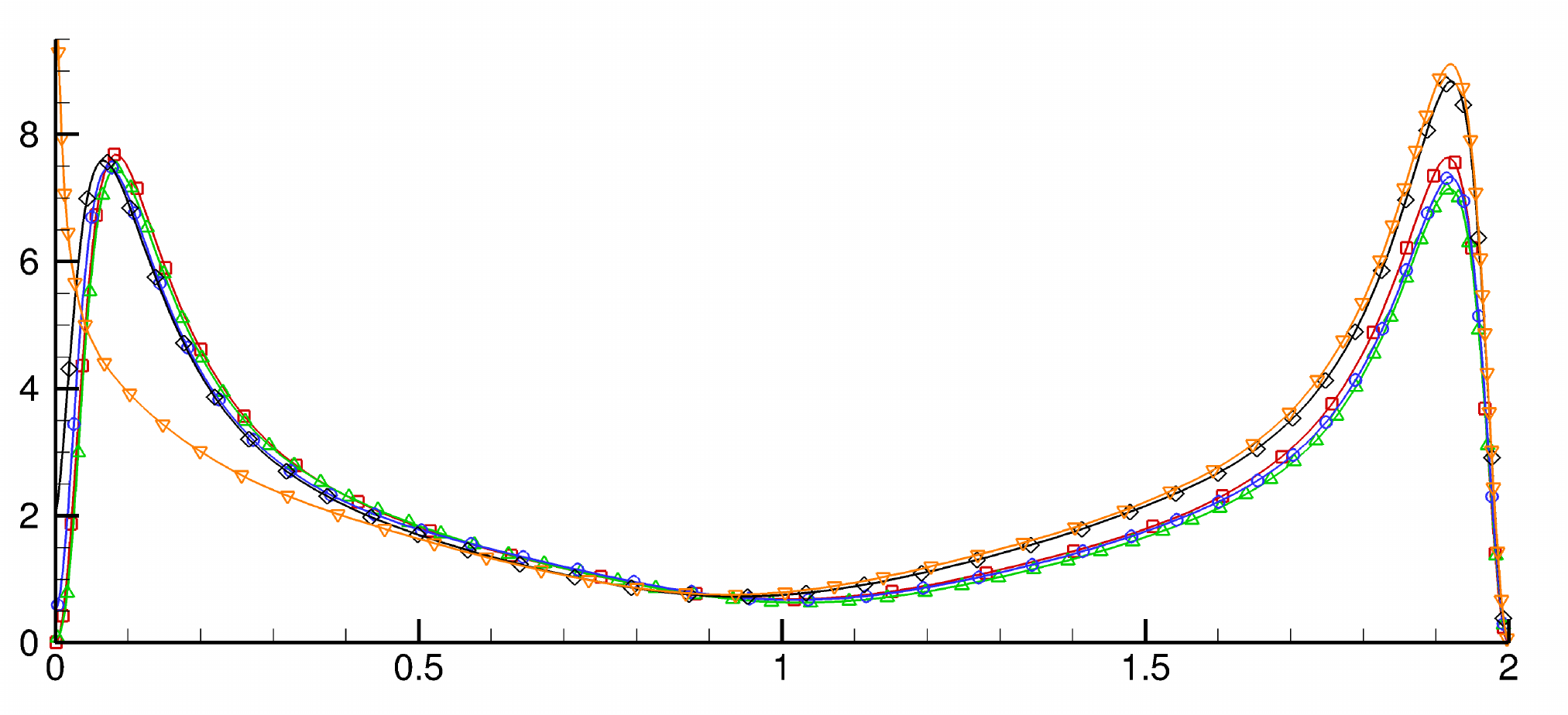}
\put(-350,170){$(b)$}
\put(-180,0){$y/\delta$}
%\put(-355,60){\rotatebox{90}{$\left<u^{\prime}u^{\prime}\right>/{(u^B_{\tau})}^2$}}
\put(-365,60){\rotatebox{90}{$u^{\prime}u^{\prime}/{(u^B_{\tau})}^2$}}
}
\centering{
\includegraphics[scale=0.6]{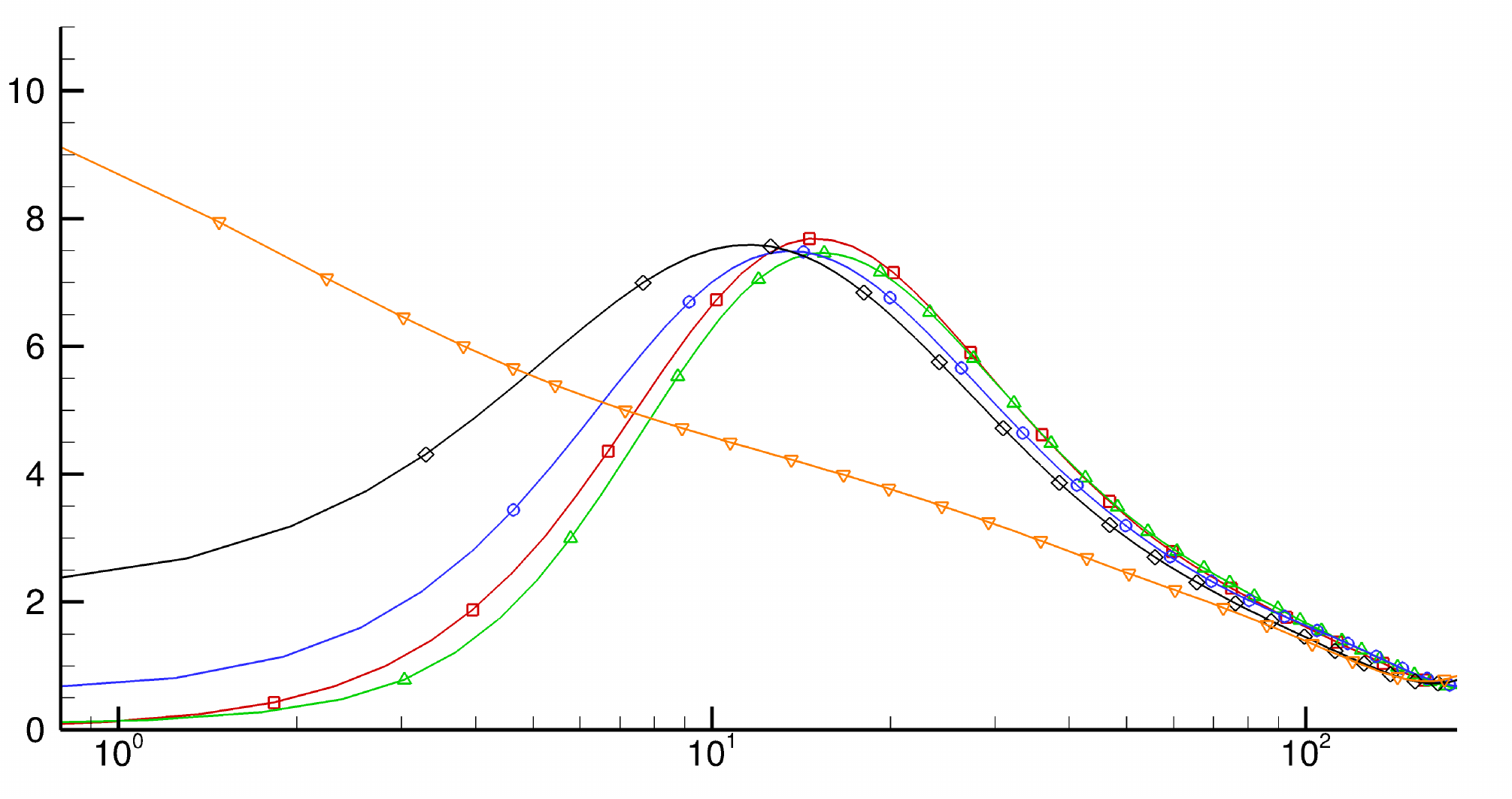}
\put(-350,185){$(c)$}
\put(-180,0){$y^+(u^B_{\tau})$}
%\put(-355,60){\rotatebox{90}{$\left<u^{\prime}u^{\prime}\right>/{(u^B_{\tau})}^2$}}
\put(-360,70){\rotatebox{90}{$u^{\prime}u^{\prime}/{(u^B_{\tau})}^2$}}
}
\caption{Streamwise component of the normal Reynolds stress normalised by (a) the average friction velocity $u_{\tau}^2$ as a function of $y/\delta$, (b) the bottom wall fiction velocity ${(u^B_{\tau})}^2$ as a function of $y/\delta$ and (c) the bottom wall fiction velocity ${(u^B_{\tau})}^2$ as a function of $y^+(u^B_{\tau})$ on a $\log$ scale. Symbols for each case are: Case T-S ($\Box$), Case T-RFW ($\vartriangle$), Case T-RI1 ($\ocircle$), Case T-RI2 ($\Diamond$), Case T-RI3($\triangledown$). The symbols are not representative of grid resolution.}
\label{fig:streamwise_rij}
\end{figure}

\begin{figure}
\centering{
\includegraphics[scale=0.6]{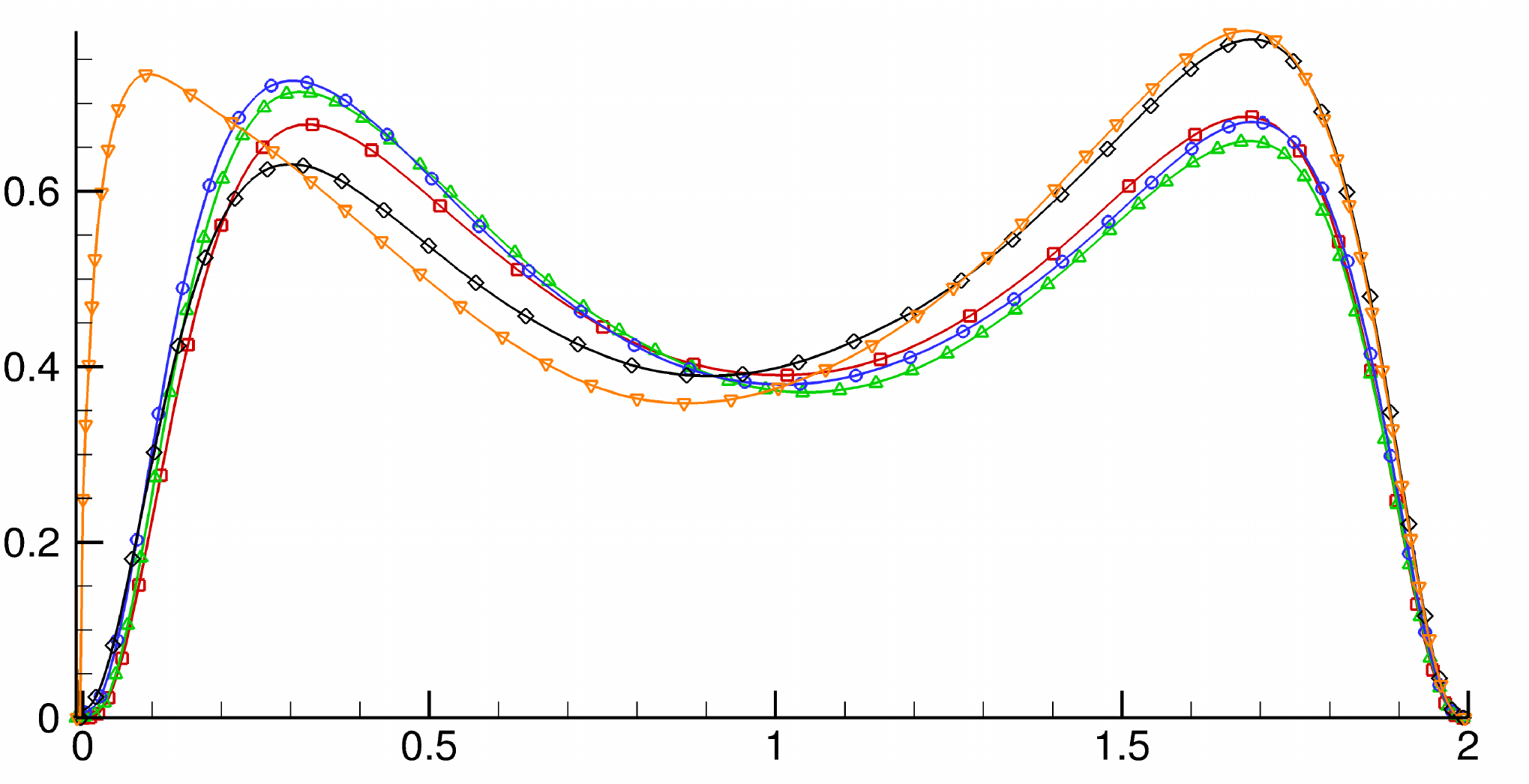}
\put(-350,180){$(a)$}
\put(-180,0){$y/\delta$}
%\put(-355,80){\rotatebox{90}{$\left<v^{\prime}v^{\prime}\right>/u_{\tau}^2$}}
\put(-365,80){\rotatebox{90}{$v^{\prime}v^{\prime}/u_{\tau}^2$}}
}
\centering{
\includegraphics[scale=0.6]{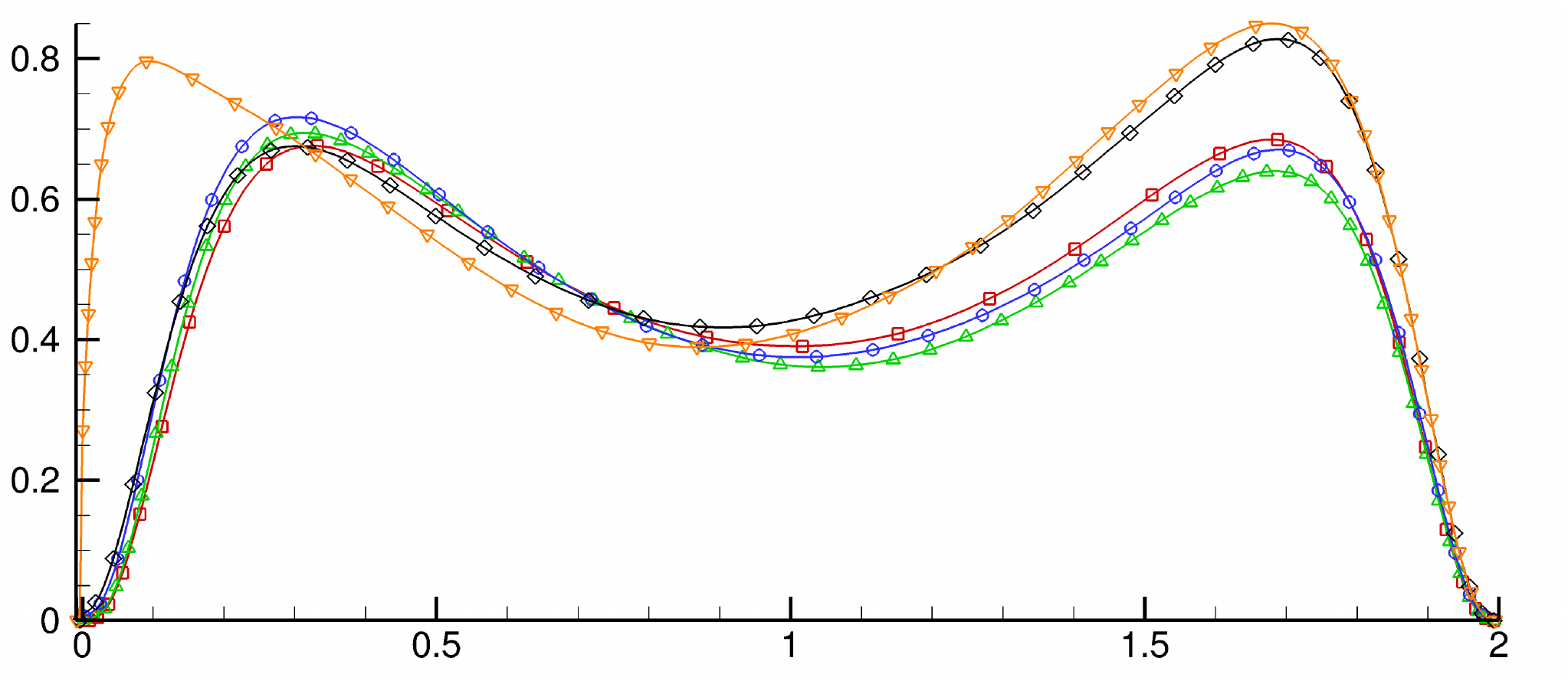}
\put(-350,180){$(b)$}
\put(-180,0){$y/\delta$}
%\put(-355,60){\rotatebox{90}{$\left<v^{\prime}v^{\prime}\right>/{(u^B_{\tau})}^2$}}
\put(-365,60){\rotatebox{90}{$v^{\prime}v^{\prime}/{(u^B_{\tau})}^2$}}
}
\centering{
\includegraphics[scale=0.6]{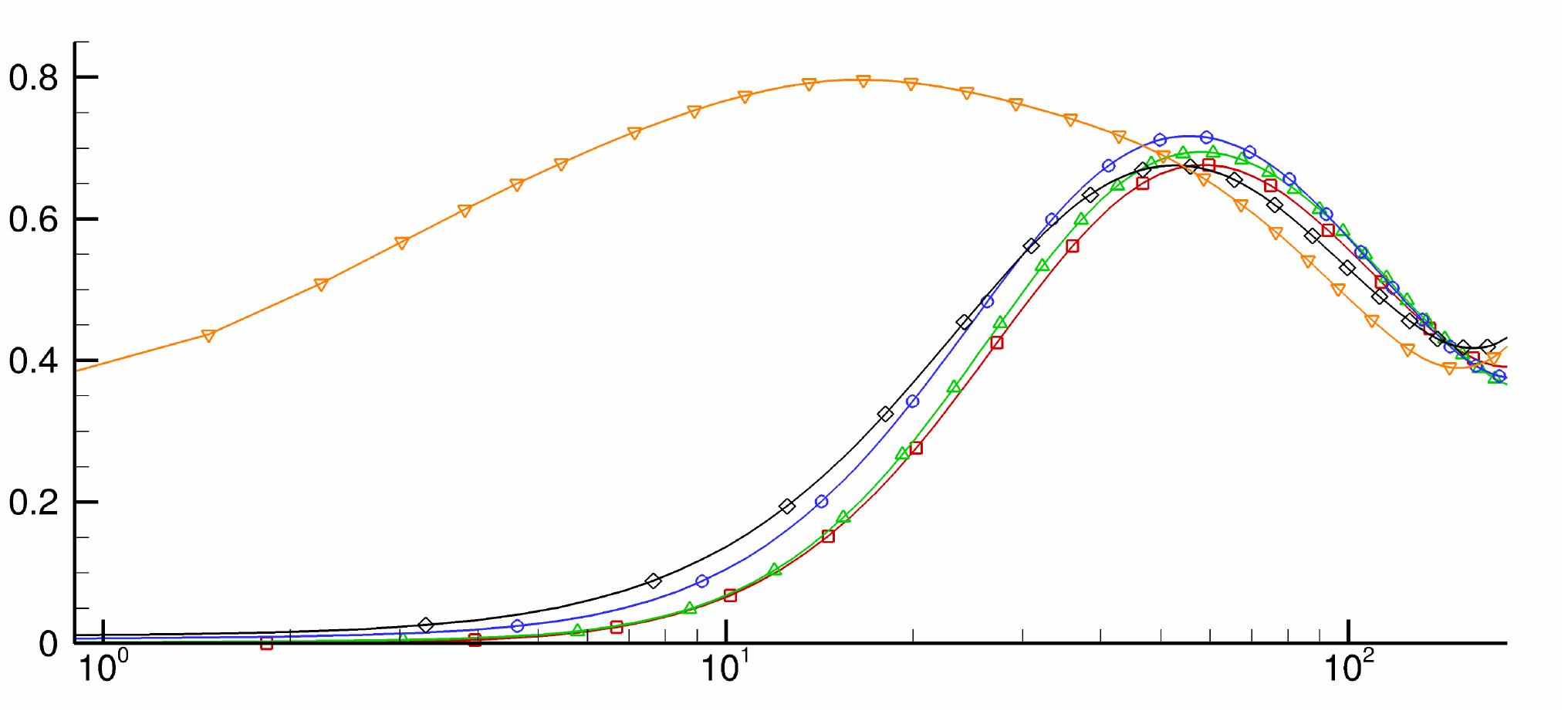}
\put(-350,180){$(c)$}
\put(-180,0){$y^+(u^B_{\tau})$}
%\put(-355,60){\rotatebox{90}{$\left<v^{\prime}v^{\prime}\right>/{(u^B_{\tau})}^2$}}
\put(-365,60){\rotatebox{90}{$v^{\prime}v^{\prime}/{(u^B_{\tau})}^2$}}
}
\caption{Wall-normal component of the normal Reynolds stress normalised by (a) the average friction velocity $u_{\tau}^2$ as a function of $y/\delta$, (b) the bottom wall fiction velocity ${(u^B_{\tau})}^2$ as a function of $y/\delta$ and (c) the bottom wall fiction velocity ${(u^B_{\tau})}^2$ as a function of $y^+(u^B_{\tau})$ on a $\log$ scale. Symbols for each case are: Case T-S ($\Box$), Case T-RFW ($\vartriangle$), Case T-RI1 ($\ocircle$), Case T-RI2 ($\Diamond$), Case T-RI3($\triangledown$). The symbols are not representative of grid resolution.}
\label{fig:wallnormal_rij}
\end{figure}

\begin{figure}
\centering{
\includegraphics[scale=0.43]{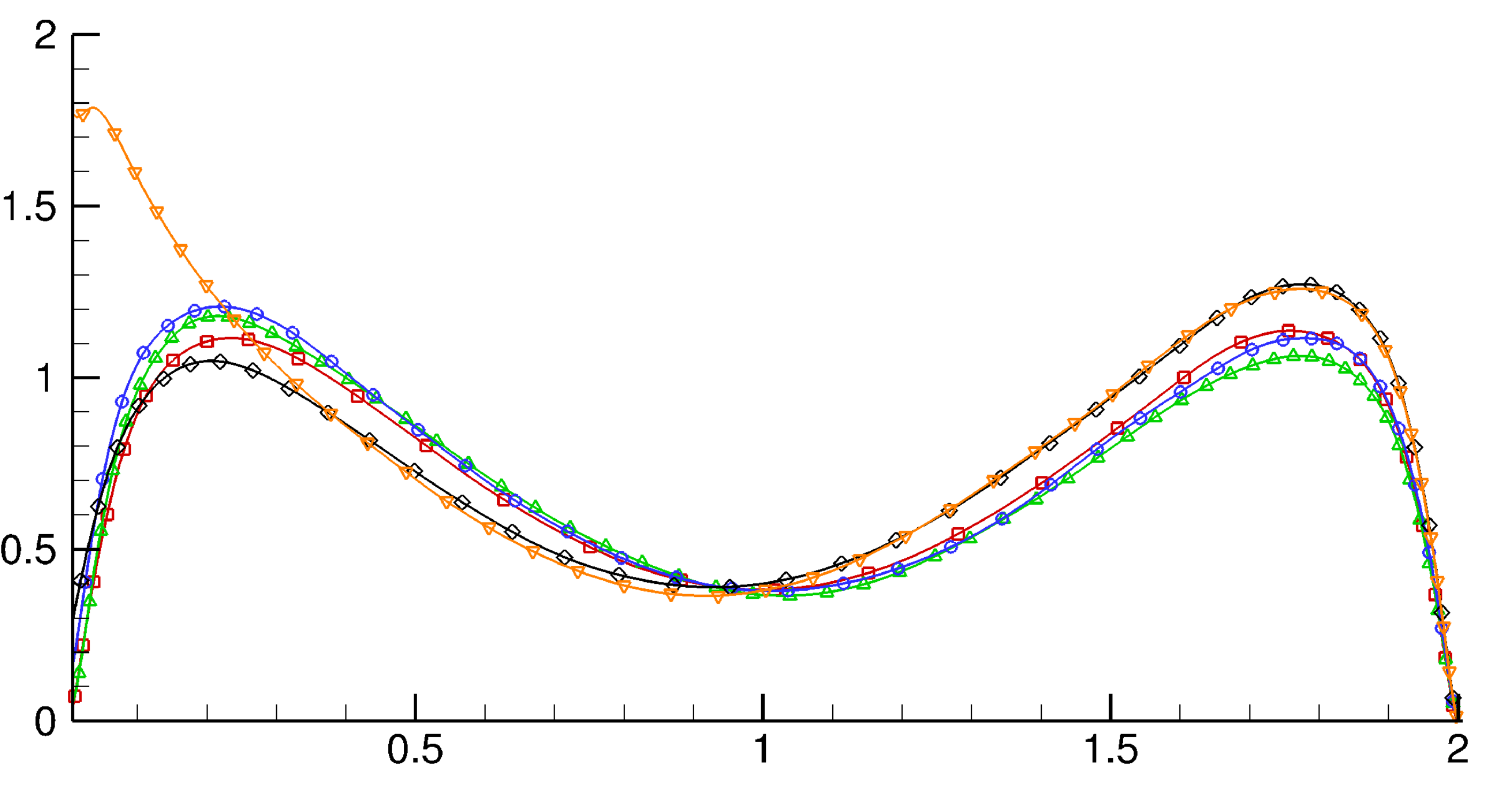}
\put(-350,190){$(a)$}
\put(-180,0){$y/\delta$}
%\put(-355,80){\rotatebox{90}{$\left<w^{\prime}w^{\prime}\right>/u_{\tau}^2$}}
\put(-360,80){\rotatebox{90}{$w^{\prime}w^{\prime}/u_{\tau}^2$}}
}
\centering{
\includegraphics[scale=0.43]{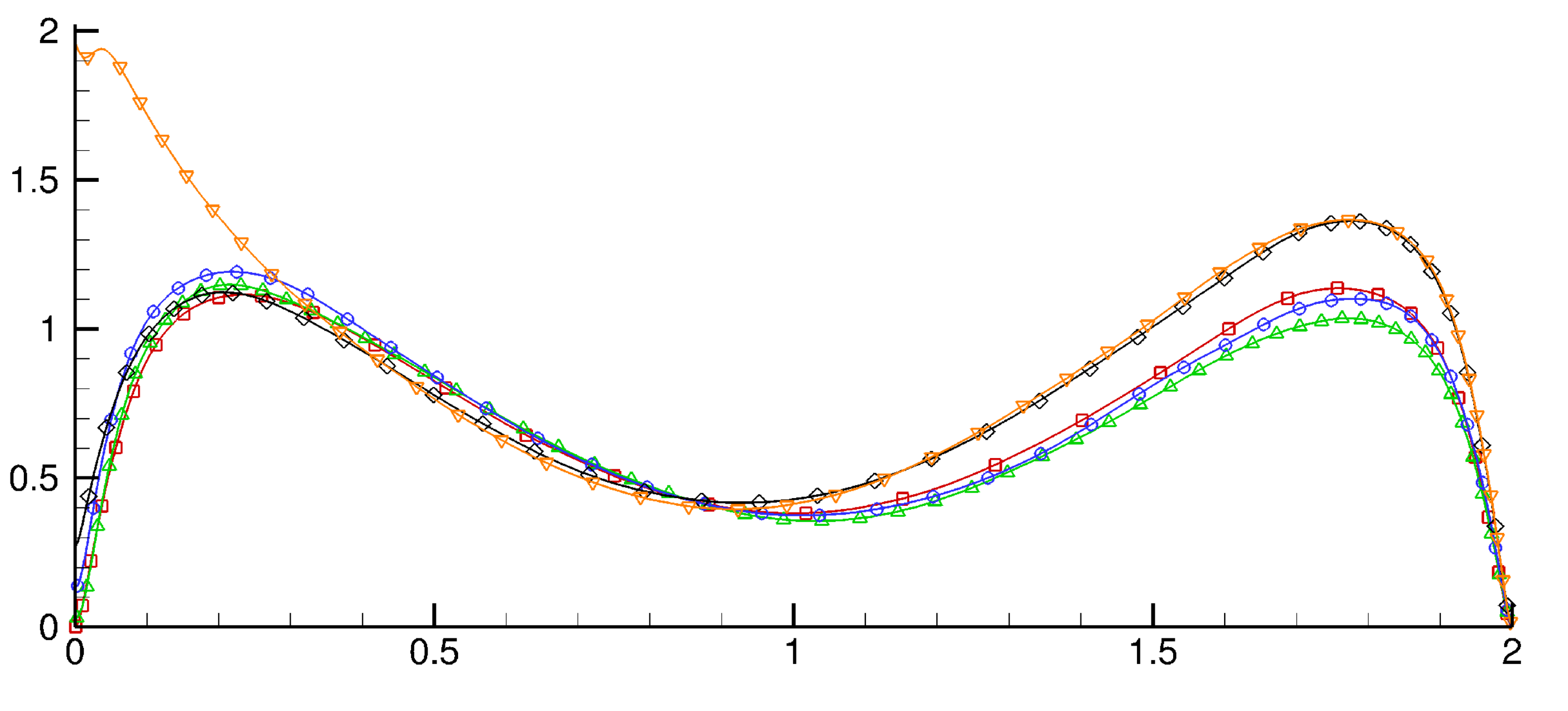}
\put(-350,180){$(b)$}
\put(-180,0){$y/\delta$}
%\put(-355,60){\rotatebox{90}{$\left<w^{\prime}w^{\prime}\right>/{(u^B_{\tau})}^2$}}
\put(-360,60){\rotatebox{90}{$w^{\prime}w^{\prime}/{(u^B_{\tau})}^2$}}
}
\centering{
\includegraphics[scale=0.43]{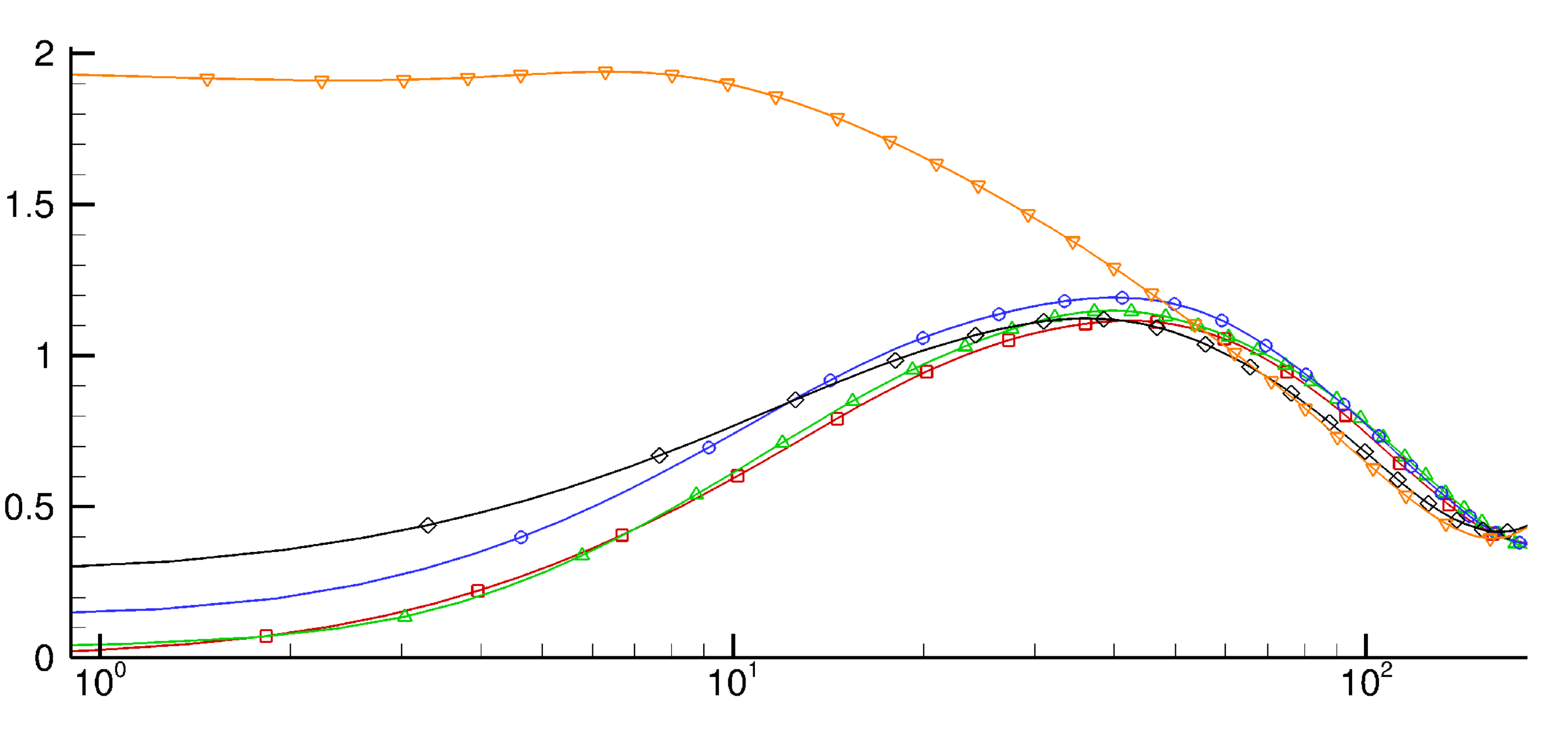}
\put(-350,175){$(c)$}
\put(-180,0){$y^+(u^B_{\tau})$}
%\put(-355,60){\rotatebox{90}{$\left<w^{\prime}w^{\prime}\right>/{(u^B_{\tau})}^2$}}
\put(-360,60){\rotatebox{90}{$w^{\prime}w^{\prime}/{(u^B_{\tau})}^2$}}
}
\caption{Spanwise component of the normal Reynolds stress normalised by (a) the average friction velocity $u_{\tau}^2$ as a function of $y/\delta$, (b) the bottom wall fiction velocity ${(u^B_{\tau})}^2$ as a function of $y/\delta$ and (c) the bottom wall fiction velocity ${(u^B_{\tau})}^2$ as a function of $y^+(u^B_{\tau})$ on a $\log$ scale. Symbols for each case are: Case T-S ($\Box$), Case T-RFW ($\vartriangle$), Case T-RI1 ($\ocircle$), Case T-RI2 ($\Diamond$), Case T-RI3($\triangledown$). The symbols are not representative of grid resolution.}
\label{fig:spanwise_rij}
\end{figure}

\begin{figure}
\centering{
\includegraphics[scale=0.6]{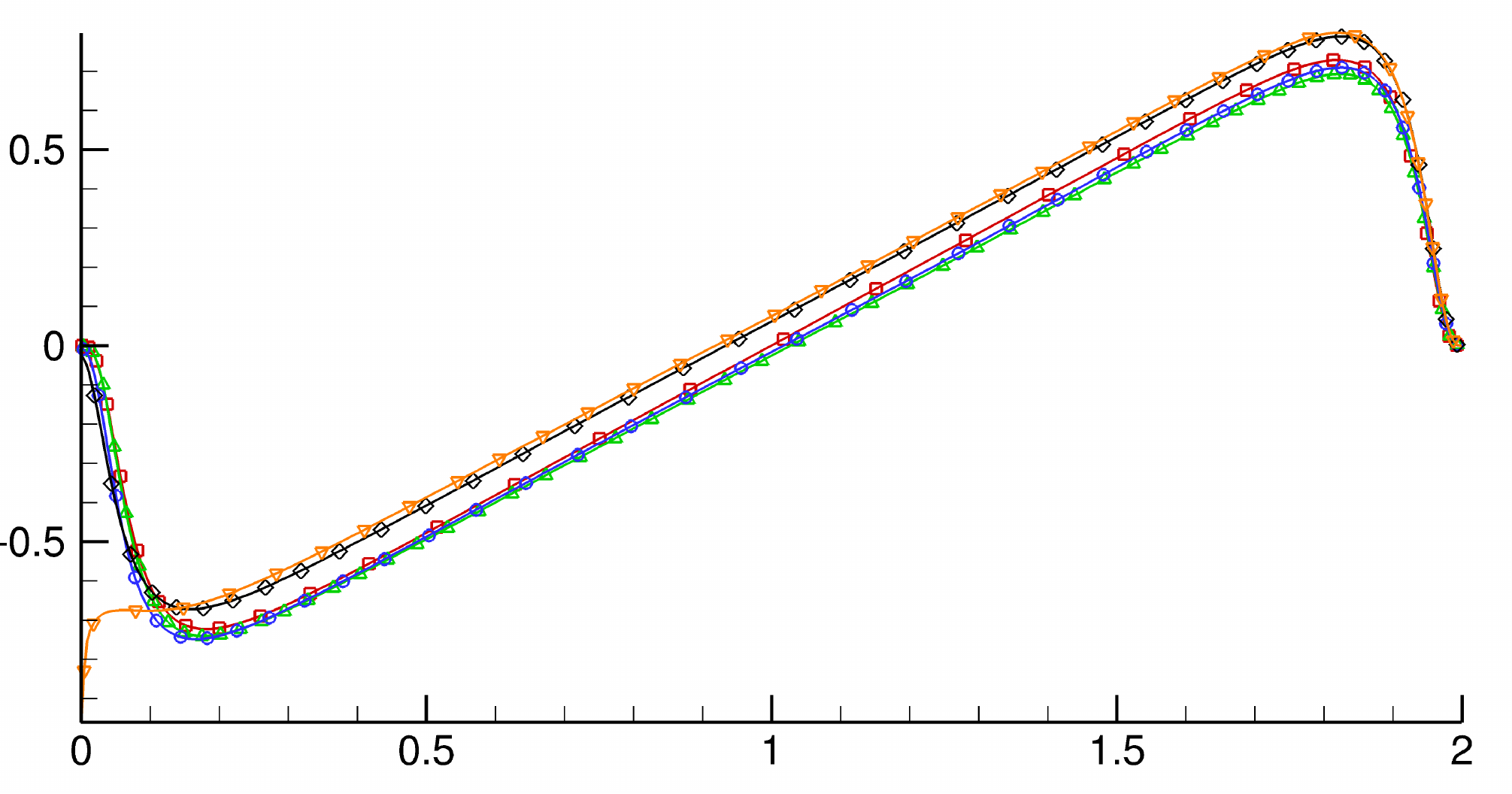}
\put(-350,180){$(a)$}
\put(-180,0){$y/\delta$}
%\put(-355,80){\rotatebox{90}{$\left<u^{\prime}v^{\prime}\right>/u_{\tau}^2$}}
\put(-360,90){\rotatebox{90}{$u^{\prime}v^{\prime}/u_{\tau}^2$}}
}
\centering{
\includegraphics[scale=0.6]{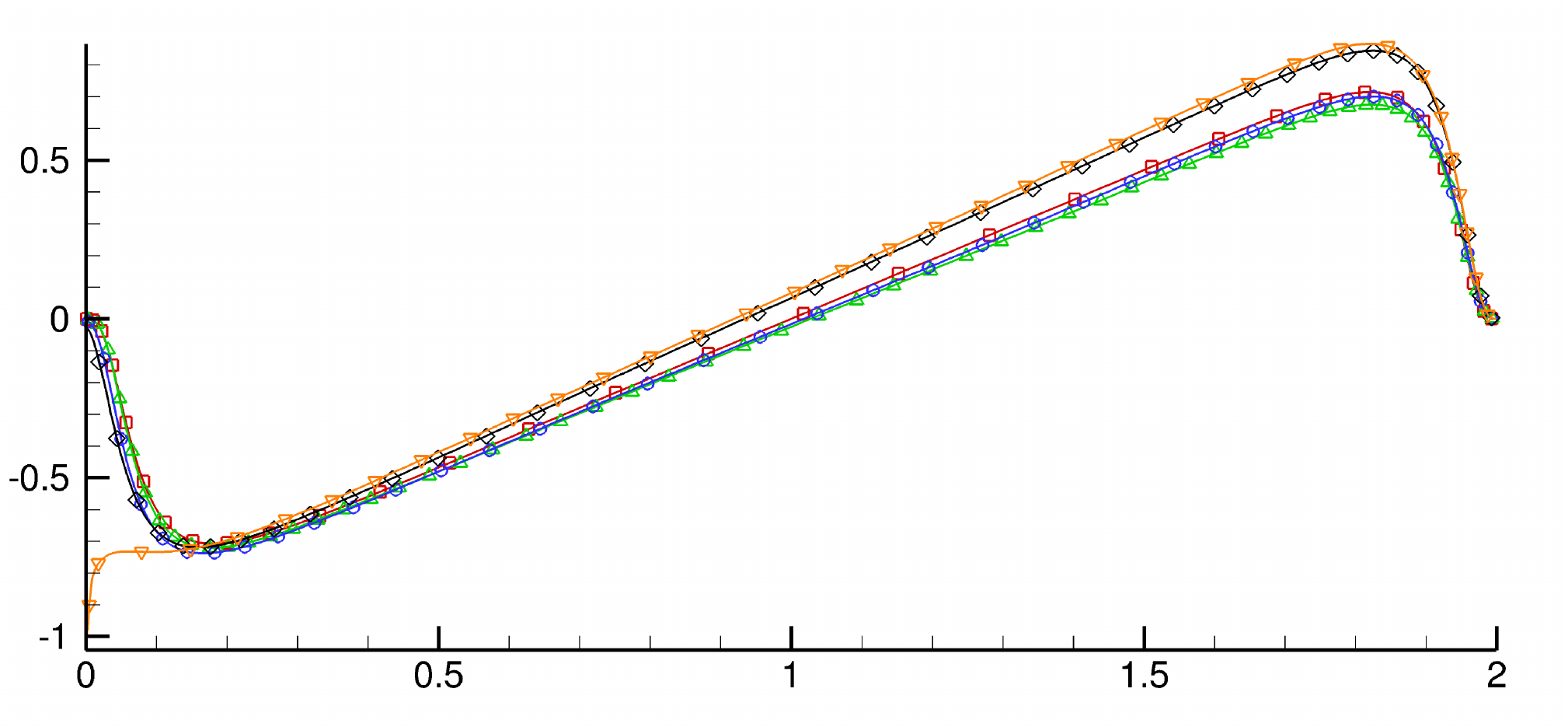}
\put(-350,160){$(b)$}
\put(-180,0){$y/\delta$}
%\put(-355,60){\rotatebox{90}{$\left<u^{\prime}v^{\prime}\right>/{(u^B_{\tau})}^2$}}
\put(-360,70){\rotatebox{90}{$u^{\prime}v^{\prime}/{(u^B_{\tau})}^2$}}
}
\centering{
\includegraphics[scale=0.62]{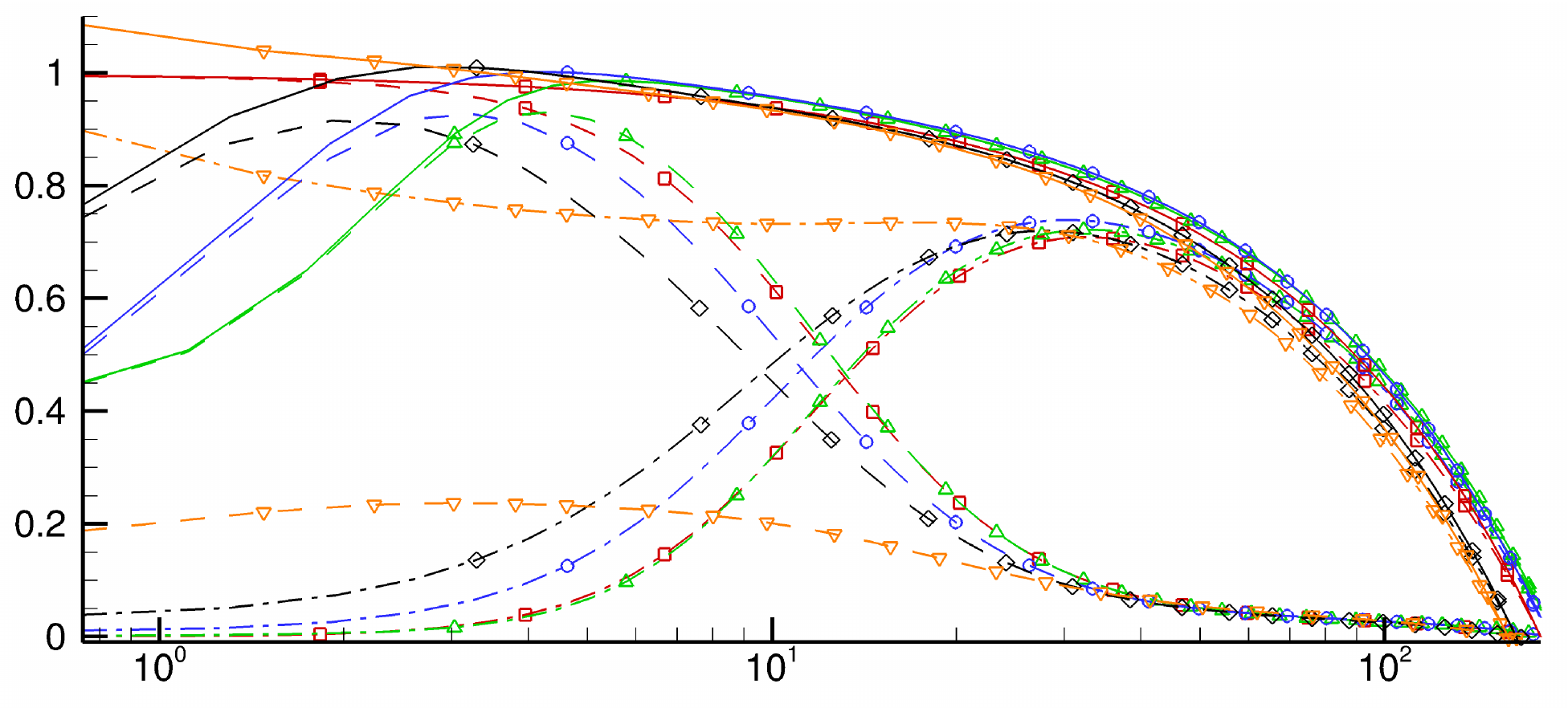}
\put(-350,160){$(c)$}
\put(-180,-5){$y^+(u^B_{\tau})$}
%\put(-355,60){\rotatebox{90}{$-\left<u^{\prime}v^{\prime}\right>/{(u^B_{\tau})}^2$}}
\put(-360,30){\rotatebox{90}{$\mu(dU/dy)^+,-(u^{\prime}v^{\prime})^+,\tau^+_t$}}
}
\caption{Reynolds shear stress normalised by (a) the average friction velocity $u_{\tau}^2$ as a function of $y/\delta$, (b) the bottom wall fiction velocity ${(u^B_{\tau})}^2$ as a function of $y/\delta$. (c) The viscous stress, Reynolds shear stress and total stress normalized by the bottom wall fiction velocity ${(u^B_{\tau})}^2$ as a function of $y^+(u^B_{\tau})$ on a $\log$ scale. Symbols for each case are: Case T-S ($\Box$), Case T-RFW ($\vartriangle$), Case T-RI1 ($\ocircle$), Case T-RI2 ($\Diamond$), Case T-RI3($\triangledown$). In (c) the solid line represents total stress $\tau^+_t$, dashed lines the viscous stress $\mu(dU/dy)^+$ and dashed-dotted lines the Reynolds shear stress $-(u^{\prime}v^{\prime})^+$. The symbols are not representative of grid resolution.}
\label{fig:shear_rij}
\end{figure}
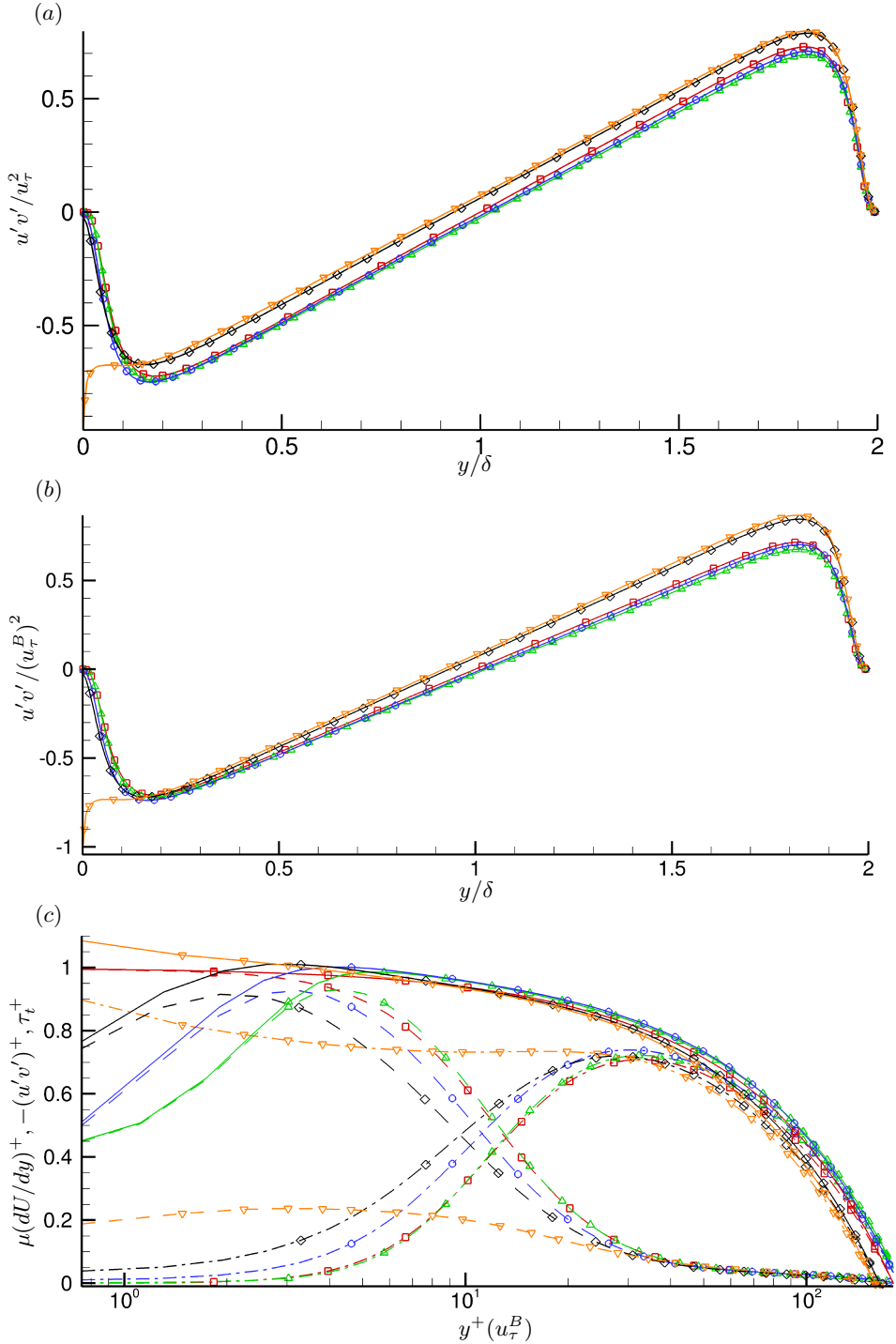

The Reynolds stress profiles are shown for the different cases as a function of the wall-normal distance. Figure \ref{fig:streamwise_rij}(a) shows the streamwise component of the Reynolds stress normalised by $u_{\tau}$. Case T-RFW seems to slightly increase the peak of $u'u'$ while simultaneously shifting it away from the SHS wall. The presence of an interface seems to dampen that effect where Case T-RI1 shows a decrease in the peak while shifting it towards the SHS wall. This effect is further amplified in Case T-RI2 where the peak is clearly damped and the profile moves closer to the wall. Case T-RI3 exhibits the largest slip and hence is closest to the SHS wall where there exists a sharp rise in the streamwise component: however the peak is larger than the other cases breaking the symmetry of the profile completely. The overall trend in damping the peak of $u'u'$ while shifting it closer to the slip wall has been observed in the literature for longitudinal grooves and posts. For reference, $u'u'$ is scaled by $u^B_{\tau}$ in figure \ref{fig:streamwise_rij}(b). We get a reasonable collapse in the profiles near the SHS wall with the exception of Case T-RI3 indicating a change in the near-wall behaviour. Figure \ref{fig:streamwise_rij}(c) shows a $\log$ plot of $u'u'$ as a function of the wall-normal distance $y^+(u^B_{\tau})$ in wall units scaled by $u^B_{\tau}$. Large differences are observed in the near-wall region due to the slip effect. The shift away from the SHS wall for Case T-RFW and the shift towards the SHS wall for Cases T-RI1 and T-RI2 are more evident. 

The wall-normal Reynolds stress component $v'v'$ is shown in figure \ref{fig:wallnormal_rij}. Initially $v'v'$ is normalised by $u_{\tau}$ as shown in figure \ref{fig:wallnormal_rij}(a). Case T-RFW amplifies the peak stress and shifts away from the slip wall. Comparing it to Case T-RI2, it is evident that the interface has a damping effect. It is interesting to note that once the interface was introduced initially, Case T-RI1 showed a further amplification in the peak from Case T-RFW and not the opposite. This is likely due to be the fact that the interface height barely covers any of the roughness, and therefore the inhomogeneity between slip and no-slip due to the interface and random rough patches cause larger fluctuations. This is of course damped out once the interface covers more of the surface and larger slip areas are present. Case T-RI3 shows the largest shift towards the SHS wall as expected but the peak stress does not not seem to follow any further damping effect with increasing slip area. Figure \ref{fig:wallnormal_rij}(b) shows $v'v'$ scaled by $u^B_{\tau}$ and no collapse in the data is observed. The $\log$ plot in figure \ref{fig:wallnormal_rij}(c) shows the velocities going to zero since at the interface an infinite surface tension is assumed. Therefore the wall-normal velocity $v_N=0$ at the interface and a no-slip boundary condition is applied elsewhere over the rough surface. Note that the profiles of Case T-RI1 and Case T-RI2 intersect around $y^+ \approx 28$ where Case T-RI2 exhibits larger wall-normal stresses for $y^+<28$ and then Case T-RI1 tends to become larger for the region above that. Case T-RFW does not exhibit an observable shift away from the wall while it remains clear that when an interface is present, the shift towards the SHS wall remains considerable.

The streamwise component of Reynolds stress is shown in figure \ref{fig:spanwise_rij}, which follows a similar trend in the behaviour to that described above. Case T-RFW shows an increase in $w'w'$ when scaled by $u_{\tau}$ as shown in figure \ref{fig:spanwise_rij}(a). The addition of an interface (Case T-RI1) further increases the peak in $w'w'$. The same reasoning applies to the spanwise component as described earlier for the wall-normal component since the roughness has no preferential direction. Therefore the interplay between slip and no-slip due to the interface and the protruding roughness holds here too. Similarly as soon as the interface covers a large portion of the roughness as in Case T-RI2, the peak is damped out.  Case T-RI3 is highly skewed towards the rough wall with a peak velocity that is much larger than the rest of the cases. Similar to what was observed in the wall-normal stresses, Cases T-RI1 and T-RI2 intersect each other as shown in figure \ref{fig:spanwise_rij}(c). The location however is smaller where $y^+ \approx 13$. Anywhere below that, Case T-RI2 exhibits higher stresses than Case T-RI1 and the opposite holds true when $y^+>13$. The spanwise slip is evident due to the presence of the interface where Case T-RI3 still exhibits the largest slip effect. 

The Reynolds shear stress is an important quantity to examine since its behaviour is closely related to turbulence levels and the structure of the near-wall turbulence. Figure \ref{fig:shear_rij}(a) shows how the presence of an interface (Case T-RI2) reduces shear while the presence of roughness enhances it. Note that since the wall-normal shear stress tends to increase when an interface is introduced (Case T-RI1), that effect translates here where we see a further enhancement in mixing instead of a reduction in the peak. Similar to the discussion above, it is not until the interface covers a large portion of the roughness that the damping effect takes place. Therefore it is evident that there are competing effects between the interface suppressing vertical velocity fluctuations and the asperities doing the opposite by enhancing them. Case T-RI3 exhibits a large gradient near the SHS wall since the wall-normal velocity is zero yet the streamwise component sees a large slip effect as was shown in figure \ref{fig:streamwise_rij}(a). As we move away from the wall, the shear stress has to balance out with the top wall given that we are running a constant pressure gradient, which explains why the profiles are parallel in that region. Figure \ref{fig:shear_rij}(b) shows the $u'v'$ component normalised by $u^B_{\tau}$, and a good collapse of the data is observed with the exception of Case T-RI3. This indicates that overall, the near-wall turbulence is not fundamentally changed for Cases T-RFW, T-RI1 and T-RI2 whereas the turbulent structures of Case T-RI3 are different. Figure \ref{fig:shear_rij}(c) shows $-u'v'$ plotted on a $\log$ scale in wall units (normalised by $u^B_{\tau}$) along with the viscous stress $\mu(dU/dy)$ and total stress $\tau_t$ which shows the total shear stress budget. We can see the cases where $-u'v' > 0$ and where $\tau_t > \mu (dU/dy)$ leading to $u^+_s < b^+_s$ as discussed earlier. Case RFW seems to collapse onto the baseline while Cases T-RI1 and T-RI2 shift towards the slip wall. The shear stresses go to zero near the wall due to the infinite surface tension that keeps the interface flat. 

Earlier we discussed how $u_s$ is not an accurate predictor of $DR$ since both fully wetted roughness and SHS result in a positive slip velocity. $\Delta U_b$ on the other hand correlates with $\Delta \tau_w$. One can show that the Reynolds shear stress is tied to the change in bulk velocity which is manifested in additional turbulent losses. This is done by applying a triple integration to the averaged transport equation for the streamwise momentum equation. This was demonstrated in \cite{hasegawa2011} and \cite{Turk2014} where the following identities are obtained:
\begin{equation}
 U_b = \frac{Re{_\tau}}{3} + u_s - \int_0^{\delta} \! (1-\frac{y}{\delta})(-u^{\prime}v^{\prime}) \, \mathrm{d}y.
 \label{eq:delta_ub_slip_shear}
\end{equation}
For the baseline case (T-S) where we have no-slip walls: 
\begin{equation}
 U_{b,0} = \frac{Re{_\tau}}{3} - \int_0^{\delta} \! (1-\frac{y}{\delta})(-u^{\prime}_0 v^{\prime}_0) \, \mathrm{d}y.
 \label{eq:delta_ub_slip_shear}
\end{equation}
This leads to the final form given by the following:
\begin{equation}
 \Delta U_b = U_b - U_{b,0} = u_s - \int_0^{\delta} \! (1-\frac{y}{\delta})(-u^{\prime}v^{\prime}+u^{\prime}_0 v^{\prime}_0) \, \mathrm{d}y,
 \label{eq:delta_ub_slip_shear}
\end{equation}
where $u^{\prime}_0 v^{\prime}_0$ denotes the shear stress of the baseline case. 
Take Case T-RFW as an example. The fully wetted roughness enhances vertical velocity fluctuations and so does having an interface at a small height location (Case T-RI1). There $-u^{\prime}v^{\prime}>-u^{\prime}_0 v^{\prime}_0$ and therefore $\int_0^{\delta} \! (1-\frac{y}{\delta})(-u^{\prime}v^{\prime}+u^{\prime}_0 v^{\prime}_0) > 0$ which happens also to be larger than the $u_s$ caused by the roughness or the presence of the interface. This gives $\Delta U_b < 0 $ indicating an increase in drag. For Cases T-RI2 and T-RI3, $u^{\prime}v^{\prime}$ is damped when compared to baseline. Therefore $-u^{\prime}v^{\prime}<-u^{\prime}_0 v^{\prime}_0$ and the integral term ends up coming out to be negative which results in $\Delta U_b > 0$ and hence drag reduction. One implication of this result is that a superhydrophobic surface might fail in the sense of reducing drag at high pressure in spite of the interface itself being stable. 

\subsubsection{Flow structures}
\begin{figure}
\centering{
\includegraphics[width=84mm]{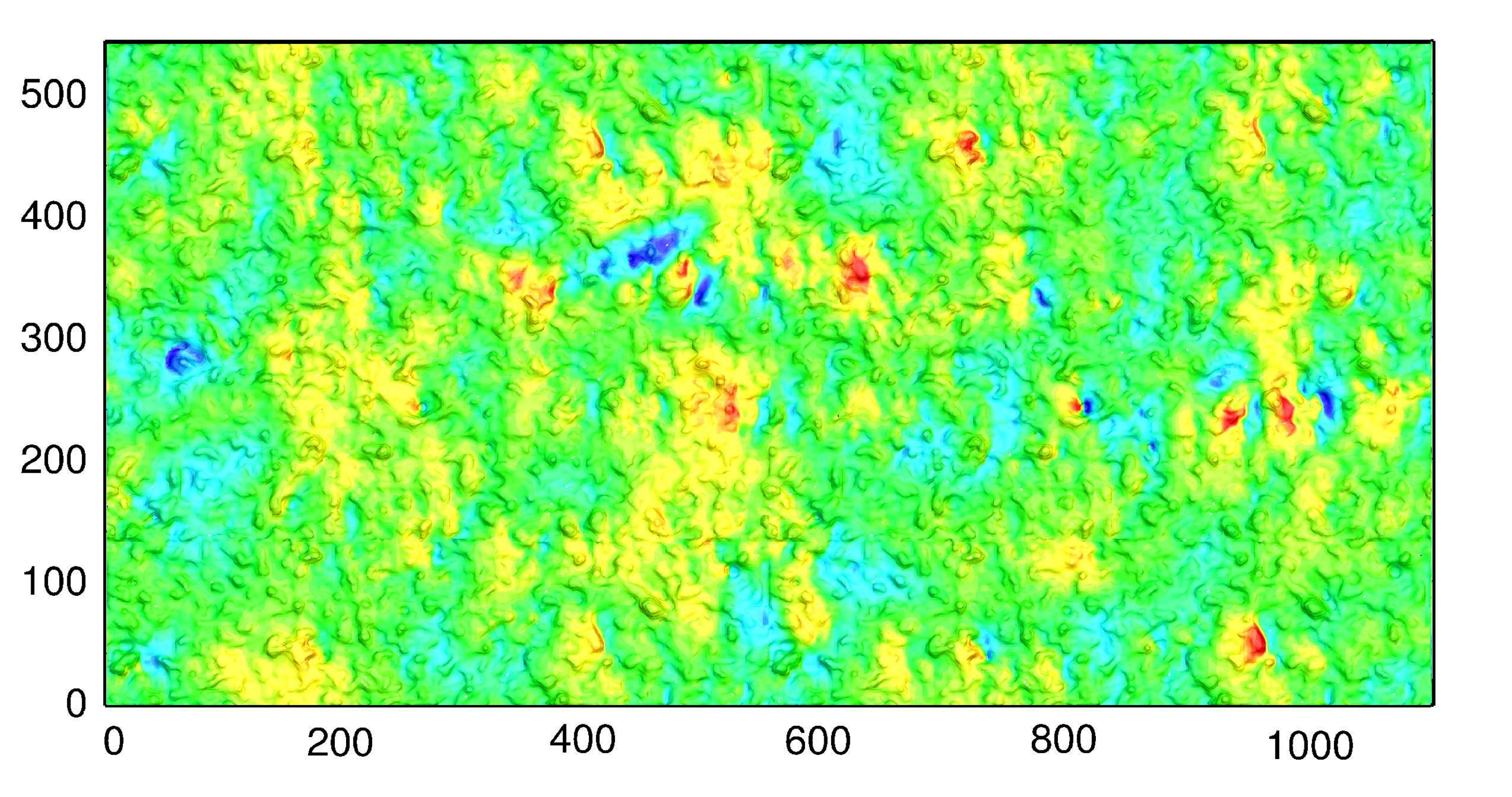} 
\includegraphics[width=80mm]{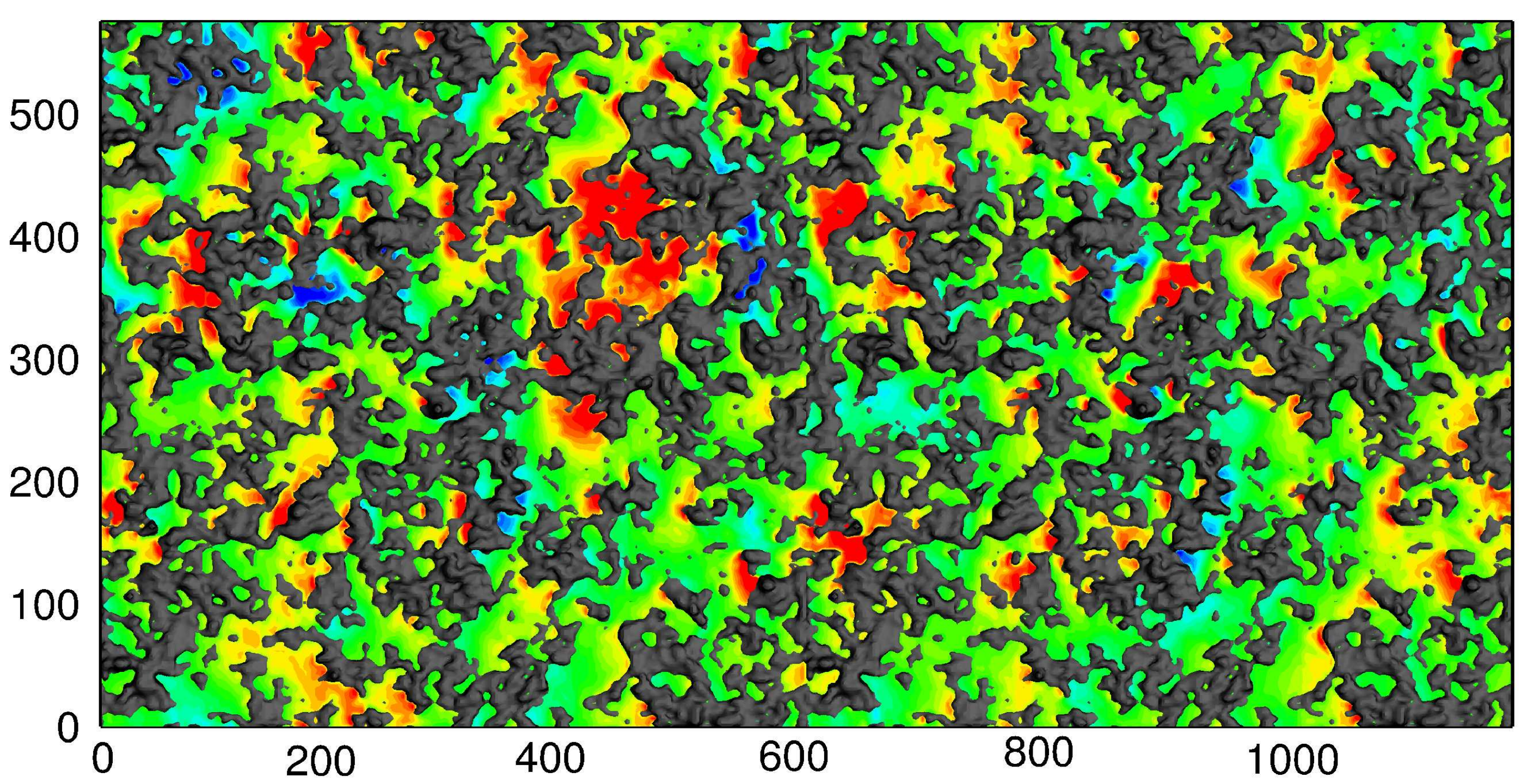} 
\includegraphics[width=80mm]{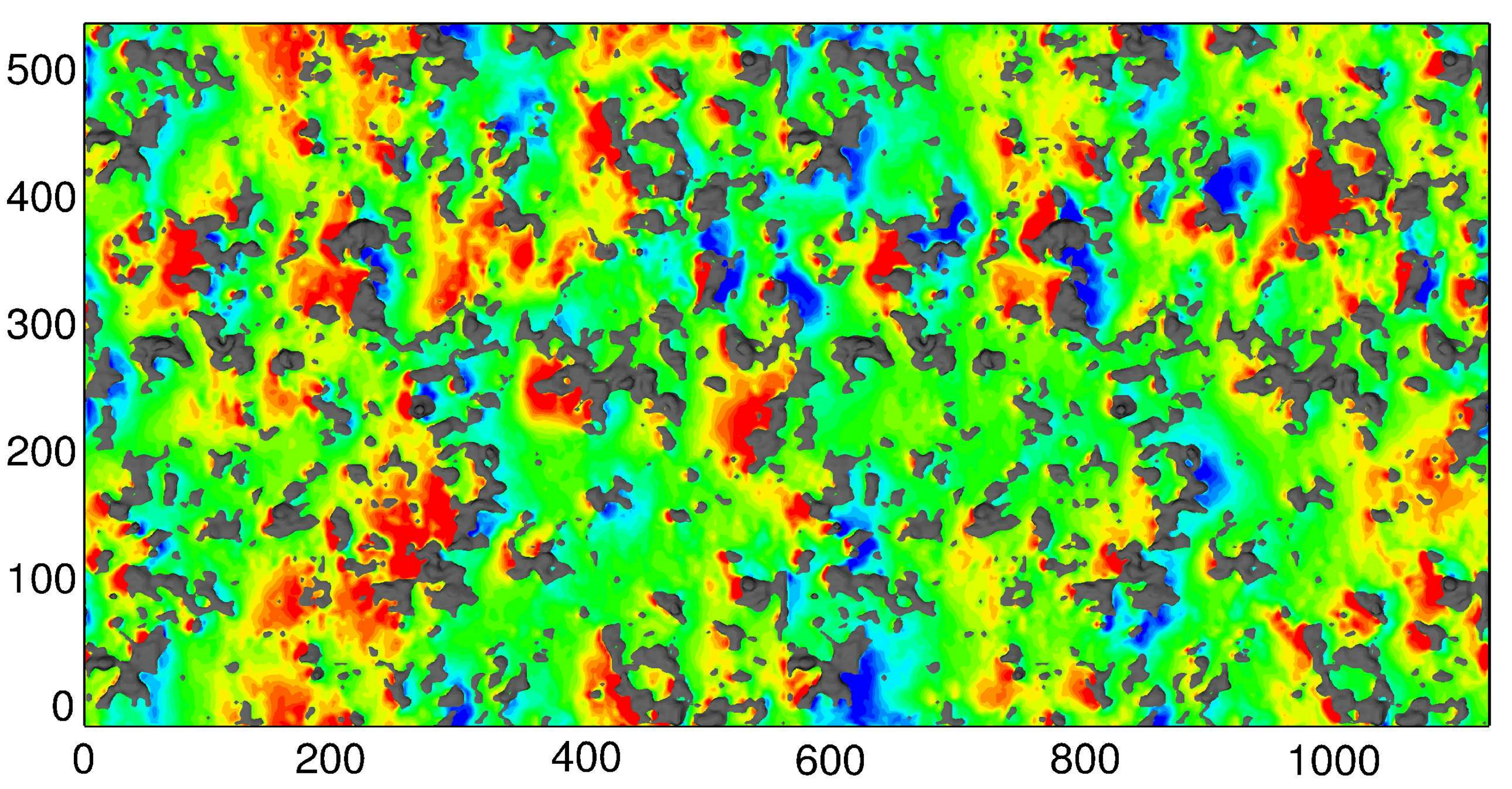} 
\includegraphics[width=83mm]{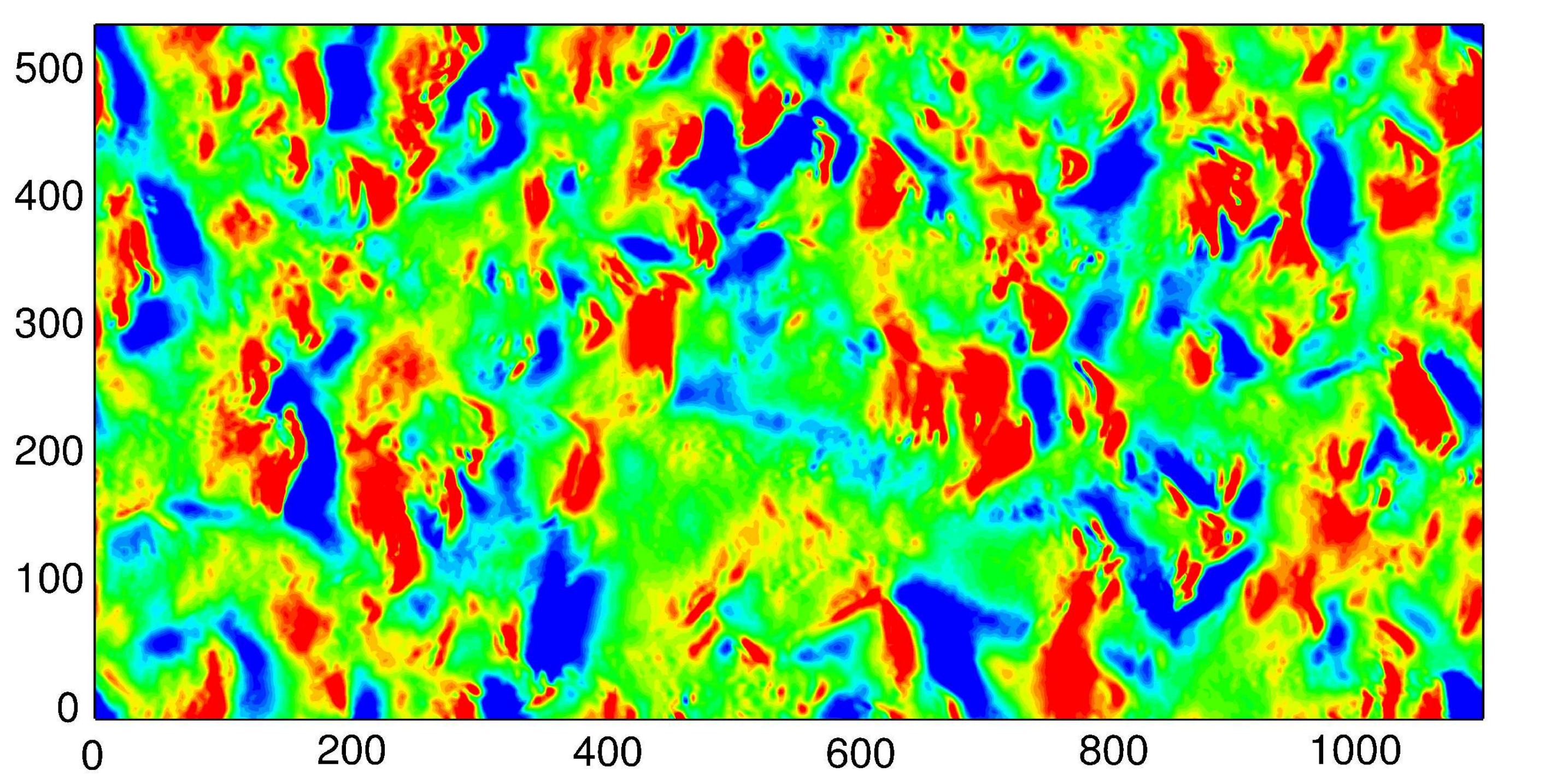} 
\put(-115,-10){$x^+$}
\put(-250,60){$z^+$}
}
\caption{Instantaneous contours of $p^+$ for (a) Case T-RFW on the roughness, (b) Case T-RI1 (c) Case T-RI2 and (d) Case T-RI3 on the interface location. Range of the contours is from $-5$ to $5$ in wall units.}
\label{fig:ws_pplus_inst}
\end{figure}

\begin{figure}
\centering{
\includegraphics[width=80mm]{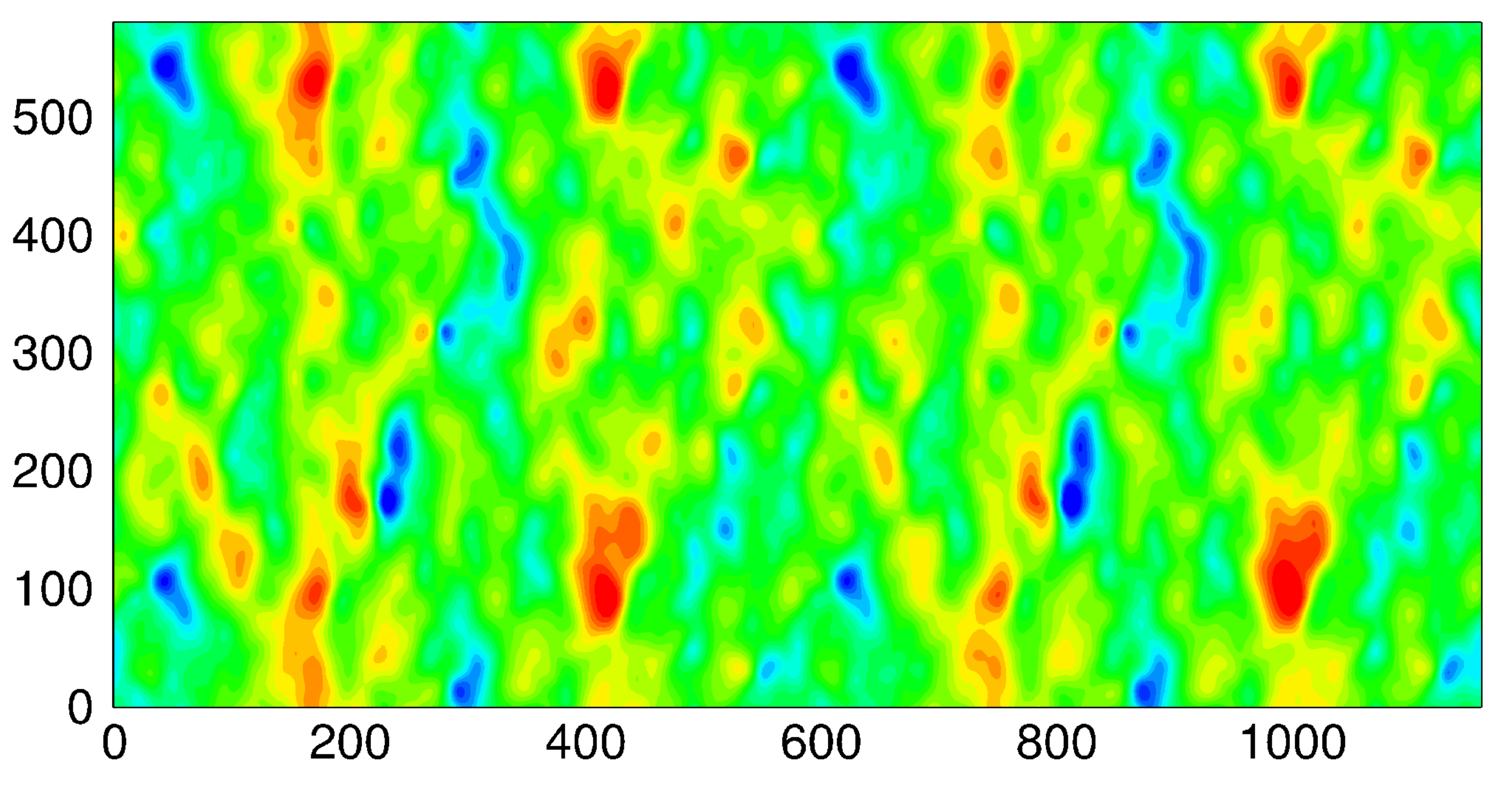}
\includegraphics[width=80mm]{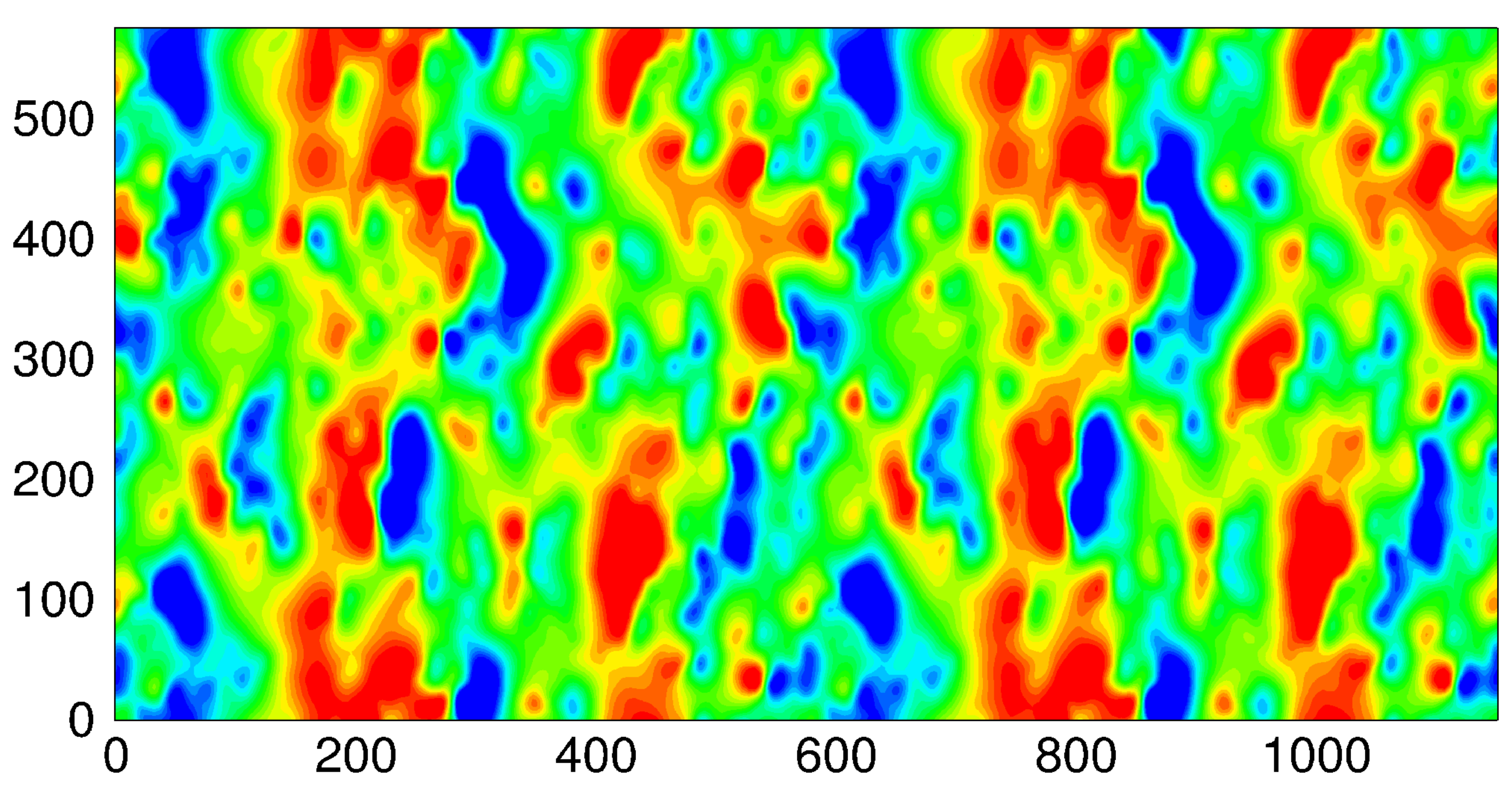}
\includegraphics[width=80mm]{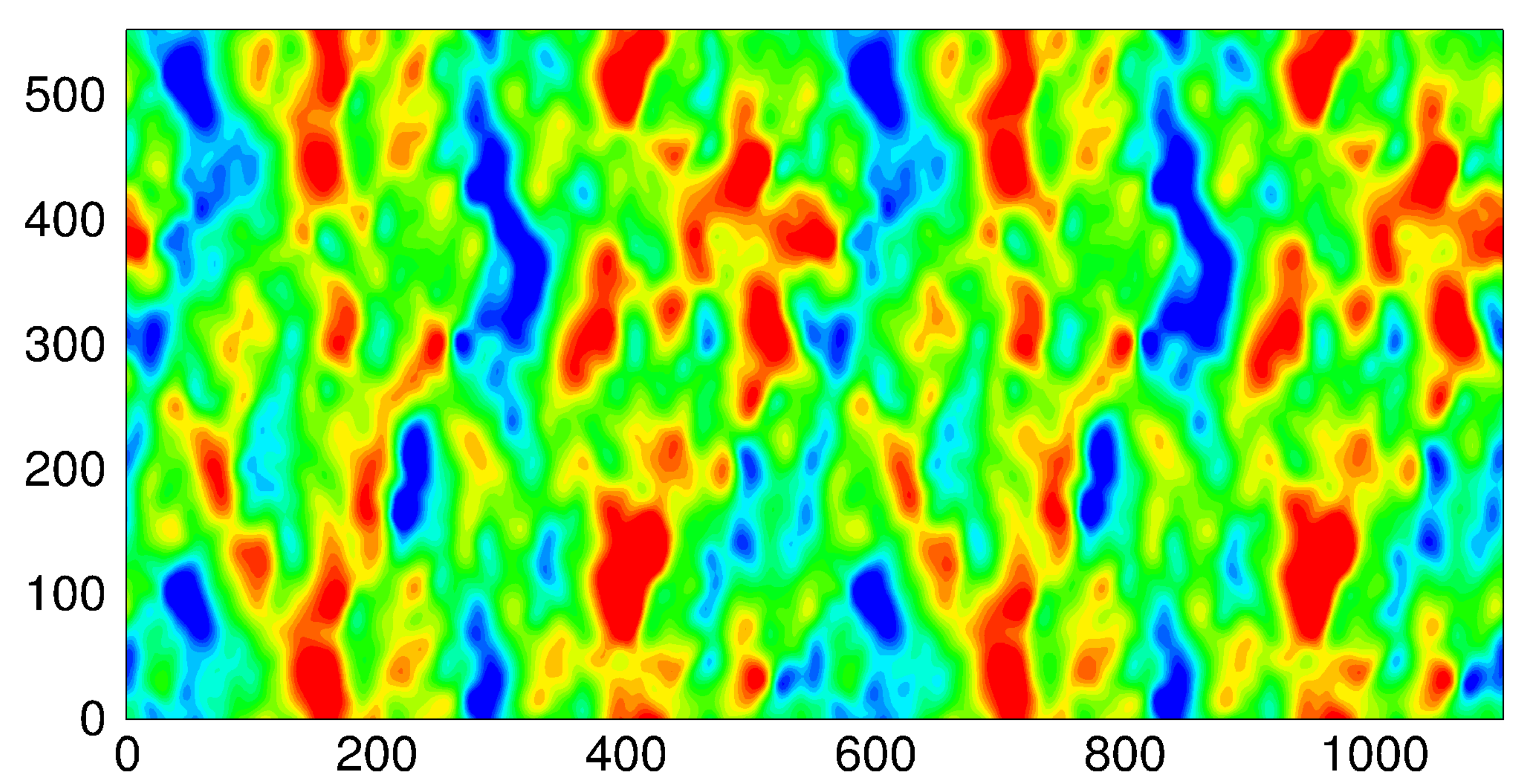}
\includegraphics[width=80mm]{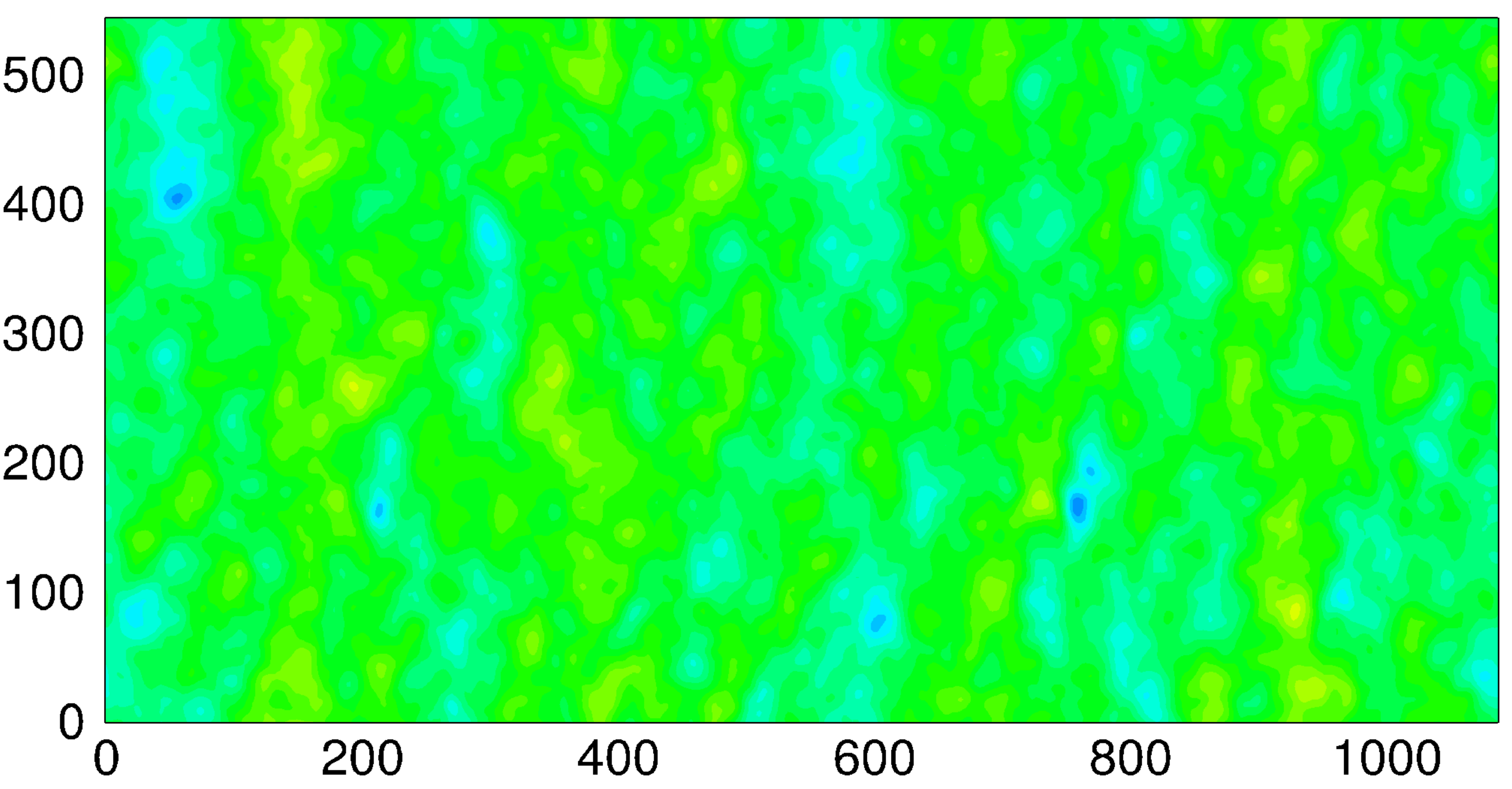}
\put(-115,-10){$x^+$}
\put(-250,60){$z^+$}
}
\caption{Time averaged contours of $p^+$ at $y^+=15$ for (a) Case T-RFW, (b) Case T-RI1 (c) Case T-RI2 and (d) Case T-RI3. Range of contours is from $-2$ to $2$ in wall units.}
\label{fig:ys_pplus_avg}
\end{figure}

\begin{figure}
\centering{
\includegraphics[width=80mm]{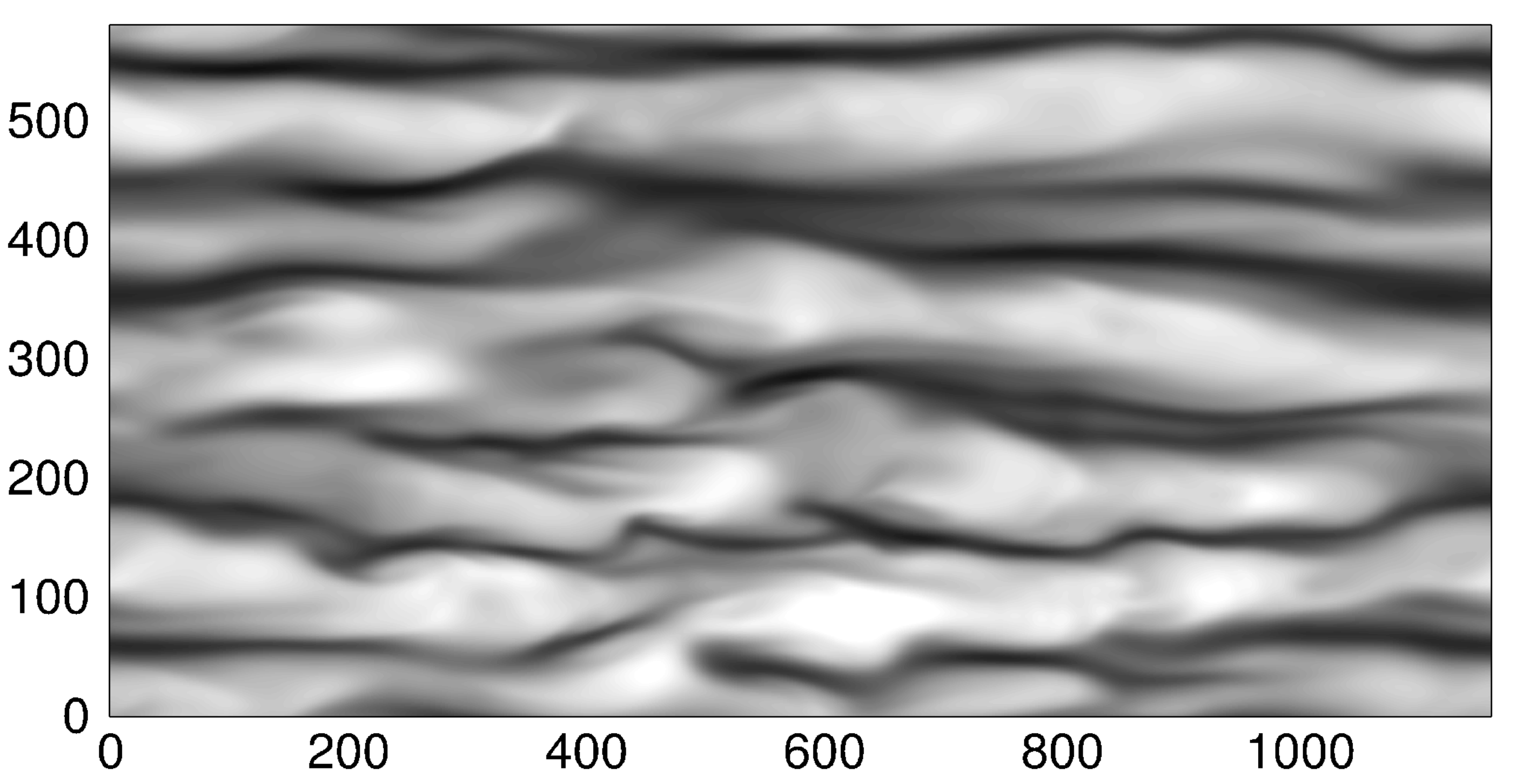}
\includegraphics[width=80mm]{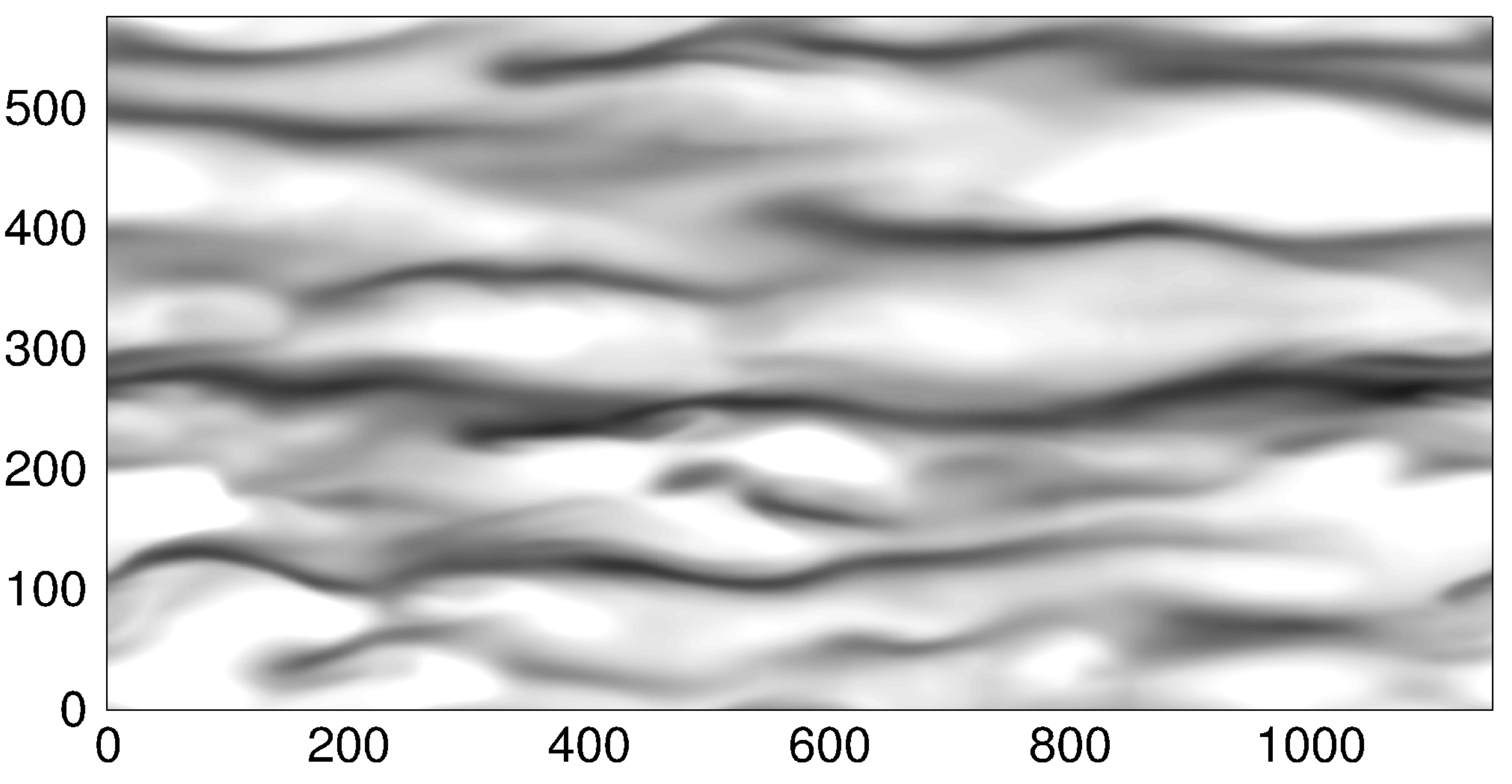}
\includegraphics[width=80mm]{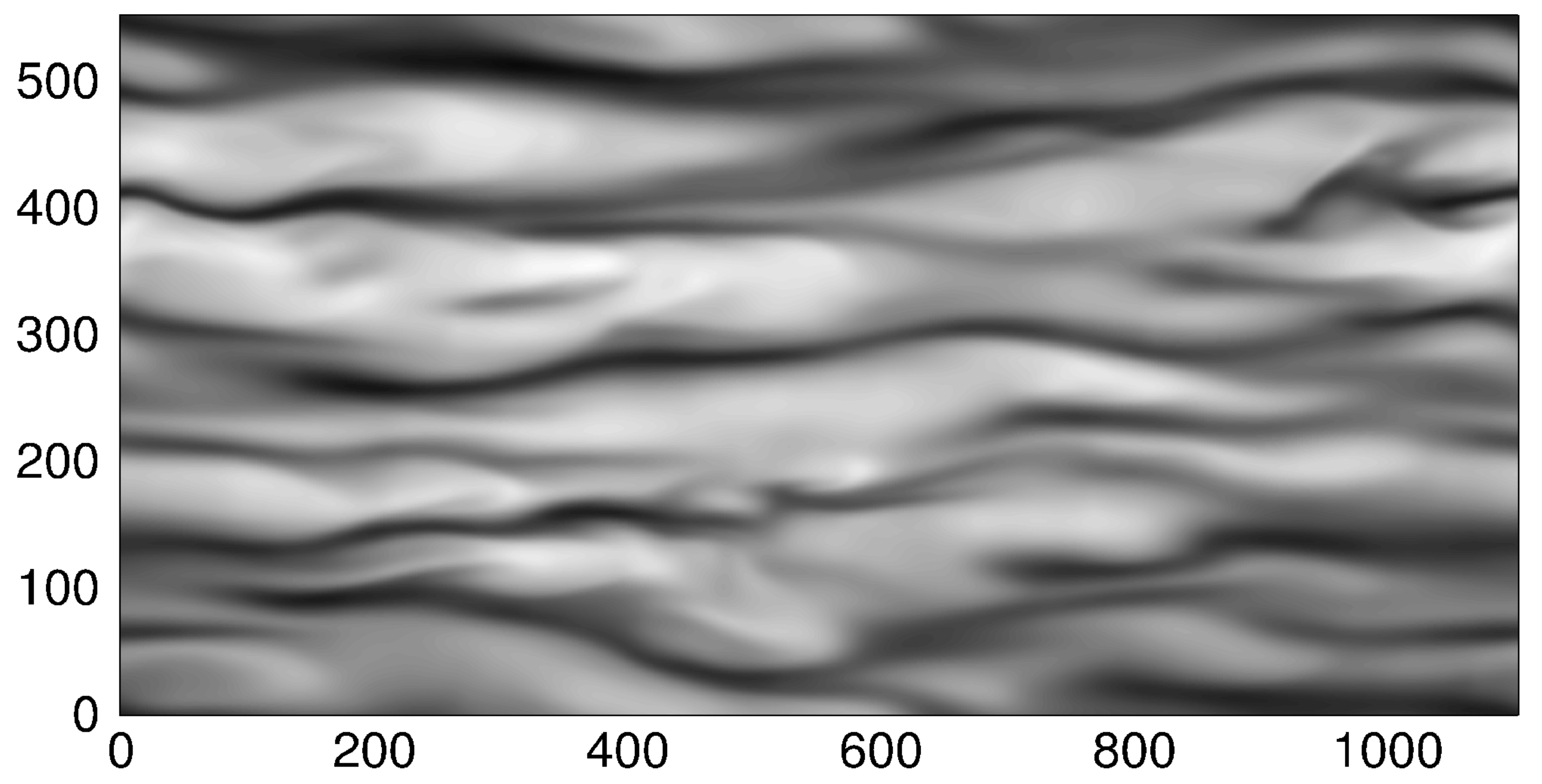}
\includegraphics[width=80mm]{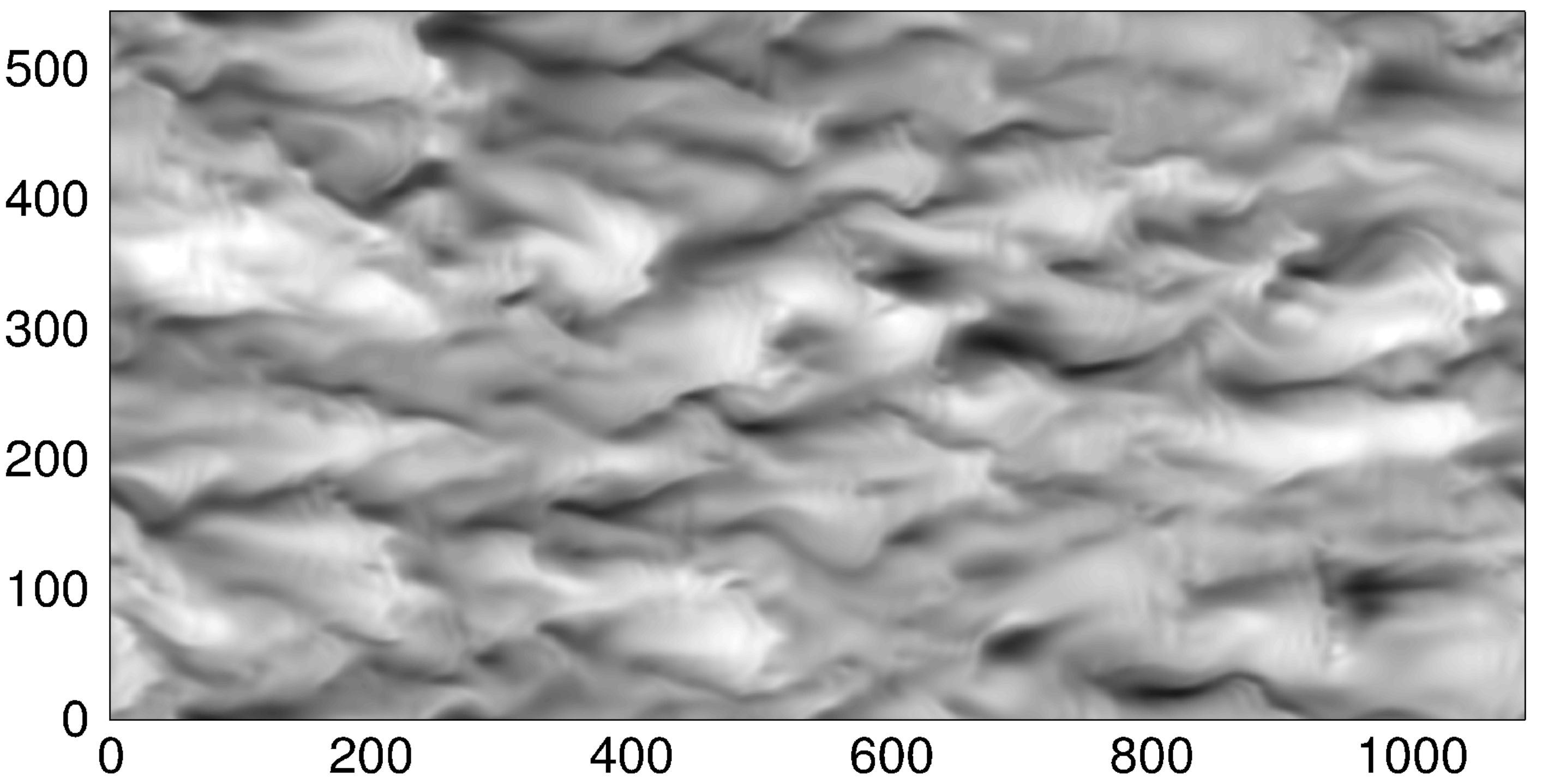}
\put(-115,-10){$x^+$}
\put(-250,60){$z^+$}
}
\caption{Instantaneous greyscale contours of $u/U_b$ at $y^+=15$ for (a) Case T-RFW, (b) Case T-RI1 (c) Case T-RI2 and (d) Case T-RI3. Range of contours is from $0.2$ to $1$.}
\label{fig:ys_u_Ub_inst}
\end{figure}

 \begin{figure}
\centering{
\includegraphics[width=80mm]{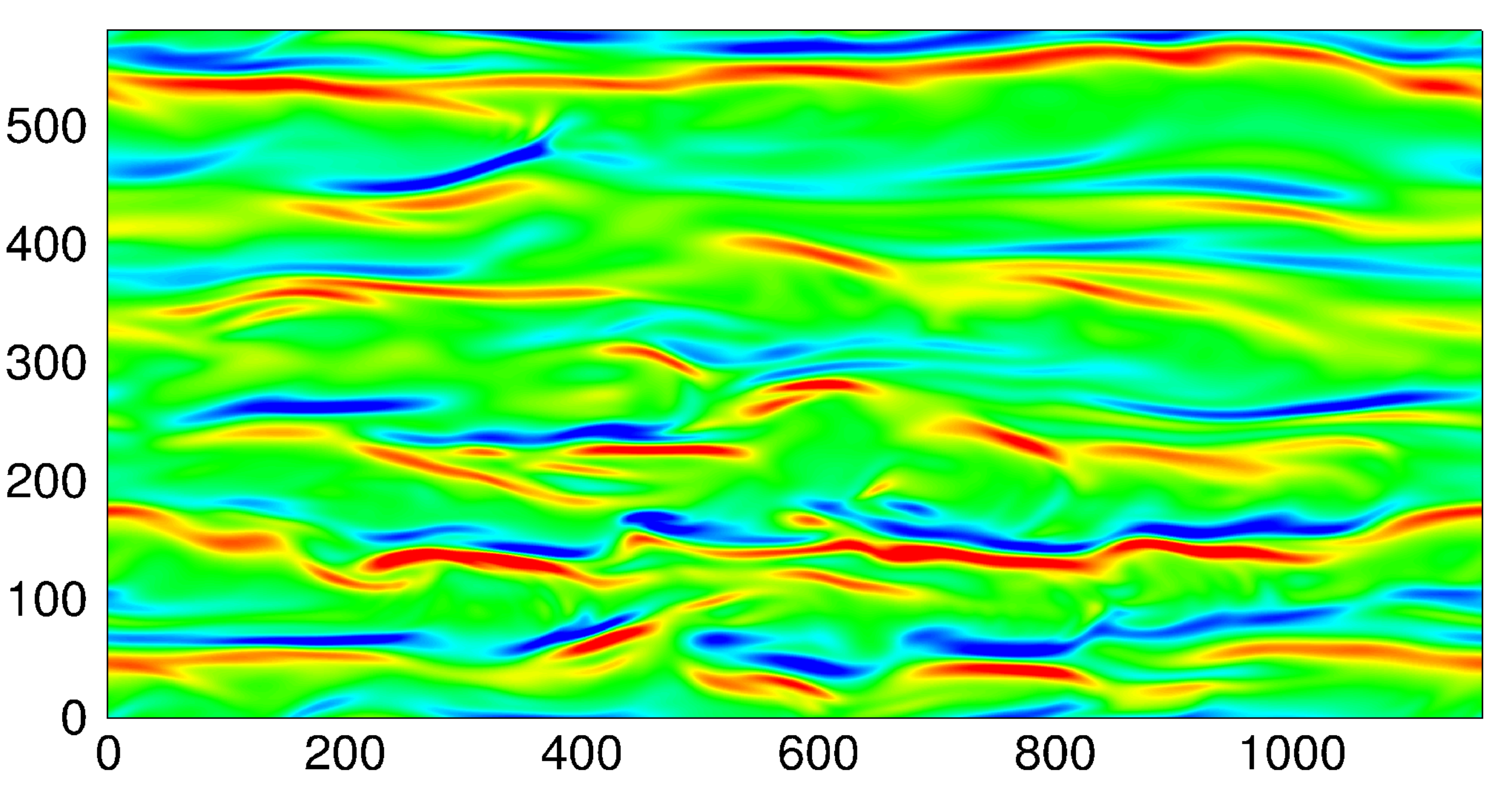}
\includegraphics[width=80mm]{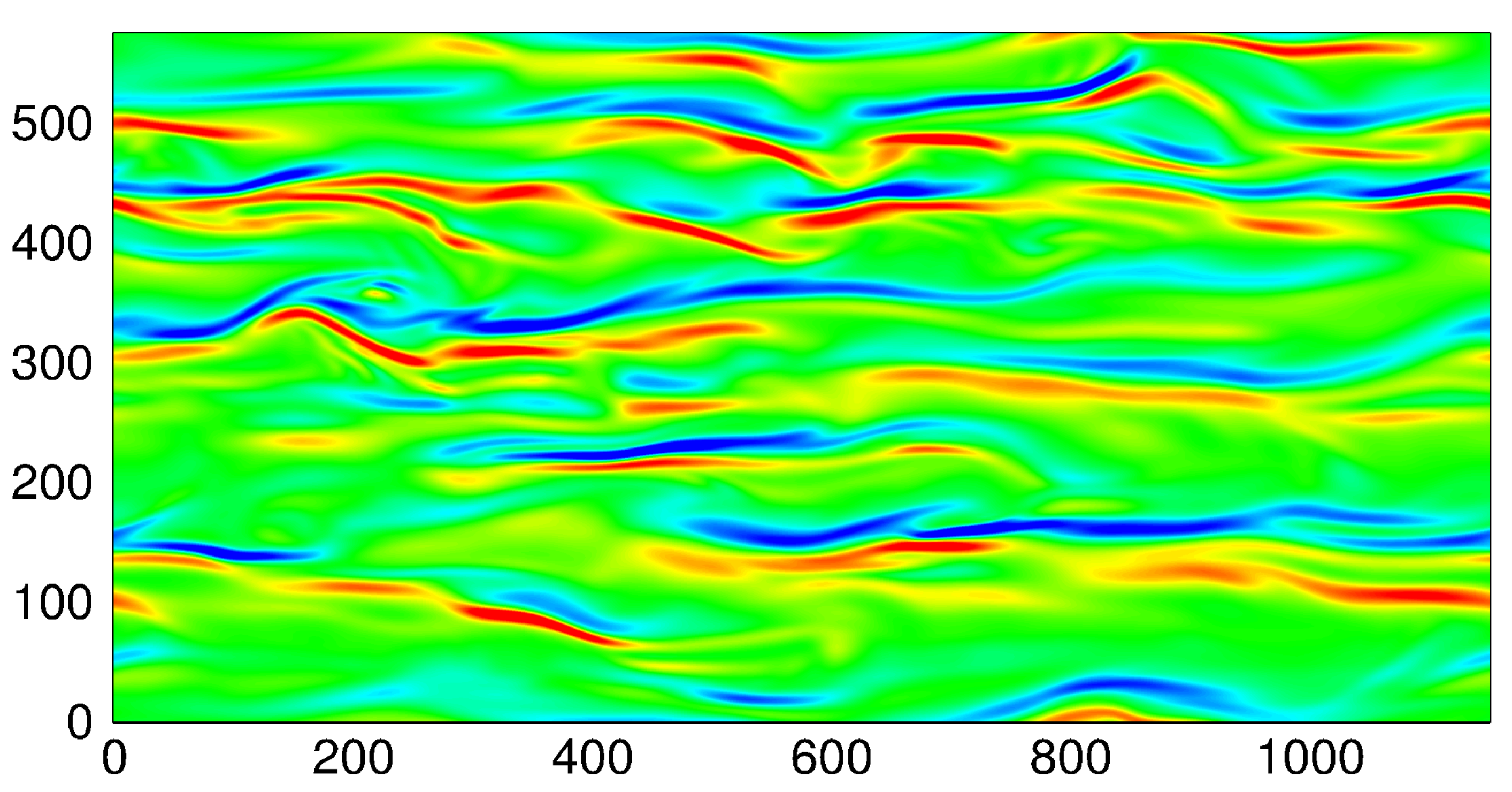}
\includegraphics[width=80mm]{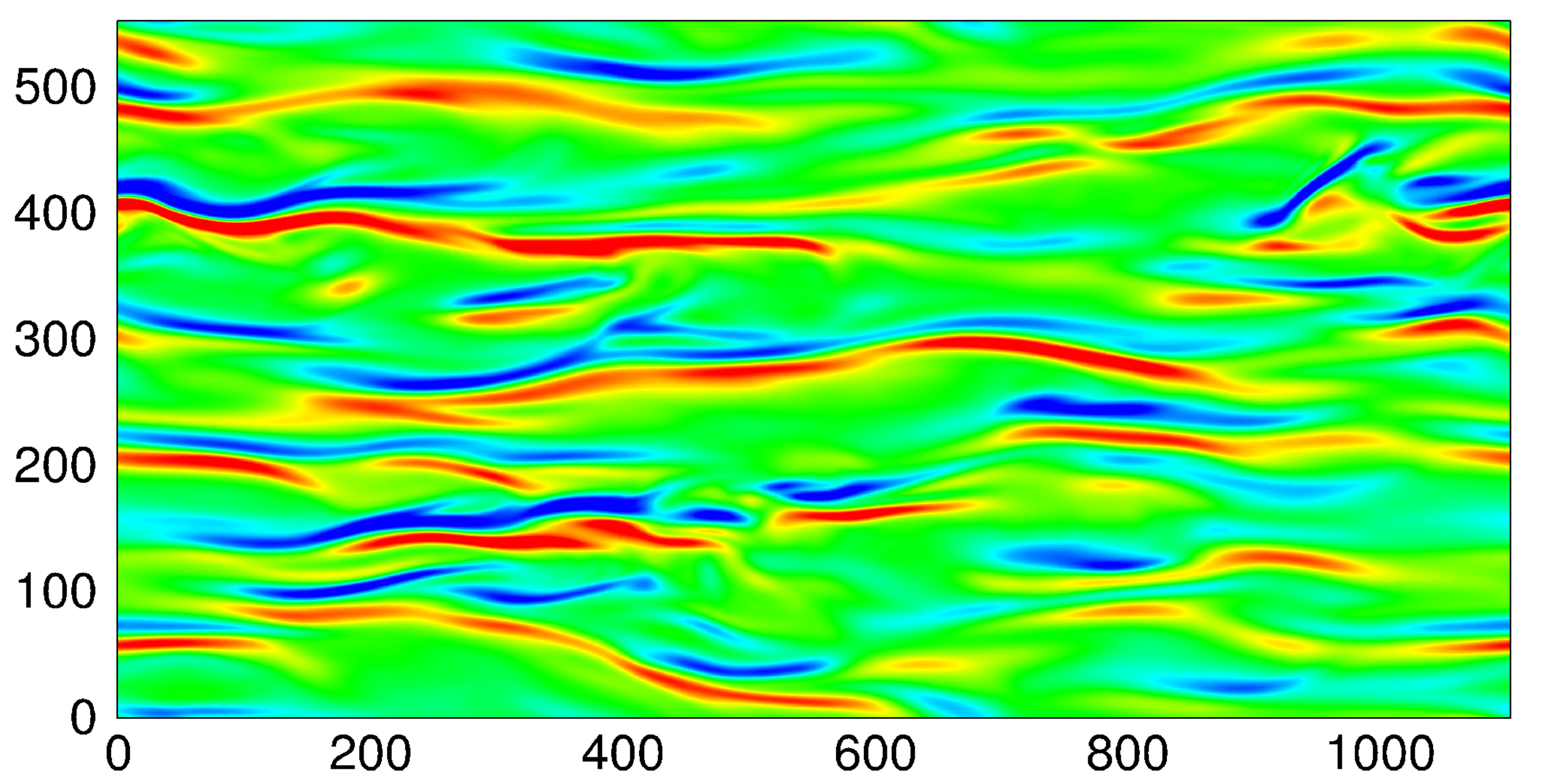}
\includegraphics[width=80mm]{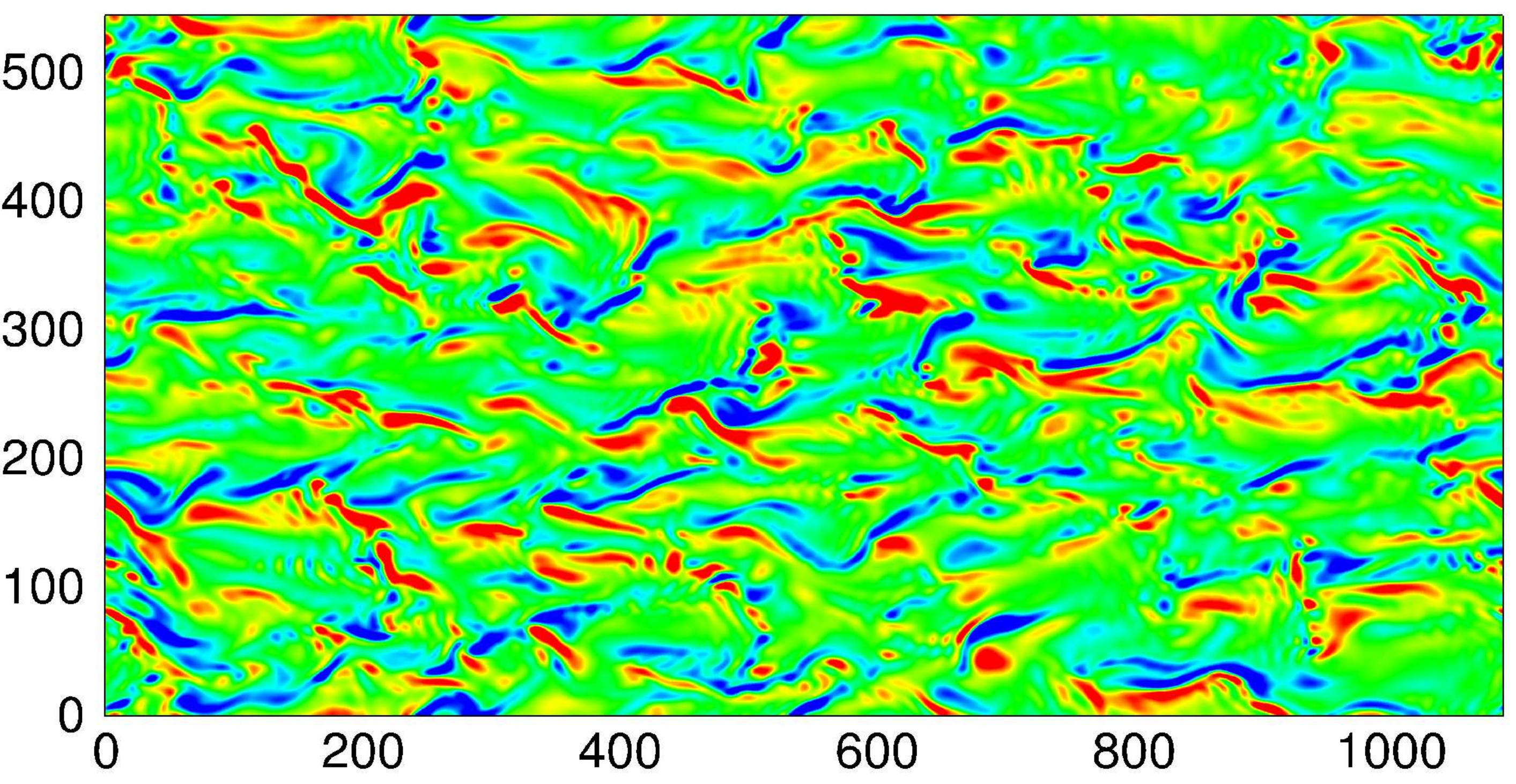}
\put(-115,-10){$x^+$}
\put(-250,60){$z^+$}
}
\caption{Instantaneous contours of $\omega^+_y$ at $y^+=15$ for (a) Case T-RFW, (b) Case T-RI1 (c) Case T-RI2 and (d) Case T-RI3. Range of contours is from $-0.5$ to $0.5$ in wall units.}
\label{fig:ys_wyplus_inst}
\end{figure}

\begin{figure}
\centering{
\includegraphics[width=90mm]{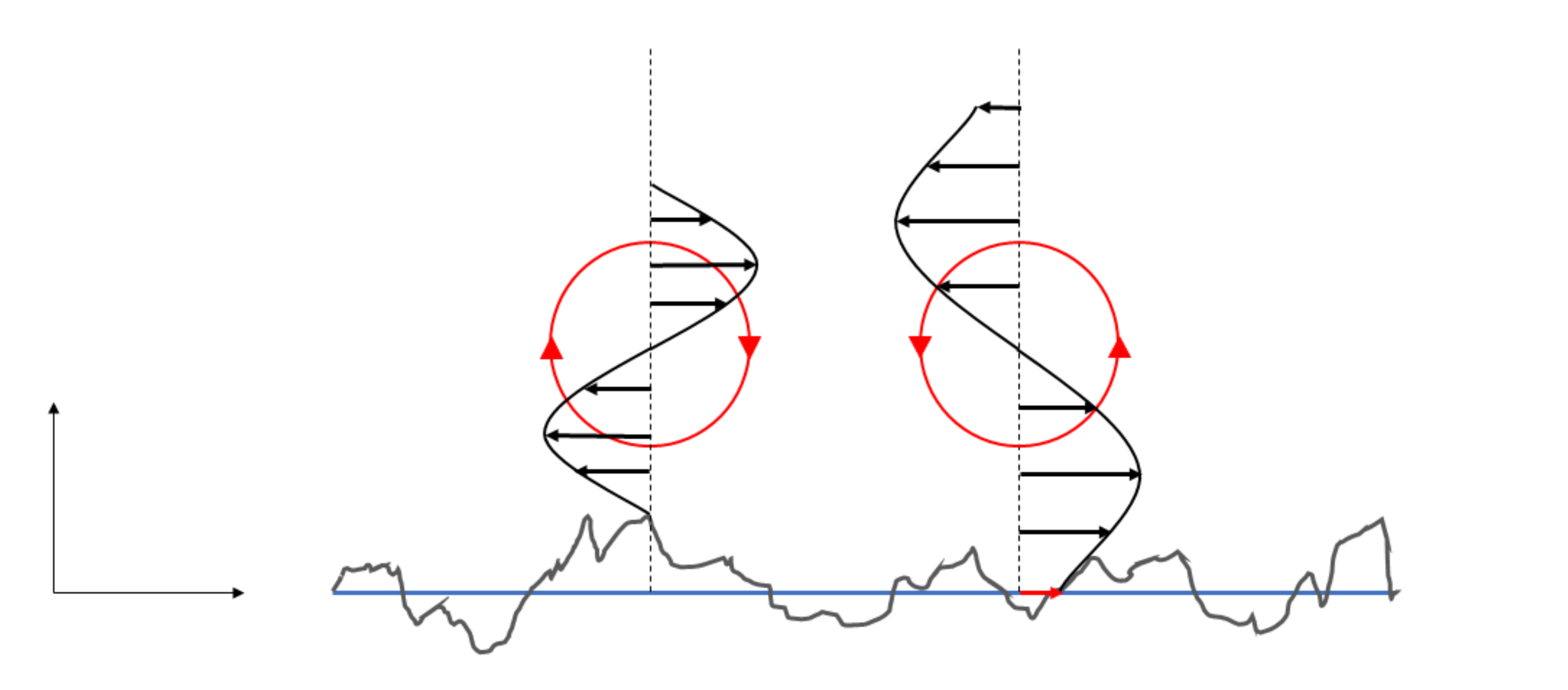}
\put(-250,50){$y$}
\put(-220,10){$z$}
\put(-80,8){$w_s$}
}
\centering{
\includegraphics[width=90mm]{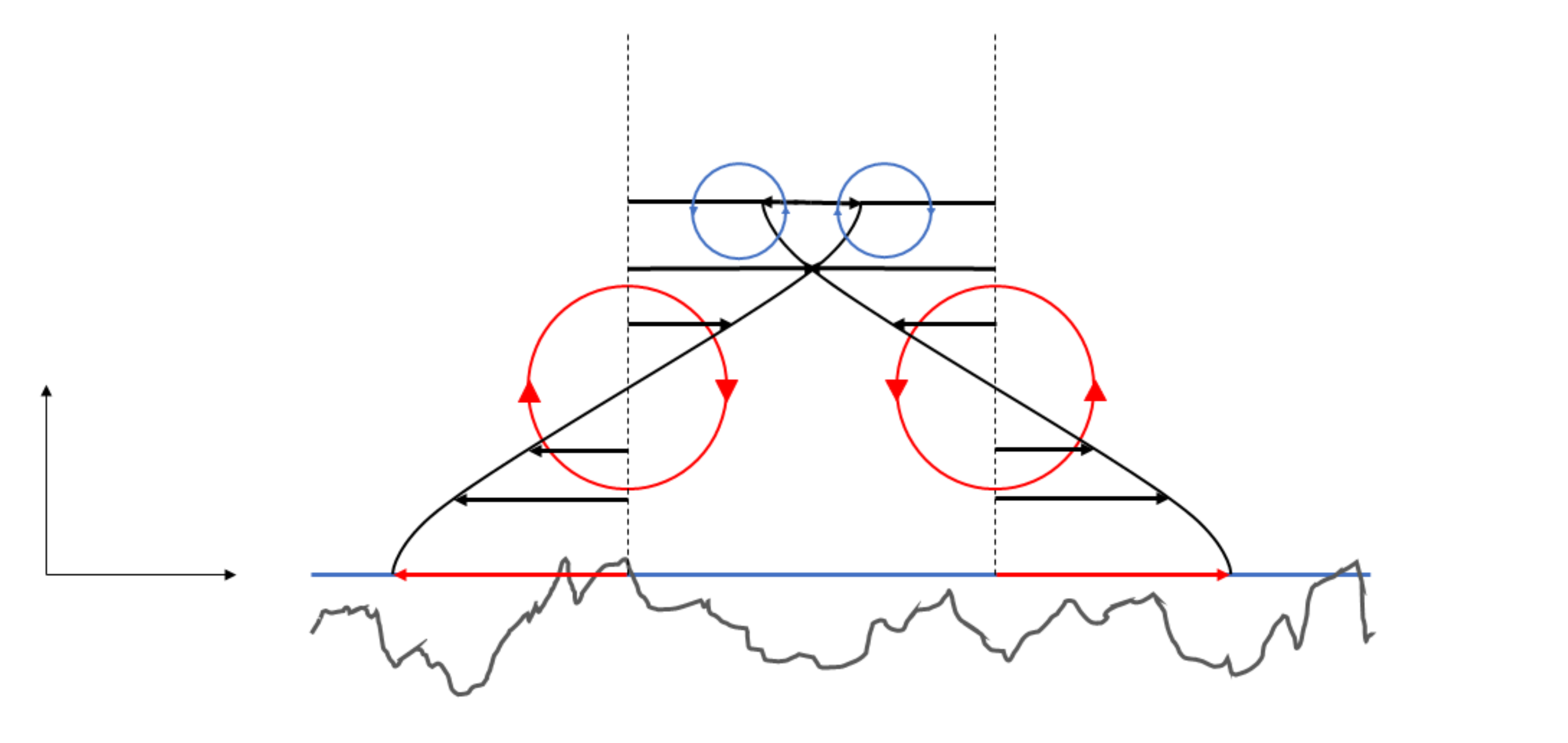}
\put(-250,60){$y$}
\put(-220,20){$z$}
\put(-190,20){$w_s$}
}
\caption{Illustration of the near-wall vortex structure. The asymmetry in slip due to the random roughness is illustrated in (a) whereas (b) shows the effect of having an interface cover most of the roughness and its effect on slip and induced secondary flow interaction.}
\label{fig:vort_illus}
\end{figure}

The near-wall shear stress $u'v'$ collapsed when scaled with the bottom wall shear velocity indicating that the near-wall behaviour remains unchanged with the exception of Case T-RI3. In this section we look more closely at the near-wall region. Figure \ref{fig:ws_pplus_inst}(a-d) shows instantaneous pressure fluctuations in wall units where $p^+=p/{(u^B_{\tau})}^2$. 
Figure \ref{fig:ws_pplus_inst}(a) shows the pressure contours along the rough wall while figures \ref{fig:ws_pplus_inst}(b)-\ref{fig:ws_pplus_inst}(d) are taken at the interface location (the reference plane) for each case. The axes are shown in wall units scaled by $u^B_{\tau}/\nu$. The contours change noticeably from figures \ref{fig:ws_pplus_inst}(a) to \ref{fig:ws_pplus_inst}(d). The instantaneous behaviour in pressure for figures \ref{fig:ws_pplus_inst}(a) and \ref{fig:ws_pplus_inst}(d) shows similar large length scales for fully wetted and fully covered roughness on the order of $100\nu/u_{\tau}$. Larger pressure intensities are visible in figure \ref{fig:ws_pplus_inst}(d) due to slip at the interface. Local contribution due to the presence of protruding roughness elements is observed in figures \ref{fig:ws_pplus_inst}(b) and \ref{fig:ws_pplus_inst}(c). Asperities cause stagnation in front of them as observed by the high pressure values. Low pressure values are seen in their wake. This effect is reduced as more asperities are covered by the interface as shown in figure \ref{fig:ws_pplus_inst}(c). This behaviour was also observed in \cite{Seo2015}.
%For Case T-RFW, the contribution of turbulent fluctuations as they advect downstream as observed in figure \ref{fig:ws_pplus_inst}(a). The typical length scale if observed to be around $\thicksim 100 \nu/u_{\tau}$ similar to those of smooth walls. Large pressure values are observed at the roughness asperities and low pressure regions in their wake, however they are less dominant than the turbulent pressure fluctuations. Once an interface is introduced, the contribution due to stagnation of flow in front of the asperities and behind them becomes more dominant as observed in figure \ref{fig:ws_pplus_inst}(b). The effect of a stable interface and the stagnation due to the asperities shows a noticeable effect, as regions of high pressure regions are much larger in length scales than those of the turbulent fluctuations. As the interface height increases, Figure \ref{fig:ws_pplus_inst}(c) shows that the effect of stagnation still dominates but the regions of high and low pressure are smaller in length scale than figure \ref{fig:ws_pplus_inst}(b) simply because more asperities are covered up. Figure \ref{fig:ws_pplus_inst}(c) shows higher intensities in the pressure contours, stagnation due to asperities does not contribute since the rough wall is fully covered. Figure \ref{fig:ws_pplus_inst}(c) is less quiescent than figure \ref{fig:ws_pplus_inst}(a) due to the nature of a fully slip regime at the interface. 

We extract slices in the wall-normal plane such that $y^+ = 15$ and examine the time-averaged pressure distribution. Figure \ref{fig:ys_pplus_avg}(a-d) shows striking visual difference between Case T-RFW, Cases T-RI1 and T-RI2, and Case T-RI3. The mean pressure is zero for a smooth channel; this is only observed in figure \ref{fig:ys_pplus_avg}(d) when the interface fully covers the roughness. Figure \ref{fig:ys_pplus_avg}(a) shows a variation in the mean pressure due to the signature of the rough wall. The contribution of stagnation pressure at the interface is overlaid with the contribution of asperities. Large variations in mean pressure interspersed across the domain are observed in figures \ref{fig:ys_pplus_avg}(b) and \ref{fig:ys_pplus_avg}(c) due to the interaction between the flow at the interface and around the asperities. 

%Figure \ref{fig:ys_pplus_avg}(a) shows the pressure footprint scaled with $\thicksim 100 \nu/u_{\tau}$; the scale of wall-pressure fluctuation seems to increase by almost $1.5$ times its original size in the spanwise direction, and is a little wider in the streamwise direction. Case T-RI2 is shown in figure \ref{fig:ys_pplus_avg}(c); when the roughness is submerged even more, the pressure length scale in the spanwise direction remains fairly large while its length in the streamwise direction shrinks due to the increase in slip velocity and less protruding roughness elements. Case T-RI3 shows a clear decrease in the size, coherence and magnitude of the pressure fluctuations when the largest slip area is achieved as evident from figure \ref{fig:ys_pplus_avg}(d). 

Since the pressure footprint can be related to near-wall velocity, we plot the instantaneous velocity $u$ normalised by $U_b$ in greyscale on the same $y$-slice. The presence of an interface affects the spanwise streak motion, and the distance between streaks can be visually seen to be around $100$ wall units in figure \ref{fig:ys_u_Ub_inst}(a). Figure \ref{fig:ys_u_Ub_inst}(b) shows regions with larger streamwise velocities where the separation distance between streaks becomes larger than $100$ wall units in the spanwise direction. The slightly larger distance between the streaks is maintained in Case T-RI2 but that coherent structure is completely destroyed in Case T-RI3 as shown in figure \ref{fig:ys_u_Ub_inst}(d). This behaviour for Case T-RI3 is likely to be due to the lower near-wall shear yielding lower values of $Sq^2/\epsilon$ where $S$, $q^2$ and $\epsilon$ are the mean shear, twice the turbulence kinetic energy and turbulent dissipation respectively. As shown by \cite{rogers1987structure} and \cite{lee1990structure} for homogeneous shear flow, only high values of $Sq^2/\epsilon$ as encountered in the near-wall region produce streaks.    

 The instantaneous wall-normal vorticity $\omega^+_y$ is shown in figure \ref{fig:ys_wyplus_inst}(a-d) to illustrate pairs of counter-rotating vortices in regions of low- and high-momentum fluid streaks. The $\omega^+_y$ contours show similar behaviour to the streamwise velocity with varying interface height.  
A striking feature in the $\omega^+_y$ plots is the slight asymmetry of counter-rotating vortex pairs. Figure \ref{fig:vort_illus}(a) schematically shows how due to the random nature of the slip no-slip behaviour over rough SHS, each vortex in a counter-rotating vortex pair experiences different slip areas producing an inhomogeneous spanwise slip $w_s$. This causes asymmetric velocity profiles to interact in the wall-normal direction. As the interface height increases, the slip area increases and less of the solid protrudes, and therefore $w_s$ becomes more homogeneous. The slip effect is however more amplified in this case (figure \ref{fig:vort_illus}b) and the spanwise velocity is much larger, penetrating further up in the wall-normal direction, with tertiary vortices set up above the pair of counter-rotating vortices. This could possibly explain why in figures \ref{fig:ys_u_Ub_inst}(d) and \ref{fig:ys_wyplus_inst}(d) we observe less coherence and more violent mixing of those structures. While such mixing is associated with drag increase, the slip velocity in the streamwise direction is more dominant which offsets this deleterious effect to give a net positive $DR$. This is explained by the change in bulk velocity as a function of slip velocity and the integral difference of shear stresses given by eq.(\ref{eq:delta_ub_slip_shear}).

\section{Summary} 
\label{sec:summ}

DNS of laminar Couette flow at $Re=740$ and turbulent channel flow at $Re_{\tau}=180$ are performed, where the bottom wall is a realistically rough SHS. The surface scan is reproduced computationally, and the surface statistics are verified again with the experiments. Simulations are also performed for a smooth wall to serve as a baseline. Simulations of the fully wetted case and an air--water interface at various heights are compared to the smooth channel. The effects of roughness and interface heights are discussed in detail. 

Simulations of laminar Couette flow show a penetration effect up to $40\%$ in the wall-normal direction due to the roughness. Various interface heights $h$ were considered and a nonlinear dependence of drag reduction $DR$ on $h$ is observed. The dependence can be categorised into three distinct regions. The drag is sensitive to the interface location in region $II$ described in section \ref{sec:mff} where $h$ is in the range of $-0.32 < h/S_q < 2.15$. The negative skewness $S_{sk} = -0.32$ of the roughness profile indicates that the surface contains more valleys than peaks and asperities. More than half of the surface roughness is filled with gas when $h$ is within the vicinity of $S_q$ since a large number of valleys become wetted. The solid fraction $\phi_s$ decreases and the gas fraction $\phi_g$ increases with increasing $h$ where fewer asperities are exposed to the outer flow. Therefore it can be shown that $DR$ is a function of $\phi_g$. To demonstrate the relation between $h$ and $\phi_g$, we first calculate $\phi_s$ by measuring the amount of non-wetted area above the interface. This is similiar to the bearing area curve (BAC). Based on the definition $\phi_g = 1- \phi_s$, the gas fraction is obtained and plotted as a function of $h$. A nonlinear fit is given by $\phi_g = 0.5\left[1+\tanh\left(0.95\frac{h}{S_q}-0.875\right)\right]$ and shows good agreement with the data. This is useful since $\phi_g$ is not known \textit{a priori} and $h$ is prescribed as an initial condition. Effective slip can be directly related to $DR$ using the definition $\frac{b_{eff}}{H}=DR/(1-DR)$. A power law using a linear regression fit is obtained and corrected for the fully wetted case. The relation is given by $\frac{b_{eff}}{H} = 0.5\left({\phi_g}\right)^{5/2}+ 0.02$ which shows a good agreement with the data and provides a useful model for the slip length given $\phi_g$. The results are compared to previous work done on structured geometry. It is observed that random rough surfaces behave like post geometries for $\phi_g<0.85$ and like transverse grooves in the upper limit. 

Based on the observations of the three distinct regions made in the laminar Couette flow, four interface heights are chosen for turbulent DNS channel flow. Simulations of a fully wetted rough case $h=S_v$ and the interface heights at $h=0$, $S_q$ and $S_p$ are performed for turbulent channel flow and the results are discussed. The mean velocity profile is normalised by two quantities: the average channel wall friction velocity $u_{\tau}$ and the bottom wall friction velocity $u^B_{\tau}$. The mean velocity profiles show the effect of roughness where a reduction in mass flux is obtained. The presence of an interface increases mass flux. The velocity profiles are offset by the slip velocities and show a good agreement in the law of the wall where the data collapse. Case T-RI3 exhibits an early departure from the law of the wall indicating a change in the turbulent structures. Various mean flow properties are extracted and plotted, where we show that $u_s$ can become a large fraction of the bulk velocity $U_b$ (up to $68$\%), and the slip length $b_s$ maintains the scaling law proposed by \cite{ybertetal_2007}. We show that $u_s$ is not a good indicator of $DR$ by itself since roughness induces a positive $u_s$ while increasing drag. A more reasonable quantity to describe $DR$ is the change in bulk velocity $\Delta U_b$ since it implicitly contains the information from $u_s$ and the additional turbulent losses in the form of a weighted Reynolds shear stress $\Delta U_b = u_s - \int_0^{\delta} \! (1-\frac{y}{\delta})(-u^{\prime}v^{\prime}+u^{\prime}_0 v^{\prime}_0) \, \mathrm{d}y$. This has implications for $DR$ where the surface might fail in reducing drag although the interface itself is stable for high pressure. $\Delta U_b$ shows a good correlation with $\Delta \tau_w$. Although $u_s$ and $b_s$ continually increase with increasing $\phi_g$, the change in wall shear stress $\Delta \tau_w$ plateaus at large gas fraction and so does the $\Delta U_b$. We discuss scaling laws for these quantities and correlate them with each other. 

The Reynolds stresses are also examined showing an overall behaviour consistent with previous work on structured geometries. Slip tends to shift the profiles towards the SHS wall whereas roughness pushes it away from wall. Asperities enhance negative shear stress and therefore momentum mixing, while the interface suppresses them. The streamwise Reynolds stress $u'u'$ and Reynolds shear stress $u'v'$ show a good collapse in the data when normalised by $u^B_{\tau}$ indicating that the near-wall turbulence remains fundamentally unchanged with the exception of Case T-RI3. This prompted a further investigation into the nature of the near-wall turbulent structures. We looked at instantaneous pressure contours in the near-wall region and at $y^+=15$ where time-averaged pressure $p^+$, instantaneous streamwise velocity $u/U_b$ and wall-normal vorticity $\omega^+_y$ are examined. 

Pressure fluctuations in the near-wall region exhibit a competing effect between large-scale turbulent fluctuations and a contribution due to stagnation pressure in front of the asperities.  
The instantaneous behaviour in pressure shows similar large-length-scale fluctuations for Case T-RFW and Case T-RI3 on the order of $100\nu/u_{\tau}$. Larger pressure intensities are visible in Case T-RI3 due to the slip effect at the interface. Local contribution due to the presence of protruding roughness elements is observed in Case T-RI1 and Case T-RI2. Asperities cause stagnation in front of them as observed by the high pressure values. Low pressure values are seen in their wake. This effect is reduced as more asperities are covered by the interface. Time-averaged pressure fluctuations show that Case T-RI3 resembles a smooth channel since the asperities are completely covered leading to a zero mean pressure variation. This is not observed for Cases T-RFW, T-RI1 and T-RI2 where the effect of asperities and the interface are clearly seen as large variations in mean pressure interspersed across the domain. At $y^+=15$, Case T-RI3 does indeed alter the near-wall turbulence where we see a complete loss of coherent streaks, as observed from the pressure fluctuations, streamwise velocity and wall-normal vorticity. A physical mechanism is proposed to explain the observed trends in flow structure.

\section*{Acknowledgements}
This work was supported by the United States Office of Naval Research (ONR) MURI (Multidisciplinary University Research Initiatives) program under Grant N00014-12-1-0874 managed by Dr Ki-Han Kim. Computing resources were provided by the Minnesota Supercomputing Institute (MSI). We are grateful to Prof. W. Choi at University of Texas Dallas and Prof. G. H. McKinley at MIT for providing us with the scanned surface data used in the present work. The authors would like to thank Dr P. Kumar and Dr Y. Li for their helpful discussions and suggestions.  

\appendix
\section{Validity of assumptions}
\label{appA}

In practice, for the superhydrophobic surface to sustain its drag reducing properties, the surface tension must be strong enough to maintain the presence of an air--water interface. This implies that the capillary pressure must be larger than the background turbulent pressure fluctuations. The balance between surface tension and the external pressure results in a meniscus shape and a contact angle at the wall contact boundaries. In the study, the interface is assumed to be flat, which it may not be, and that the interface is always sustained. We investigate the range of validity of our assumptions by using scaling arguments of the driving mechanisms in interfacial physics and comparing their orders of magnitudes.  The asterisk is used to denote dimensional quantities.\\

\subsection{Small interface deflection approximation}
Let $s^*$ represent the interface deflection and $w^*$ the average cavity width of the rough surface. The Young-Laplace equation gives:
\begin{equation}
 \Delta p^*_{c} = \frac{2\sigma^*}{R^*},
 \label{eq:younglaplace}
\end{equation}
where $\Delta p^*_{c}$ is the capillary pressure across the interface. Assuming the interface is pinned at the contact points of the cavity width, then $R^*$ is the radius of the interface. We can then relate $w^*$ to $s^*$ given that $w^*$ represents the chord of a circular segment such that $w^*=2\sqrt{s^*(2R^*-s^*)}$. Substitute for $R^*$ using eq. (\ref{eq:younglaplace}) to obtain the following relation:
\begin{equation}
 \frac{s^*}{w^*} \approx \frac{w^*\Delta p^*_{c}}{8\sigma^*}.
 \label{eq:s/w}
\end{equation}
 For a flat interface, $s^*/w^* \ll 1$ where $s^*/w^*$ represents the ratio of interface deflection to cavity width. Assume the maximum deflection to be no larger than $10$ \% such that the maximum deflection (contact) angle is less than $\thicksim$ $3^{\circ}$ so we obtain 
 \begin{equation}
  \frac{w^*\Delta p^*_{c}}{8\sigma^*} < 0.1 ,
 \end{equation}
 which gives 
 \begin{equation}
  w^* < \frac{0.8\sigma^*}{\Delta p^*_c} .
 \end{equation}
 Therefore the maximum sustained pressure given a cavity width is
 \begin{equation}
 \label{eq:pc_w}
  \Delta p^*_{c} < \frac{0.8\sigma^*}{w^*} .
 \end{equation}
 
 \subsection{Interface stability approximation}
 In a realistic environment, the turbulent pressure fluctuations play an important role in determining whether the interface breaks or remains intact. In order for the surface to maintain its drag reducing properties, capillary pressure must be strong enough to maintain the air--water interface and overcome turbulent pressure fluctuations. Using similar scaling arguments as before, we know that the turbulent pressure fluctuations scale as follows:
 \begin{equation}
 \label{eq:prms}
  p^*_{rms} \thicksim \textit{O} (\rho^* u^{*2}_{\tau}),
 \end{equation}
 and the capillary pressure as
 \begin{equation}
  \Delta p^*_{c} \thicksim \textit{O} \bigg( \frac{\sigma^*}{w^*} \bigg).
 \end{equation}
 In a stable configuration, $\Delta p^*_c \gg p^*_{rms}$ must be satisfied. Therefore we obtain the following relation:
 \begin{equation}
  \textit{O} \bigg( \frac{\sigma^*}{w^*} \bigg) \gg \textit{O} (\rho^* u^{*2}_{\tau}).
 \end{equation}
 The above equation can be rearranged such that 
 \begin{equation}
  w^* \ll \textit{O} \bigg(\frac{\sigma^*}{\rho^* u^{*2}_{\tau}} \bigg),
 \end{equation}
 which gives an upper bound on the friction velocity
 \begin{equation}
  u^*_{\tau} \ll \textit{O} \Bigg( \sqrt{ \frac{\sigma^*}{\rho^* w^*} } \Bigg).
 \end{equation}
  Therefore in terms of $Re_{\tau}$, 
  \begin{equation}
  \label{eq:retau_intstab}
  Re_{\tau} \ll \textit{O} \Bigg( \sqrt{ \frac{\rho^* \sigma^* \delta^{*2}}{\mu^{*2} w^*} } \Bigg),
 \end{equation}

%  or in wall units as,
%  \begin{equation}
%   \bigg( w^* \frac{\rho^* u^*_{\tau}}{\mu^*} \bigg) \ll \textit{O} \bigg(\frac{\sigma^*}{\mu^* u^{*}_{\tau}} \bigg),
%  \end{equation}
%  \begin{equation} 
%    w^+  \ll \textit{O} (Ca^{-1}),
%  \end{equation}
%  where $Ca$ is the ratio of viscous to capillary stresses known as the capillary number. If we multiply both sides by $Re_{\tau}$ we get the following: 
%  \begin{equation} 
%    Re_{\tau}  \ll \textit{O} \bigg[\frac{1}{w^+} \bigg( \frac{\rho^* \sigma^* \delta^*}{\mu^{*2}} \bigg) \bigg].
%  \end{equation}
 
 \subsection{Range of validity}
 For the following analysis, we take water as a reference fluid at standard conditions: $\rho^*=997$ $\mathrm{kg/m^3}$, $\mu^*=8.94 \times 10^{-4}$ $\mathrm{Pa \cdot s}$ and $\sigma^*=7.2 \times 10^{-2}$ $\mathrm{N/m}$. In our numerical simulation, the surface $S_q$ is approximately $1/90  $th of the channel half-height $\delta^*$ which gives $\delta^* \thicksim \textit{O}(10^{-4} \mathrm{m})$. From a design perspective, there exists a top down approach (the present study) where a surface is given and we estimate the range of validity of $Re_{\tau}$. In a bottom up approach, we can find the upper limit of the maximum allowable $w^*$ that sustains an interface given an $Re_{\tau}$.    
 \subsubsection{Top down approach} 
 In our numerical experiment, $w^*$ is of \textit{O}$(10 \mathrm{\mu m})$ therefore eq. (\ref{eq:pc_w}) yields $\Delta p^*_c < 5.7$ $\mathrm{kPa}$ suggesting that the interface can sustain pressures up to that value before the assumption of flat interfaces breaks down. Set the calculated pressure as the upper limit for $p^*_{rms}$ and substitute eq. (\ref{eq:prms}) in eq. (\ref{eq:pc_w}) to obtain 
 \begin{equation}
 u^*_{\tau} < \textit{O} \Bigg( 0.894\sqrt{\frac{\sigma^*}{\rho^* w^*}} \Bigg).
 \end{equation}
 In terms of $Re_{\tau}$ we have
 \begin{equation}
 \label{eq:retau_intdef}
  Re_{\tau} < \textit{O} \Bigg( 0.894\sqrt{\frac{\rho^* \sigma^* \delta^{*2}}{\mu^{*2} w^*} } \Bigg).
 \end{equation}
 Therefore, the assumption of a flat interface is valid for $Re_{\tau} < 270$. For the assumption of a stable interface, we use eq. (\ref{eq:retau_intstab}) to obtain $Re_{\tau} \ll 300$. It is clear from these results that the assumption of a flat interface puts a more stringent requirement on the allowable $Re_{\tau}$ which can also be seen by comparing eq. (\ref{eq:retau_intstab}) to eq. (\ref{eq:retau_intdef}).
 
\subsubsection{Bottom up approach}
Given a range of $Re_{\tau}$, we can estimate the largest allowable cavity width between roughness peaks. It is helpful to define terms in wall units such that eq. (\ref{eq:retau_intstab}) is rewritten as 
\begin{equation}
\label{eq:retau_wplus}
  Re_{\tau} \ll \textit{O} \bigg[ \frac{1}{w^+} \bigg( \frac{\rho^* \sigma^* \delta^*}{\mu^{*2}} \bigg) \bigg].
\end{equation}
Therefore $w^+$ for interface stability is given as
\begin{equation}
\label{eq:retau_wplus}
  w^+ \ll \textit{O} \bigg[ \frac{1}{Re_{\tau}} \bigg( \frac{\rho^* \sigma^* \delta^*}{\mu^{*2}} \bigg) \bigg].
\end{equation}
The maximum sustained capillary pressure can also be written in wall units, 
\begin{equation}
  \Delta p^+_{c} < \frac{0.8Ca^{-1}}{w^+}, 
\end{equation}
where $Ca=\mu^* u^*_{\tau}/\sigma^*$ is the ratio of viscous to capillary stresses known as the capillary number.  For small interface deflections, $w^+$  is therefore
\begin{equation}
 w^+  < \frac{0.8Ca^{-1}}{\Delta p^+_{c}}.
\end{equation}
An example of such bottom up calculation is given in table \ref{tab:bottomup}. %Note that the critical $w^+$ for failure are consistent with the scaling law $We^+ U^+_s L^+=\textit{O}(1)$ obtained by \citet{Seo2015} where $U^+_s$ is the slip velocity and $L^+$ is the equivalent cavity width $w^+$ in the structured roughness such as posts and grooves.

\begin{table}
\begin{center}
\begin{tabular}{lccccc}
 $Re_{\tau}$& $w^+_{IS}$($\ll$) & $Ca$ & $\Delta p^+_c$($\ll$) & $w^+_{SD}$($<$)\\
 \hline
 $180$	& $50$ &  $1.9 \times 10^{-2}$ & $0.8$ & $40$&\\
 $395$	& $23$ &  $4.3 \times 10^{-2}$ & $0.81$ & $18.4$&\\
 $1000$ &  $9$ &  $1.4 \times 10^{-3}$ & $64$ & $7.2$&\\
 $10,000$ & $0.9$ &  $1.4 \times 10^{-4}$ & $6.4 \times 10^3$ & $0.72$&\\
 $100,000$ & $0.09$ &  $1.4 \times 10^{-5}$ & $6.4 \times 10^5$ & $0.072$&\\
\end{tabular}
\caption{Maximum allowable average cavity widths in wall-units $w^+$ for a range of $Re_{\tau}$. $w^+_{IS}$ and $w^+_{SD}$ represent the average cavity width satisfying the interface stability and small deflection conditions respectively. The maximum allowable capillary pressure $\Delta p^+_{c}$ in wall-units is also shown.}
\label{tab:bottomup}
\end{center}
\end{table}
 
 It is important to note that for this analysis, $w^+$ represents an average cavity width of the random rough surface. It does not say anything about the largest value that is prone to failure first. As the height of the interface increases, $w^+$ increases and more asperities are covered up. Therefore the most realistic numerical simulations would be with an interface below $S_q$ of the roughness where typically the gas fraction $\phi_g < 0.6$. As mentioned earlier, the goal behind our numerical experiment was to investigate the effect of the interface height on the drag reducing properties of SHS. Also it is worth mentioning that although the above analysis gives the upper limit of allowable $Re_{\tau}$ for a given $w^+$, it is known from the literature that adding hierarchical structures to the same size posts can resist destabilisation. Hence, for the same geometry, the maximum allowable $Re_{\tau}$ can be larger due to the added multiscale roughness. 

\section{Surface statistics}
\label{appB}
\begin{figure}
\centering{
\includegraphics[width=70mm]{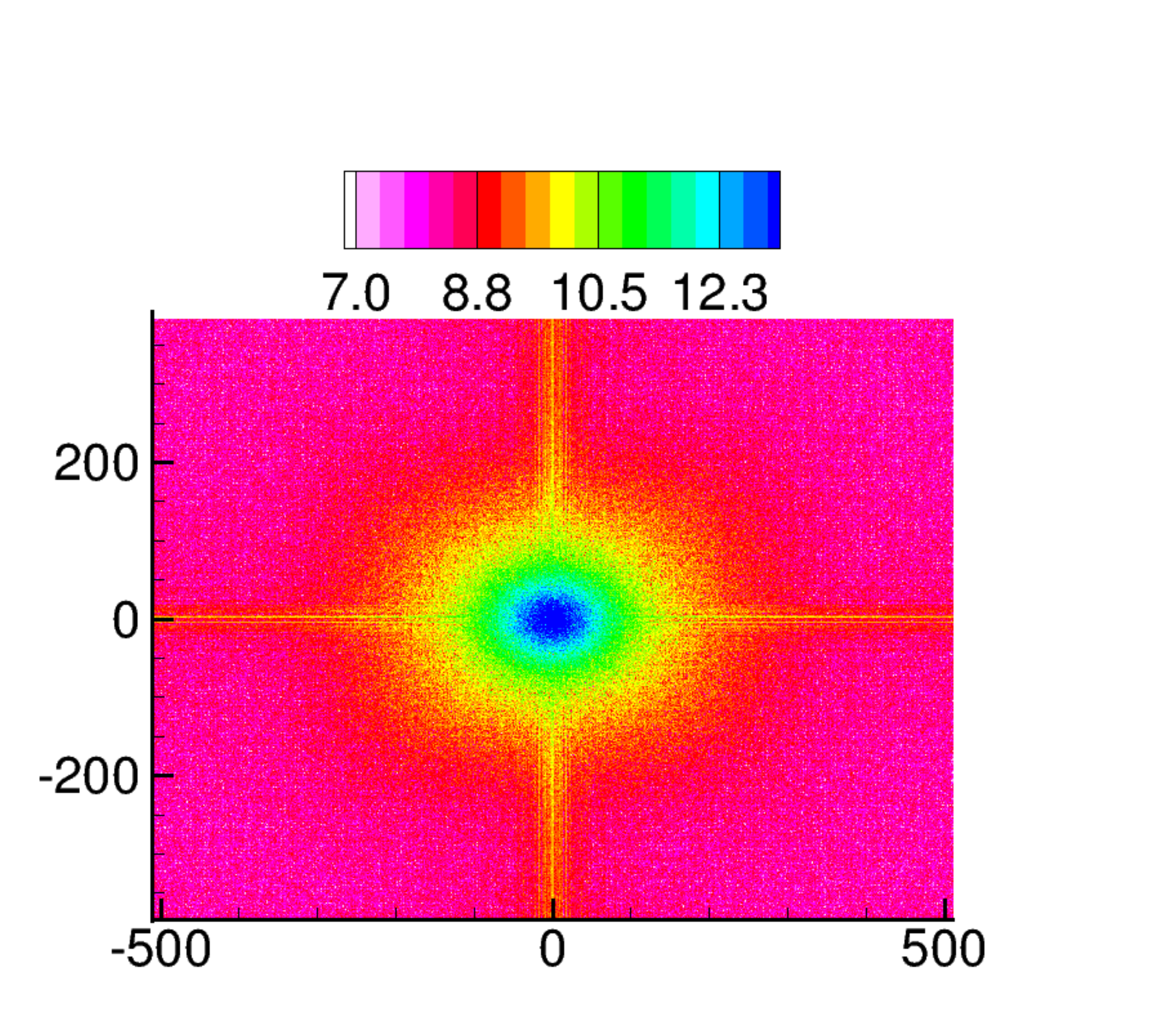}
\put(-110,0){$k_x$}
\put(-200,70){$k_z$}
}
\caption{Power spectral density (PSD) contour plot where $k_x$ and $k_z$ are the wavenumbers in the streamwise and spanwise directions respectively.}
\label{fig:pdf_psd}
\end{figure}

\begin{figure}
\centering{
\includegraphics[width=110mm]{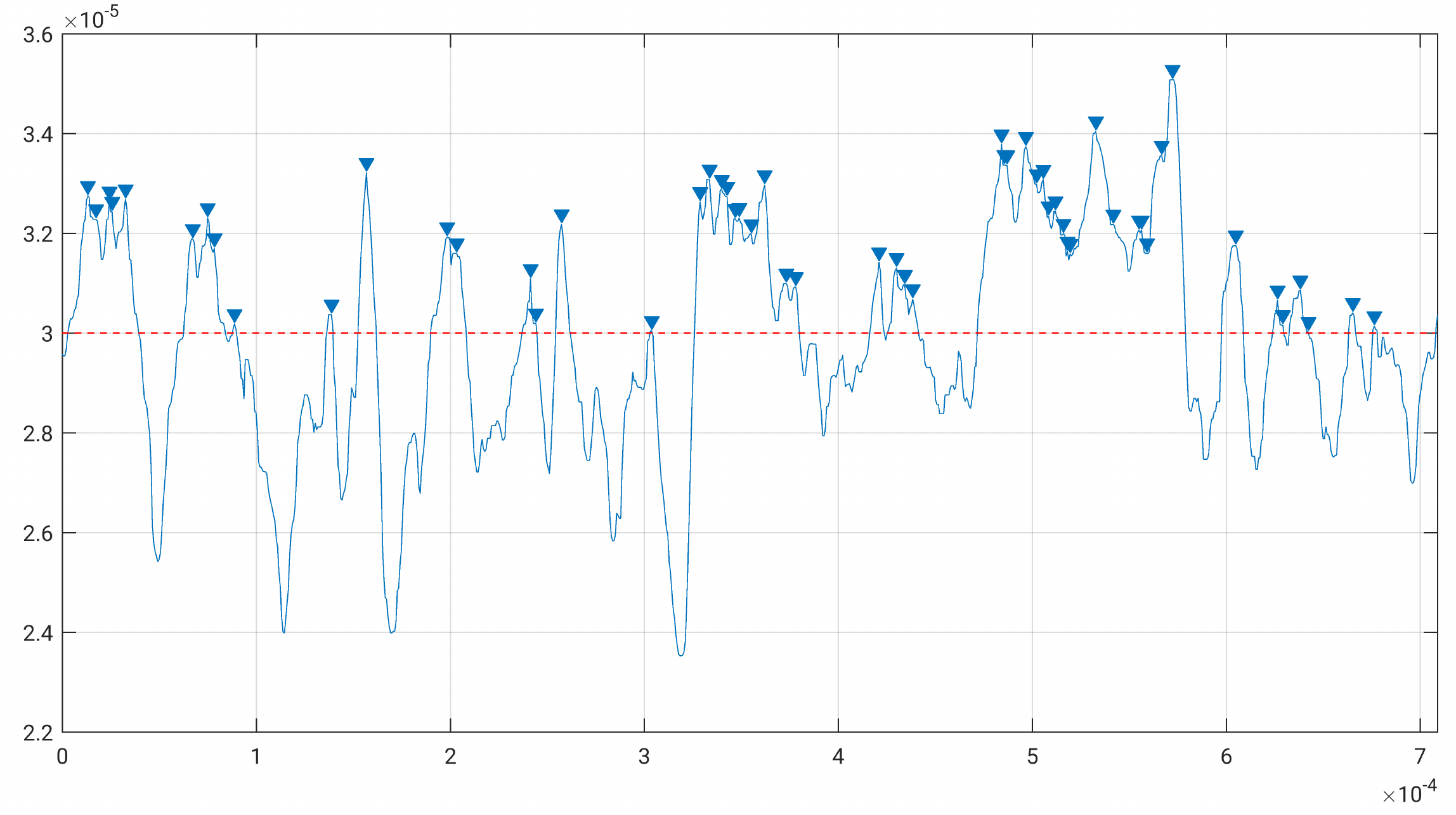}
\put(-155,5){$x$}
\put(-320,95){$h$}
}
\caption{A 2-D slice of a random spanwise location of the original surface roughness (solid blue line) illustrating the identified peaks (solid blue triangles) using the arithmetic mean elevation (dashed red line) as a threshold. The mean distane between peaks is used to obtain the average roughness gap $L$.}
\label{fig:peak_count}
\end{figure}

The power spectral density (PSD) of the surface height obtained from the scan is shown in figure \ref{fig:pdf_psd}, where  the visible cross-pattern is due to the aliasing effects at the non-periodic boundaries of the unfiltered surface.
The original surface statistics are verified with the experimental values provided through private communication. The surface statistics are reported in table \ref{tab:stats}. Table \ref{tab:scaled_masked_stats} compares the values of the original surface statistics for the scaled turbulent channel roughness with the step-wise masked representation as used by the solver. Figure \ref{fig:peak_count} illustrates the peaks as they are identified given a threshold. 
\begin{table}
 \begin{center}
\begin{tabular}{l|{c}|{c}|{c}|r}
Parameter & Description & Formula & Value \\
\hline
$S_{a}$      & Average Roughness Height & $\frac{1}{N_{x}N_{z}}\sum\limits_{k=1}^{N_{z}} \sum\limits_{i=1}^{N_x} |h_{i,k}|$ &  1.59 ${\mu}$m \\
%\hline
$S_{q}$      & RMS Roughness Height & $[\frac{1}{N_{x}N_{z}}\sum\limits_{k=1}^{N_{z}} \sum\limits_{i=1}^{N_x} h_{i,k}^2]^{1/2}$  & 2.03 ${\mu}$m \\
%\hline
$S_{v}$      & Maximum Valley Depth & $min(h_{i,k})$ &  -10.0 ${\mu}$m \\
%\hline
$S_{p}$      & Maximum Peak Height &  $max(h_{i,k})$ &  8.31 ${\mu}$m \\
%\hline
$S_{z,max}$      & Maximum Peak to Valley Height& $max(h_{i,k})-min(h_{i,k})$ &  18.38 ${\mu}$m \\
%\hline
$S_{z,5\times5}$  & Mean Peak to Valley Height  & $\frac{1}{25}\sum\limits_{i=1}^{5\times5} S_{z,i}$ & 12.75 ${\mu}$m \\
%\hline
$S_{sk}$     & Skewness  &  $\frac{1}{N_{x}N_{z}S_{q}^3}\sum\limits_{k=1}^{N_{z}} \sum\limits_{i=1}^{N_x} h_{i,k}^3$ &  -0.32 \\
%\hline
$S_{ku}$     & Kurtosis (Flatness) & $\frac{1}{N_{x}N_{z}S_{q}^4}\sum\limits_{k=1}^{N_{z}} \sum\limits_{i=1}^{N_x} h_{i,k}^4$  &  3.47\\
%\hline
$S_{dq}$     & RMS Slope of Roughness & $[\frac{1}{N_{x}N_{z}} \sum\limits_{k=1}^{N_{z}} \sum\limits_{i=1}^{N_x} [{\Delta_{i}^2 + \Delta_{k}^2}]]^{1/2} $  &  0.547\\
%\hline
$S_{w}$      & Wenzel Roughness & $\frac{1}{N_{x}N_{z}} \sum\limits_{k=1}^{N_{z}} \sum\limits_{i=1}^{N_x} [{1+\Delta_{i}^2 + \Delta_{k}^2}]^{1/2} $ &  1.129\\
%\hline
$S_{m}$	     & Mean Peak Spacing & $\frac{1}{Np-1} \sum \limits_{k=1}^{Np-1} (P_{k+1}-P_k)$ & 10.64 ${\mu}$m \\
%\hline
$\Delta_i$     & Directional Derivative & ${\partial {h}}/{\partial {x_i}}$ & $\frac{1}{2dx_i}(h_{i+1}-h_{i-1})$ \\
\hline
\end{tabular}
\caption{Statistical parameters of the scanned surface used in the present work. $N_x$ and $N_z$ are the number of points in the streamwise and spanwise directions respectively. $N_p$ denotes the total number of peaks, $P_k$ the peak location and $h$ the roughness height.}
\label{tab:stats}
\end{center}
\end{table}

\begin{table}
\begin{center}
\begin{tabular}{l|{c}|{c}|{c}|r}
Parameter & Original Surface & Step-wise Surface & Error (\%) \\
\hline
$S_{a}$      & $7.0585 \times 10^{-3}$ & $7.0613 \times 10^{-3}$ &  0.04 \\
%\hline
$S_{q}$      &  $8.972 \times 10^{-3}$ & $9.0211 \times 10^{-3}$  & 0.547  \\
%\hline
$S_{v}$      &  $-4.256 \times 10^{-2}$ & $-4.243 \times 10^{-2}$ &  0.305  \\
%\hline
$S_{p}$      &  $3.626 \times 10^{-2}$ & $3.724 \times 10^{-2}$ &  2.7 \\
%\hline
$S_{z,max}$  &  $7.883 \times 10^{-2}$ & $7.967 \times 10^{-2}$ &  1.065  \\
%\hline
$S_{z,5\times5}$  & $6.1725 \times 10^{-2}$  & $6.22172 \times 10^{-2}$ & 0.785 \\
%\hline
$S_{sk}$     & $-0.3347$ & $-0.3234$ &  3.37 \\
%\hline
$S_{ku}$     & $3.494$ & $3.484$  & 0.286 \\
%\hline
$S_{dq}$     & $0.3985$ &  $0.4059$ &  1.856 \\
%\hline
$S_{w}$      & $1.07312$ & $1.0757$ &  0.24 \\
\hline
\end{tabular}
\caption{Comparison of the statistical parameters of the original surface scaled for the turbulent channel flow with the step-wise distribution of the surface used in the present work.}
\label{tab:scaled_masked_stats}
\end{center}
\end{table}

\bibliographystyle{jfm}

\bibliography{final_submit}

\end{document}